\documentclass[superscriptaddress,showpacs,aps,twocolumn]{revtex4-1}
\usepackage{amsmath}
\usepackage{graphicx}
\usepackage{dcolumn}
\usepackage{bm}
\usepackage{epsfig}
\usepackage{color}
\usepackage{tikz}
\usepackage{comment}
\newcommand \beq{\begin{eqnarray}}
\newcommand \eeq{\end{eqnarray}}
\newcommand \bea{\begin{eqnarray}}
\newcommand \eea{\end{eqnarray}}

\newcommand \kvec{{\bf k}}

\newcommand \qvec{{\bf q}}

\newcommand\rvec{{\bf r}}

\newcommand\Gvec{{\bf G}}

\def\simge{\mathrel{%
       \rlap{\raise 0.511ex \hbox{$>$}}{\lower 0.511ex \hbox{$\sim$}}}}
\def\simle{\mathrel{
       \rlap{\raise 0.511ex \hbox{$<$}}{\lower 0.511ex \hbox{$\sim$}}}}

\begin{document}
    \title{Electronic band gaps from Quantum Monte Carlo methods}
\author{Yubo Yang}
\affiliation{Department of Physics, University of Illinois, Urbana, Illinois 61801, USA}
\author{Vitaly Gorelov}
\affiliation{Maison de la Simulation, CEA, CNRS, Univ. Paris-Sud, UVSQ, Universit{\'e} Paris-Saclay, 91191 Gif-sur-Yvette, France}
\author{Carlo Pierleoni}
\affiliation{Maison de la Simulation, CEA, CNRS, Univ. Paris-Sud, UVSQ, Universit{\'e} Paris-Saclay, 91191 Gif-sur-Yvette, France}
\affiliation{Department of Physical and Chemical Sciences, University of L'Aquila, Via Vetoio 10, I-67010 L'Aquila, Italy}
\author{David M. Ceperley}
\affiliation{Department of Physics, University of Illinois, Urbana, Illinois 61801, USA}
\author{Markus Holzmann} 
\affiliation{Univ. Grenoble Alpes, CNRS, LPMMC, 3800 Grenoble, France}
\affiliation{Institut Laue Langevin, BP 156, F-38042 Grenoble Cedex 9, France}

\date{\today}

\begin{abstract}
We develop a method for calculating the fundamental electronic gap of semiconductors and insulators using grand canonical Quantum Monte Carlo simulations.
We discuss the origin of the bias introduced by supercell calculations of finite size and show
how to correct the leading and subleading finite size errors either based on observables accessible in the finite-sized simulations or from DFT calculations.
Our procedure is applied to solid molecular hydrogen and compared to experiment for carbon and silicon crystals.
\end{abstract}


\maketitle

\section{Introduction}
Insulator and semiconductors are characterized by a non-vanishing {\it fundamental gap}
\cite{book}, 
defined in terms of the ground state energies of a system of fixed ions as the number of electrons is varied:
\beq
\Delta_{N_e} = E_0(N_e+1)+ E_0(N_e-1)- 2 E_0(N_e)
\label{gap}
\eeq
where $E_0(N_e)$ is the ground state energy of an $N_e$ electron system.

Within density functional theory (DFT), it is common to interpret the eigenvalues of the
Kohn-Sham equations as excitation energies, the gap being the minimum excitation energy. However, the resulting band gap
within the local density approximation (LDA) is typically found too small \cite{Perdew}.
This qualitative failure can be alleviated
either by hybrid functionals or by adding corrections based on GW many-body perturbation theory,
although the precise value depends on the underlying functional and approximation scheme involved \cite{book}.
In principle, the fundamental gap can be calculated from any method for ground state energies
based on the above formula.
High precision methods for correlation energies as, for example, provided
by quantum Monte Carlo (QMC) \cite{rev1,rev2,rev3,Ruggeri18} or coupled cluster methods
\cite{Shepherd13,Gruber18} can be used.  
In this paper, we propose a new method for accurate calculations of the fundamental gap
within explicitly correlated methods and demonstrate its use with fixed-node Diffusion Monte Carlo (DMC) benchmark studies on solid H$_2$,  C, and Si.

Methods based on correlated many-body wave functions
are usually applied to finite sized systems, e. g.
limited to supercells containing only few unit cells.
QMC calculations of single particle excitations for adding and removing electrons 
\cite{Kent98,Kent99,Mitra15,McMinis15} crucially rely on the imposed extrapolation law
(e.g. $1/L$ in \cite{McMinis15} opposed to $1/L^3$ in \cite{Mitra15} where $L$ denotes the linear extension of the supercell). This introduces considerable uncertainty in the results.
Heuristically, single particle excitations are expected to converge slowly for electronic systems,
inversely proportional to $L$,
due to the interaction of charges across the periodic boundaries \cite{Payne,Engel}.
Extrapolations with respect to the size of the supercells are then
essential to obtain reliable values of the gap in the thermodynamic limit.

Most of the QMC calculations 
\cite{Ceperley87,Mitas94,Williamson98,Towler2000,Kolorenc08,Ma13,Wagner14,Yu15,Zheng18,Frank19} 
have therefore addressed charge neutral, particle-hole excitations,
where faster convergence with respect to the size of the supercell is expected.
Although the comparison with experiment is appealing \cite{rev3}, 
a later, more extended DMC study \cite{Hunt} of simple semiconductor materials
with larger supercells observed a $1/L$ dependance of the gap on the size of the supercell
for both, charged single particle and charge-neutral particle-hole excitations.
In addition, 
fixed-node energy differences are not constrained to be upper bounds for particle-hole excitations
\cite{Foulkes99} since orthogonality to the ground state cannot be strictly guaranteed.
Furthermore, all QMC calculations so far have addressed excitations at selected
symmetry points contained inside the supercell of the simulation.
The fundamental gap was then estimated indirectly by introducing 
a ``scissor operator'' \cite{Azadi2017} which assumes a rigid shift of
the underlying DFT band structure over the whole Brillouin zone.

In this paper, we show that twisted boundary conditions within the grand canonical ensemble
can be used
to  determine the fundamental gap from QMC without relying on the ``scissor'' approximation.
We prove that to leading order, finite size effects due to two-body correlations are of order $1/L$,
and are related to the dielectric constant of the material. Such effects
can be understood and corrected for by using
the long wavelength properties of the electronic
structure factor. 
For that, we extend the approach described in Ref.~\cite{fse,finitesize} 
which discusses the correction
of finite size effects on the ground state energy
based on information contained in the static correlation functions of the finite system.
Using the static structure factor from simulation, it is possible to obtain estimates of finite size corrections 
for the band gap, and its asymptotic functional form 
without the need for
explicit studies at different sizes or referring to DFT or to experimental information
external to the QMC calculation.

The paper is organized as follows. In section~\ref{sec:gctabc}, we describe the main ideas behind our new band gap method based on the grand canonical ensemble. In section~\ref{sec:fse}, we derive finite size corrections to energy differences based on an explicit many-body wavefunction and exact diagrammatic relations. In section~\ref{sec:methods}, we describe the computational methods used to calculate the fundamental gap. In section~\ref{sec:results}, we show results for H$_2$, C, and Si crystals
and compare with available experimental values of the gap in section~\ref{sec:compare}.
Finally in section~\ref{sec:conclude}, we summarize general features of the method
 and outline possible extensions and applications.

\section{Grand-Canonical twist averaged boundary condition (GCTABC)\label{sec:gctabc}}

In the following, we consider $N_e$ electrons in a perfect crystal,
neglecting both zero point  and thermal motion of the ions.
A uniform background charge (depending on $N_e$) is added to assure global 
charge neutrality when adding or subtracting electrons to a charge neutral system.
The fundamental gap, Eq.~(\ref{gap}), is unaffected by the background energy, 
because the background
charge needed when adding an electron cancels against the one needed when removing an electron.
Periodic boundary conditions of the charge densities are used to eliminate surface effects.  

The energetic cost of adding an electron to the system at fixed volume, $V=L^3$, defines the
chemical potential
\beq
\mu_{N_e}^+= E_0(N_e+1)-E_0(N_e).
\eeq
A non-vanishing gap implies a discontinuity in the chemical potential from Eq.~(\ref{gap}).

It is convenient to work in the grand-canonical ensemble. There, the
chemical potential $\mu$ is treated as an independent variable and we minimize $E_0(N_e)-\mu N_e$ with respect to $N_e$ at zero temperature and fixed volume. Insulators then represent an incompressible electronic state; for values of $\mu$
within the gap, $\partial N_e/\partial \mu=0$.

To reduce finite size effects, we employ twisted boundary conditions on many-body wave function.
as an electron is moved across the supercell, e.g. by moving an electron a distance equal to the size of the box in the $x$ direction:
\beq
\Psi (\rvec_1+L_x\hat{x}) = e^{i\theta_x} \Psi(\rvec_1).
\eeq
the phase of the many-body wavefunction changes by $\theta$.
The ground state energy  then depends on the twist angle, $E_0(N_e,\theta)$. 
Twist averaging
can significantly accelerate the convergence 
to the thermodynamic limit \cite{Lin01}.
Within the grand-canonical ensemble \cite{fse,finitesize}, the optimal number
of electrons ${\bar N}_e(\theta)$ will depend  on $\theta$ for given chemical potential $\mu$. To fix nomenclature, we define the
\emph{mean electronic density}
\begin{equation} \label{eq:ne}
n_e(\mu)=(M_\theta V)^{-1}\sum_\theta  {\bar N}_e(\theta),
\end{equation}
and the \emph{ground-state energy density}
\begin{equation} \label{eq:e0}
e_0(\mu)=(M_\theta V)^{-1} \sum_\theta E_0({\bar N}_e(\theta),\theta).
\end{equation}
$n_e$ is determined by minimizing the free energy density
\begin{equation}
f=\frac{1}{M_\theta V} \sum_\theta \min_{N_e} \left[ E_0(N_e,\theta) - \mu N_e \right],
\end{equation}
where the sum is over a uniform grid containing $M_\theta$ twist angles.
For any single electron theory the electronic density $n_e(\mu)$
and the ground state energy density $e_0(\mu)$ coincide exactly with the corresponding
thermodynamic limit values for a sufficiently large value of $M_\theta$, e.g. when the sum over twists becomes an integral over the 
Brillouin zone.  Size effects remaining after twist averaging are due to electron-electron correlations.

\begin{figure*}
\begin{minipage}[b]{\columnwidth}
\includegraphics[width=\columnwidth]{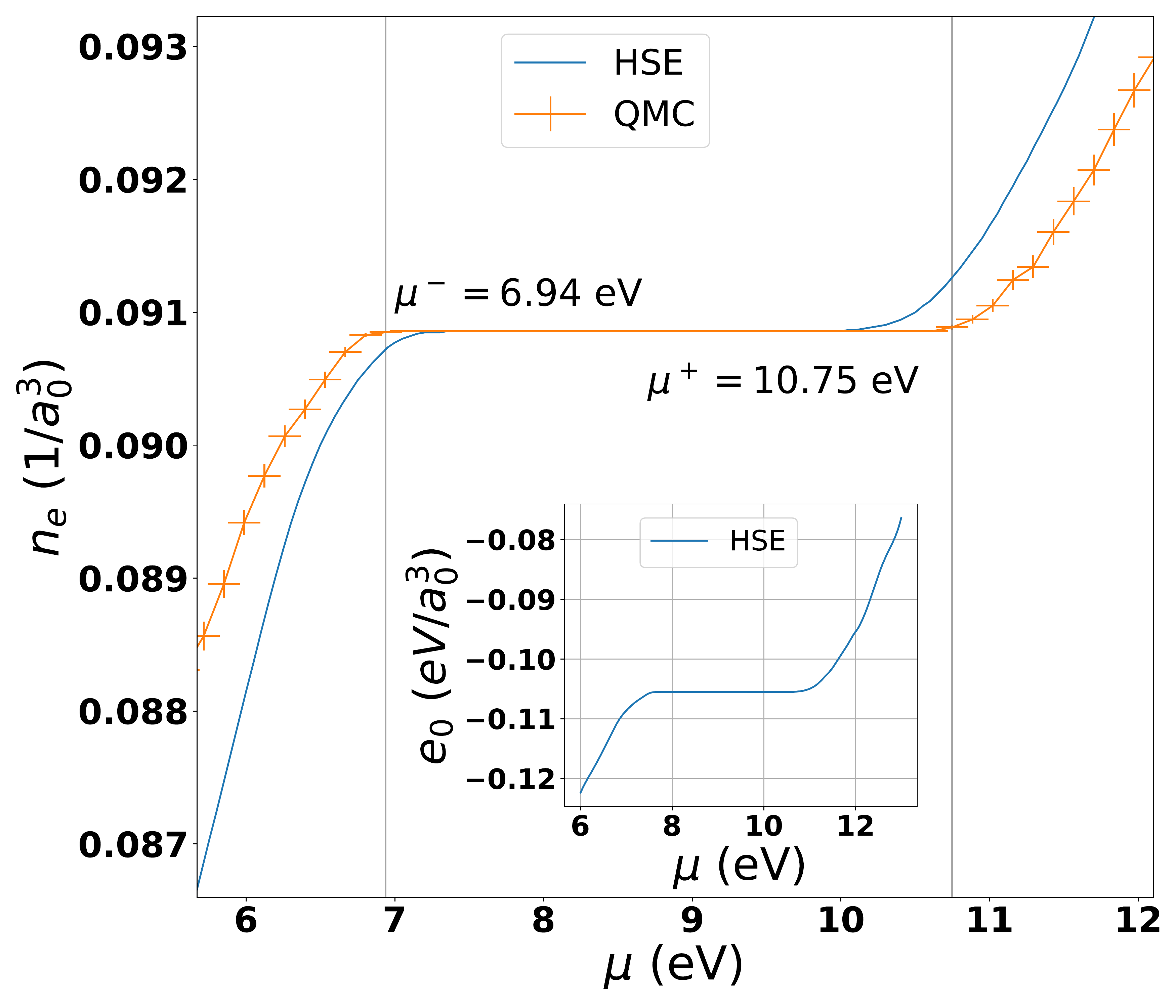}
(a) 
\end{minipage}
\begin{minipage}[b]{\columnwidth}
\includegraphics[width=\columnwidth]{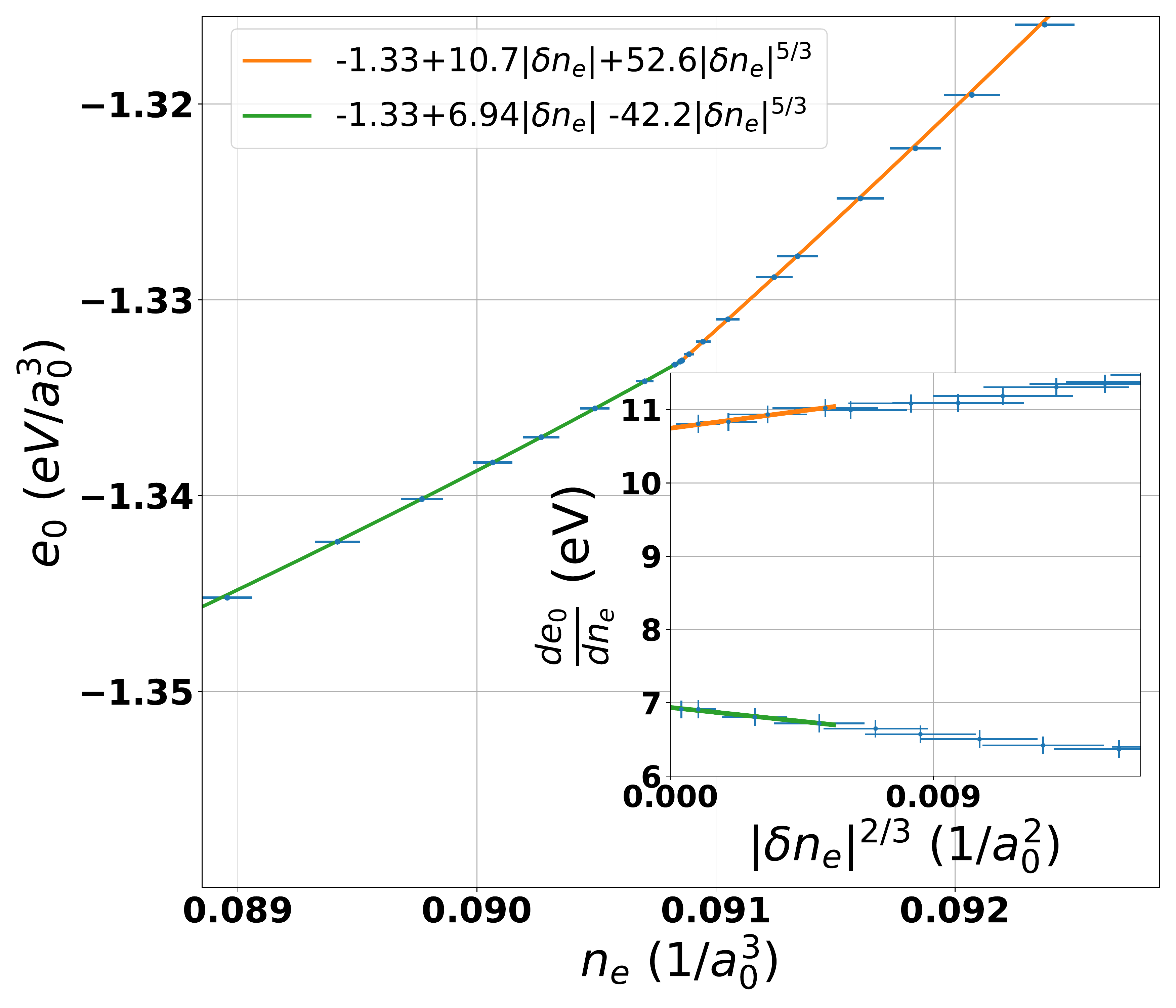}
(b) 
\end{minipage}
\caption{GCTABC analyses of the C2/c-24 structure of solid hydrogen at $r_s=1.38$ (234GPa). (a) 
the electron density $n_e$ as a function of the chemical potential $\mu$ obtained from HSE functional
in comparison to QMC, the inset illustrates the energy density as a function of $\mu$ from HSE functional.
(b) energy density $e_0$ as a function of $n_e$ using QMC, the inset shows the derivative discontinuity where $\delta n_e$ is the change of the electronic density with respect to the insulating state.
Size corrections as discussed in the text are included.
 \label{fig:H_C2CP250}}
\end{figure*}

Figure \ref{fig:H_C2CP250} illustrates $e_0(\mu)$ and $n_e(\mu)$ for solid molecular hydrogen, computed from HSE functional and from QMC (see section \ref{sec:methods} for details). The value of the  band
gap can be directly extracted from the width of the incompressible region.
Alternatively, if we
eliminate $\mu$ in favor of $n_e$, and plot $e_0$ as a function of $n_e$,
the fundamental gap is obtained by the discontinuity of the derivative, according to 
Eq.~(\ref{gap}).

The definition of the fundamental gap can apply to different symmetry sectors.
For a perfect crystal, the total momentum of the electrons modulo 
reciprocal lattice vectors is conserved, i.e. the crystal momentum. By requiring the total crystal momentum of the electrons to be fixed,
e.g. using Bloch type orbitals in the Slater-determinant,
the full band structure in the Brillouin zone can be mapped out.
For a spin-independent Hamiltonian, one can also impose the total spin 
to determine the fundamental gap in each spin sector. 
In practice, the charge gap in the spinless sector can
be determined by  adding or removing pairs of electrons. The extensions of our definitions and formulas to this case are straightforward,
e.g. $\Delta_{N_e}=[E_0(N_e+2)+E_0(N_e-2)-2E_0(N_e)]/2$. We follow this procedure of spin neutral excitations in the remainder of the paper.

\section{Finite size effects\label{sec:fse}}

\subsection{Potential energy}

A key quantity in understanding size effects is the long wave length behavior of the static structure factor,
$S_{N_e}(\kvec)=\langle \rho_{-\kvec} \rho_{\kvec} \rangle/N_e$ where $\rho_\kvec=\sum_j e^{i \kvec \cdot \rvec_j}$ is the Fourier transform of the instantaneous electron density.
The structure factor for a homogenous system obeys the bound \cite{Ceperley87,book}
\beq
S_{N_e}(k) \le \frac{\hbar k^2}{2m \omega_p}\left( 1 - \frac{1}{\epsilon_k} \right)^{1/2},
\label{davidbound}
\eeq
where $\omega_p=4 \pi \hbar^2 e^2 n_e/m$ is the plasma frequency and $\epsilon_k$ the static dielectric constant for wavevector $k$ (To simplify the notations, we will supress the dependence on the wave vector in the following). This inequality is derived by applying the plasmon-pole approximation to the sum rules of the dynamic structure factor $S(k, \omega)$. It implies that the structure factor must vanish quadratically as
$k \rightarrow 0$ \cite{Nozieres}.
Equality will be obtained if $S(k,\omega)$ reduces to a single delta function at small $k$.
The $1/N_e$ finite-size corrections of the energy per electron is a direct consequence of 
this behavior of $S_{N_e}(k)$ \cite{fse}. However, these leading order corrections are not sufficient for excitation energies, since the energy gap is of the same order as finite size corrections to the total energy.

As we will show below, the key to understanding size effects of energy differences is encoded in the change of $S_{N_e}(k)$ as electrons are added or removed. In particular, the limiting behavior of $S_{N_e\pm 1}(k)$ as $k\rightarrow0$ will provide the dominant finite size correction.

For concreteness, we will assume a Slater-Jastrow form for the ground state wave function
$\Psi_0 \propto D \exp[-U]$. The determinant, $D$, is built out of Bloch orbitals, $\phi_{\qvec n}(\rvec)$ with $\qvec$ inside the first Brillouin zone, $n$ is the band index, and $U$ is a general, symmetric $n$-body correlation factor \cite{nbody}. For simplicity we assume it is two body: $U=\sum_{i<j} u(r_i,r_j)$.  
Let us consider the action of $e^{i \kvec \cdot \rvec_j}$ on a single particle orbital
$\phi_{\qvec n}(\rvec_j)$ in the Slater determinant of the ground state.
In the limit of small $\kvec$,  
this can be approximately written as $ \phi_{\qvec+\kvec n}(\rvec_j)$.
Expanding the determinant in terms of its cofactors $\frac{\delta D}{\delta \phi_{\qvec n}(\rvec_j)}$ and making the excitation we have
\beq
\rho_\kvec \Psi_0
\propto \sum_j \sum_{\qvec,n} \frac{\delta D}{\delta \phi_{\qvec n}(\rvec_j)} e^{i \kvec \cdot \rvec_j}
 \phi_{\qvec n}(\rvec_j) e^{-U}.
\eeq
and the resulting determinant after summation over $j$
vanishes for small $k$ if the Bloch orbital $(\qvec+\kvec,n)$
is already occupied in the ground state determinant.
Considering $N_e \pm 1$ electron wave functions, $\Psi_0(N_e \pm 1; \pm \qvec,  m)$ where $N_e$ corresponds to the insulating state
with fully occupied bands in the Slater determinant, and $\qvec m$ denotes the additional particle/
hole orbital, we get
\beq
\lim_{k \to 0} \rho_\kvec \Psi_0(N_e \pm 1;\qvec, m)  
\sim \pm \Psi_0(N_e \pm 1;\qvec + \kvec, m)
\label{limitk}
\eeq
for $k \ne 0$ where different sign for particle or hole excitations on the r.h.s. is chosen to match 
the most common sign convention, e.g. of Ref.~\cite{Kohn58}. The limit $k \to 0$ is discontinuous since 
$\rho_{\kvec=0} \Psi_0(N_e \pm 1;\qvec, m)\equiv (N_e \pm 1)\Psi_0(N_e \pm 1;\qvec, m)$.

Kohn \cite{Kohn57,Kohn58} has pointed out that in the insulating state the matrix elements
\bea
\lim_{\qvec' \to \qvec} \langle \Psi_0(N_e \pm 1;\qvec, m) | \rho_{\qvec-\qvec'} |
\Psi_0(N_e \pm 1;\qvec', m) \rangle = \pm \frac1{\epsilon}
\nonumber
\label{kappastar}
\\
\eea
approach the inverse dielectric constant, $\epsilon^{-1}$, up to a sign.

Substituting Eq.~(\ref{limitk}) into Eq.~(\ref{kappastar}), suggests 
the following finite size behavior of the static structure factor of insulators
\beq
\lim_{k \to 0} S_k^{\pm}
= \alpha_{\pm} +{\cal O}(k^2),
\label{sqeps}
\\
S_k^{\pm} \equiv (N_e \pm 1) S_{N_e \pm 1}(k) - N_eS_{N_e}(k),
\eeq
where $\alpha_\pm$ is proportional to $\epsilon^{-1}$. However, $\alpha_\pm$ in general differs from $\epsilon^{-1}$ unless Eq.~(\ref{limitk}) is an exact equality.

Figure \ref{fig:c-si-skpm} shows the behavior of $S_k^{\pm}$  
for carbon and silicon crystals. Note that these functions extrapolate to a nonzero value as $k \rightarrow 0$.

The long wavelength behavior of the structure factor, Eq.~(\ref{sqeps}), then gives rise to size
corrections to excitation energies through the potential energy term
\beq
\left[ \int \frac{d^3 k}{(2 \pi)^3}  -   \frac{1}{V} \!\sum_{\kvec \ne 0} \right]
 \frac{v_k}{2} S_k^{\pm}
\simeq  \alpha_{\pm} \frac{|v_M|}{2}, 
\label{eq:vk-skeps}
\eeq
where we have defined the Madelung constant as
\beq
v_M= 
\left[ \frac{1}{V} \sum_{\kvec \ne 0} - \int \frac{d^3 k}{(2 \pi)^3}\right] v_k
\sim L^{-1} \sim N_e^{-1/3}.
\label{vMdef}
\eeq
For the Coulomb potential, $v_M$ is proportional to $L^{-1}$, the inverse linear extension of the simulation cell. The negative proportionality constant  depends on the boundary conditions, e.g. cell geometry, and can be calculated by the Ewald image technique \cite{Ewald}.

\subsection{Kinetic energy}

Following Ref.~\cite{finitesize}, we now discuss the kinetic energy
contribution $\hbar^2 [\nabla U]^2/2m$ which arises from electron correlation.
For a two-body Jastrow $U=\sum_\kvec u_k \rho_\kvec  \rho_{-\kvec}/2 V$, and we are only
interested in the long-wave length limit, $k \to 0$, of the electron-electron 
correlation, with wave vectors smaller than the reciprocal lattice vectors
of the crystal, $\Gvec$.
Isolating the singular contributions involving $\rho_{k=0} \equiv N_e$ 
in the spirit of the rotating (random) phase 
approximation (RPA) we have
\bea
\left\langle \left[\nabla U \right]^2 \right\rangle
&=& - \frac{1}{V^2} \sum_{\kvec \ne 0, \kvec' \ne 0}
(\kvec \cdot \kvec') u_k u_{k'} \langle \rho_{\kvec+\kvec'} \rho_{-\kvec} \rho_{-\kvec'} \rangle
\nonumber \\
&\simeq&
\frac{1}{V^2} \sum_{\kvec \ne 0}
 N_e
k^2 u_k^2 \langle \rho_\kvec \rho_{-\kvec} \rangle.
\eea
Therefore,
for systems with explicit long-range correlations $u_k \sim k^{-2}$, the kinetic energy will 
contribute also to the leading order size corrections with 
\begin{eqnarray}\label{eq:uk-sqeps}
\left[ \int \frac{d^3 k}{(2 \pi)^3}  -  \frac{1}{V} \sum_{\kvec \ne 0} \right]
 \frac{n_e \hbar^2 k^2u_k^2}{2m}  S_k^\pm
\simeq  \alpha_\pm c \frac{|v_M|}{2},
\end{eqnarray}
where $c= \lim_{ k \to 0} n_e \hbar^2 k^2u_k^2/(mv_k)$ is approximately given by the ratio of the $1/N_e$ finite-size
corrections of the kinetic to potential energy of the ground state energy per particle due to
two-body correlations \cite{finitesize}.

\subsection{Total gap corrections from Coulomb singularity}

Up to now, we have shown how the long range behavior of the structure factor and Jastrow factor can give rise
to a $1/L$ correction to the excitation gap with a proportionality factor determined
by the structure factor changes.
In the following, we will further demonstrate that, given the trial wave functions coincide with
the exact ground state wave function for $N_e$ and $N_e \pm 1$ electrons,
this proportionality factor is indeed given by the dielectric constant 
\beq
\Delta_\infty - \Delta_V = \frac{|v_M|}{\epsilon}+ {\cal O}\left( \frac1V \right),
\label{gapepsilon}
\eeq
as phenomenologicaly 
assumed in previous work \cite{Engel,Hunt}.

We prove this by an independent argument based on commutation relations.
Let us denote the exact insulating ground state of the $N_e$ electron system as $|\Psi_0^{N_e} \rangle$, its energy as $E_0^{N_e}$, and the exact excited state of the $N_e\pm 1$ electron system as $|\Psi_\kvec^{N_e\pm 1} \rangle$ with energy $E_\kvec^{N_e\pm 1}$; $\kvec$ indicates that the additional/subtracted electron adds/subtracts the crystal momentum $\kvec$.
We have
\beq
E_\kvec^{N_e+ 1} - E_0^{N_e}
= \frac{ \langle \Psi_\kvec^{N_e+1} | \left[ H,a_\kvec^\dagger \right] |\Psi_0^{N_e} \rangle}
{\langle \Psi_\kvec^{N_e+1} | a_\kvec^\dagger |\Psi_0^{N_e} \rangle}
\eeq
for particle and
\beq
E_\kvec^{N_e- 1} - E_0^{N_e}
= \frac{ \langle \Psi_\kvec^{N_e-1} | \left[ H,a_\kvec \right] |\Psi_0^{N_e} \rangle}
{\langle \Psi_\kvec^{N_e-1} | a_\kvec |\Psi_0^{N_e} \rangle}
\eeq
for hole excitations.
In second quantization, the Hamiltonian, 
$H=T+V_{ee}$, is given by
\bea
T &=& \sum_\kvec \left[ \frac{\hbar k^2}{2m} a_\kvec^\dagger a_\kvec 
+ \sum_\Gvec u(\Gvec) a_{\kvec+\Gvec}^\dagger a_\kvec \right],
\\
V_{ee}&=&\frac{1}{2 V} \sum_{\qvec \ne 0}
v_q \left[\rho_\qvec \rho_{-\qvec} - N_e\right],
\eea
where $a_\kvec$ is the annihilation operator for plane wave states of wave vector $\kvec$,
$u(\Gvec)$ the periodic crystal potential, and $v_q$ is the Coulomb potential
between electrons, $\rho_\qvec= \sum_\kvec a_{\kvec+\qvec}^\dagger a_\kvec$,
and $N_e=\sum_k a_\kvec^\dagger a_\kvec$.

The commutator involving the single-particle energy term is
\beq
\left[ T, a_\kvec^\dagger \right]
= \frac{\hbar^2 k^2}{2m} a_\kvec^\dagger + \sum_\Gvec u(\Gvec) a_{\Gvec+\kvec}^\dagger.
\eeq
There are corresponding terms for hole excitations, but none of these terms 
involve singular contributions responsible for anomalous size effects, so that
these terms do not contribute at leading order.
However,
\beq
\left[ V_{ee}, a_\kvec^\dagger \right]
= \frac{1}{V} \sum_{\qvec \ne 0} v_q 
\left[\rho_\qvec a_{\kvec-\qvec}^\dagger -1 \right]
\label{Veecom}
\eeq
and
\beq
\left[ V_{ee}, a_\kvec \right]
= -\frac{1}{V} \sum_{\qvec \ne 0} v_q 
\rho_\qvec a_{\kvec+\qvec}
\eeq
involve terms approaching the Coulomb singularity, $v_q \sim q^{-2} \to \infty$ for $q \to 0$.

From these terms we get the leading order size corrections by noting that
\bea
\lim_{\kvec, \qvec \to 0}
\frac{\langle \Psi_{\kvec}^{N_e+1} | \rho_\qvec a_{\kvec-\qvec}^\dagger | \Psi_0^{N_e} \rangle}
{\langle \Psi_{\kvec}^{N_e+1} |
 a_{\kvec}^\dagger | \Psi_0^{N_e} \rangle}
=
\frac{1}{2}\left[\frac{1}{ \epsilon} +1 \right]
\eea
and
\bea
\lim_{\kvec, \qvec \to 0}
\frac{\langle \Psi_{\kvec}^{N_e-1} | \rho_\qvec a_{\kvec+\qvec} | \Psi_0^{N_e} \rangle}
{\langle \Psi_{\kvec}^{N_e-1} |
 a_{\kvec} | \Psi_0^{N_e} \rangle}
=
-\frac12 \left[1+ \frac{1}{\epsilon}\right].
\eea
Both relations can be obtained \cite{footnote} by extending Kohn's diagrammatic approach 
\cite{Kohn58}. 
Integrating around the $v_q$ singularity for small $q$ in Eq.~(\ref{Veecom}),
we obtain the leading order finite size corrections. As before, this involves the Madelung constant,
Eq~(\ref{vMdef}).
In the particle channel  we get
$\frac{|v_M|}{2} \left(\frac{1}{\epsilon} - 1 \right)$ and
in the hole channel, $\frac{|v_M|}{2} \left(\frac{1}{\epsilon} + 1 \right)$.
The corrections independent of $\epsilon$ correspond to the change in the background charge which
cancel for the fundamental gap and we obtain Eq.~(\ref{gapepsilon}).

Previous, heuristic approaches \cite{Hunt} have suggested that one can use
experimental or DFT values of the dielectric constant for finite-size extrapolation.
Our approach further suggests that this value can be determined from
the QMC structure factor extrapolated to zero wave vector
\beq
\frac{2}{\epsilon} \equiv (1+ c) \lim_{N_e \to \infty} \lim_{k \to 0}
\left[S_k^+ + S_k^- \right],
\eeq
with the singular behavior of the Jastrow factor determining $c$.
We emphasize that the order of the limits involved above is crucial.

\begin{figure*}
\begin{minipage}[b]{\columnwidth}
\includegraphics[width=\columnwidth]{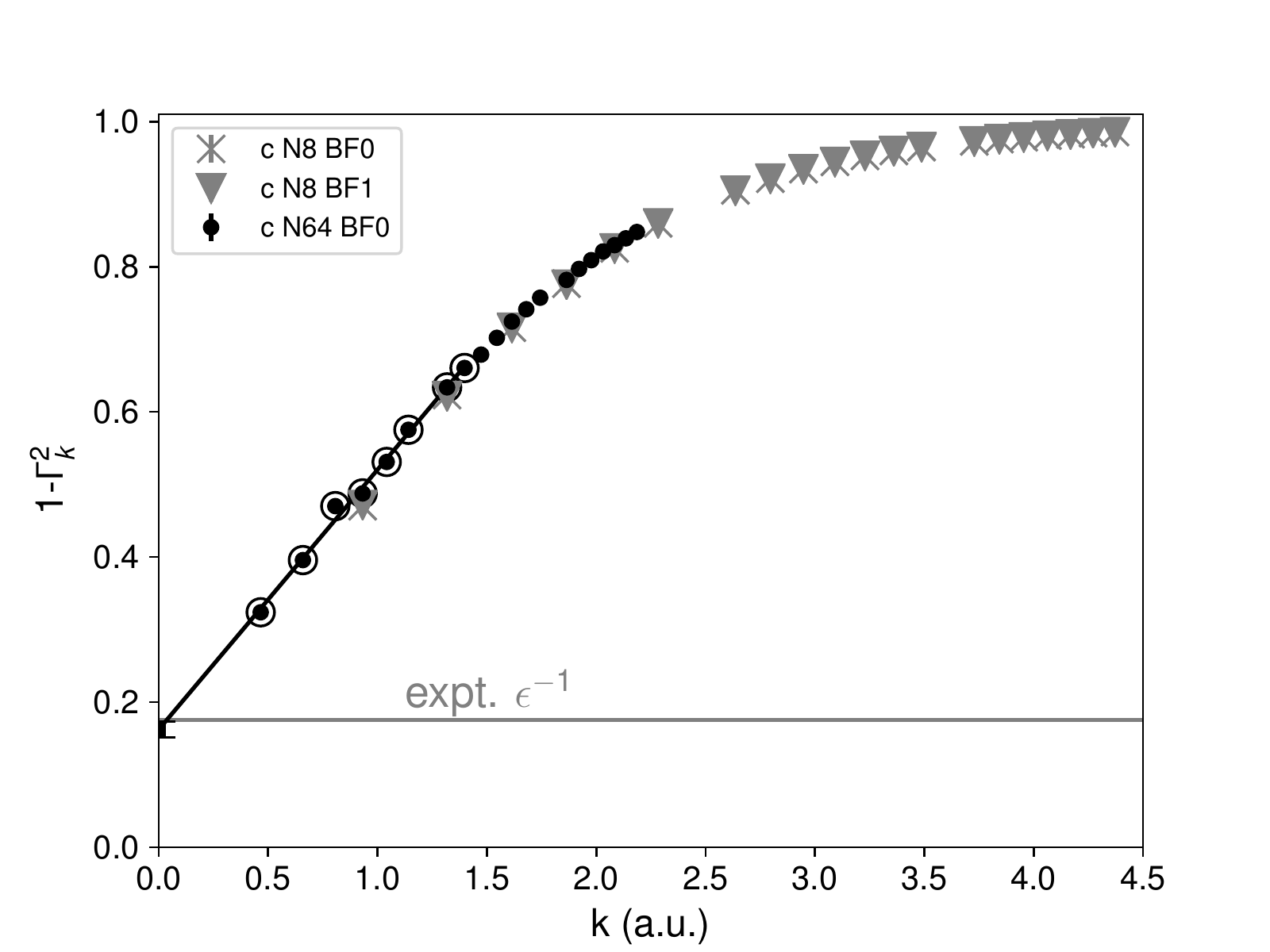}
(a) carbon
\end{minipage}
\begin{minipage}[b]{\columnwidth}
\includegraphics[width=\columnwidth]{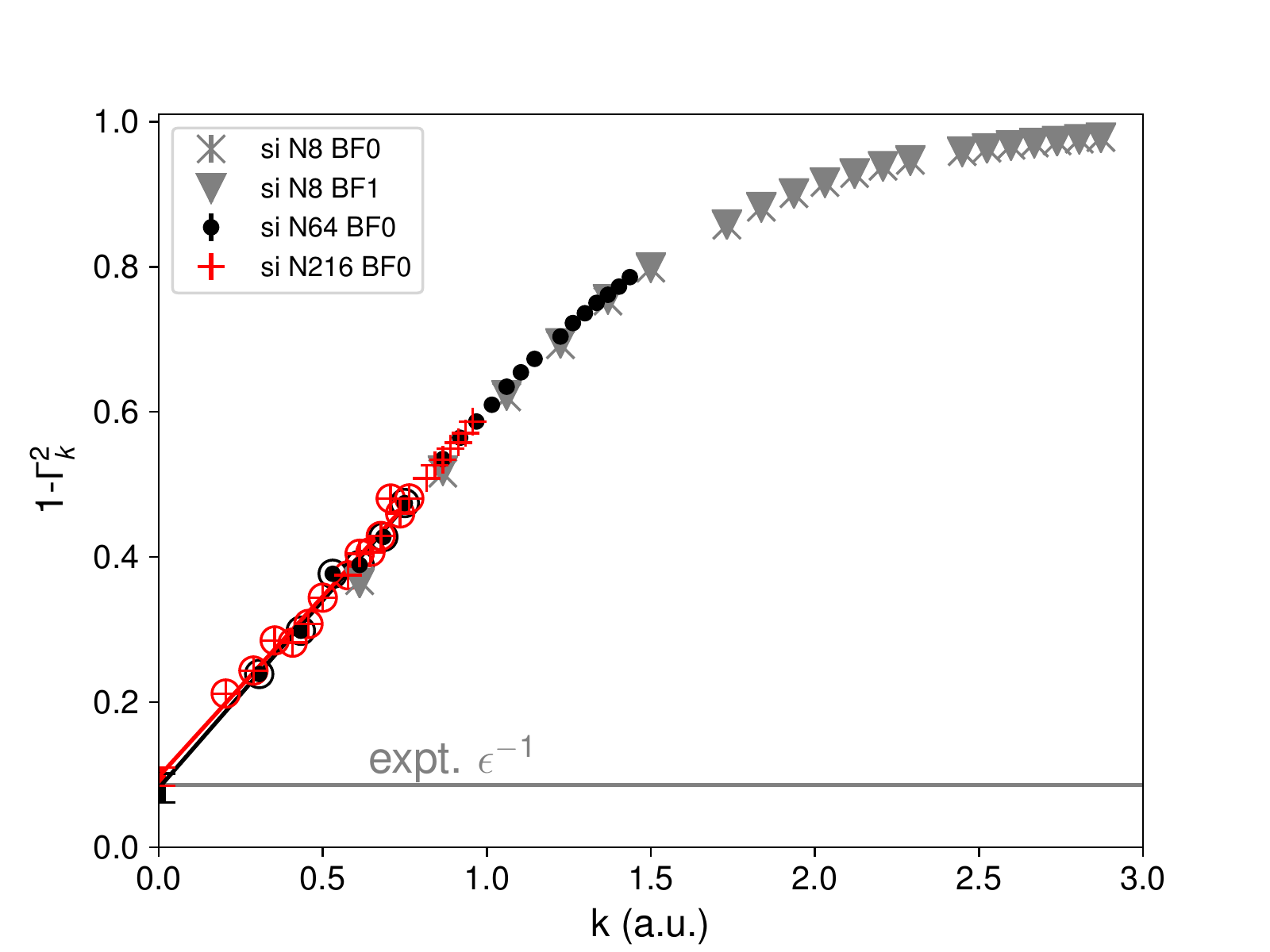}
(b) silicon
\end{minipage}
\caption{Upper bound to the inverse dielectric constant  eq.~(\ref{davidbound}), where $\Gamma_k\equiv\frac{2m\omega_pS_{N_e}(k)}{\hbar k^2}$. Lines are fits to the low-k data. 
The horizontal lines mark experimental inverse dielectric constants.
\label{fig:c-si-gk}}
\end{figure*}

An independent estimate is
based on the inequality of Eq.~(\ref{davidbound}). We can
bound and estimate the value of dielectric constant 
using the structure factor of the insulating ground state. By extrapolating $1-\Gamma_k^2$ vs. $k$ to $k=0$ we obtain an upper bound to the inverse dielectric constant, where $\Gamma_k\equiv 2m\omega_p S_{N_e}(k)/\hbar k^2$.
This involves only the extensive part
of the density-density correlations, thus, it is less sensitive to noise
and has much smaller statistical uncertainty.
In Fig.~\ref{fig:c-si-gk}, we show that for C and Si, this inequality gives accurate values of the dielectric
constant.

\subsection{Twist correction of two particle correlations}

The above size effects explain the leading order $1/L$ correction to the single particle gap.
However, as we will see in our results, 
the asymptotic region, where this law
can be reliably applied, may still be difficult to reach for
currently used system sizes and next-to-leading order effects are important.
Here, we show that an important part can be corrected for, by further restoring the full symmetry
properties in the contribution of the direct Coulomb interaction.

For inhomogeeous systems, it is convenient to separate the mean density from its
fluctuating components in the
static structure factor \cite{finitesize}, i.e.
\bea
S_{N_e}(\kvec) &= &\frac{1}{N_e} \langle \rho_\kvec \rangle \langle \rho_{-\kvec} \rangle + \delta S_{N_e}(\kvec) 
\\
\delta S_{N_e}(k) &=& \frac{1}{N_e} \left\langle \left( \rho_\kvec - \langle \rho_\kvec \rangle \right)
\left( \rho_{-\kvec} - \langle \rho_{-\kvec} \rangle \right) \right\rangle
\eea
For crystals with periodic density distributions, the Fourier components
of the mean density, $\langle \rho_\kvec \rangle$, only contribute for reciprocal lattice vectors, $\kvec \in \Gvec$. The long wave length behavior of the structure factor is entirely
due to the fluctuating part $\delta S_{N_e}(k)$, which therefore contains the leading order
size effects \cite{finitesize}. However, the mean 
single particle density,
$\langle \rho(\rvec) \rangle=V^{-1} \sum_\kvec \langle \rho_\kvec \rangle e^{i \kvec \cdot \rvec}$, of the finite system may significantly differ from  the infinite one, particularly
in cases where the supercell is not compatible with the full symmetry group of the crystal.

Averaging over twisted boundary conditions is designed to restore the symmetry of the crystal and
thus accelerate the convergence of single particle
densities to the thermodynamic limit. In the following, we denote the twist averaged  expectation value by
\beq
\overline{ {\cal{O} }} \equiv \frac{1}{M_\theta} \sum_{\theta} 
\langle {\cal{O}} \rangle_{N_e, \theta}
\eeq
where we have explicitly indicated the $N_e$ and $\theta$ dependence on the expectation value
on the r.h.s. For any single particle theory, $\overline{\rho( \rvec )}$ approaches its thermodynamic
limit for calculations at fixed $N_e$  by
averaging over a dense grid of twist angles ($M_\theta \to \infty$). 
Within many-body calculations, twist-averaging \cite{Lin01} takes over large part of this
property to any observable {\em linear} in the density. Here, we extend this approach to
correct also the quadratic expression entering the two-body contributions
of the total energy.

For the potential energy, this correction to the twist converged QMC calculation is
\bea
\delta V_{N_e}^s &= &
 \frac1{2V} \sum_\kvec v_k \delta C(\kvec) \nonumber \\
&&\delta C(\kvec) =
\overline{\rho_\kvec } \, \overline{\rho_{-\kvec}}
- \overline{\rho_\kvec \rho_{-\kvec}} .
\label{deltaV}
\eea
For the ground state energies, this correction provides only a small improvement over our previous correction~\cite{fse,finitesize}.

For the gap, many terms entering Eq.~(\ref{deltaV}) cancel and the expression can be simplified. Let us consider the case of adding/removing one electron at twist $\phi$ to
the insulating ground state, denoting $\Pi_\kvec^{\pm}$ the difference of the respective densities
\begin{equation}
\Pi_\kvec^{\pm} \equiv
\langle \rho_\kvec  \rangle_{N_e \pm 1,\phi} - \langle \rho_\kvec  \rangle_{N_e,\phi}
\end{equation}
In the thermodynamic limit, the density of the ground state system with $N_e$ electrons
coincides with the twist averaged ground state density $\overline{\rho_\kvec}$, whereas
we obtain $\overline{\rho_\kvec}+\Pi_\kvec^{\pm}$
for the density of the $N_e \pm 1$ electron system.
Inserting into Eq.~(\ref{deltaV}), we obtain the correction for the difference between the two states
\begin{equation}
\delta V_{N_e \pm 1,\phi}^s - \delta V_{N_e}^s
= \frac{1}{V} \sum_{\kvec\in \Gvec} v_k \text{Re}
\left[ \left( \overline{\rho}_\kvec
- \langle \rho_\kvec \rangle_{N_e,\phi} \right) \Pi_{-\kvec}^{\pm}
\right]
\label{twistadd}
\end{equation}
where only wave vectors of the reciprocal crystal lattice contribute
to the sum.
The corresponding finite size correction for the gap,
denoted by $\delta \Delta_s$ in the following,
is order $1/N_e$ or smaller, mainly determined by the changes
of the ground state densities at the first Bragg-peaks due to twist averaging.

Equation~(\ref{twistadd}) can be understood quite intuitively: it corrects the
direct Coulomb interaction between the electron/hole in the
excited state ($\Pi^\pm$) with the unexcited electrons. The density
of those electrons is expected to change
by $\overline{\rho}_\kvec - \langle \rho_\kvec \rangle_{N_e,\phi}$
in the thermodynamic limit.

Converged ground state densities are naturally calculated within
GCTABC.  It is straightforward to apply the correction Eq.~(\ref{twistadd})
to all excitation energies.
Alternatively, the corresponding
 DFT densities may be used. This removes the stochastic error at the cost of
introducing a small bias in the next-to-leading order size correction.

\begin{figure*}
\begin{minipage}[b]{\columnwidth}
\includegraphics[width=\columnwidth]{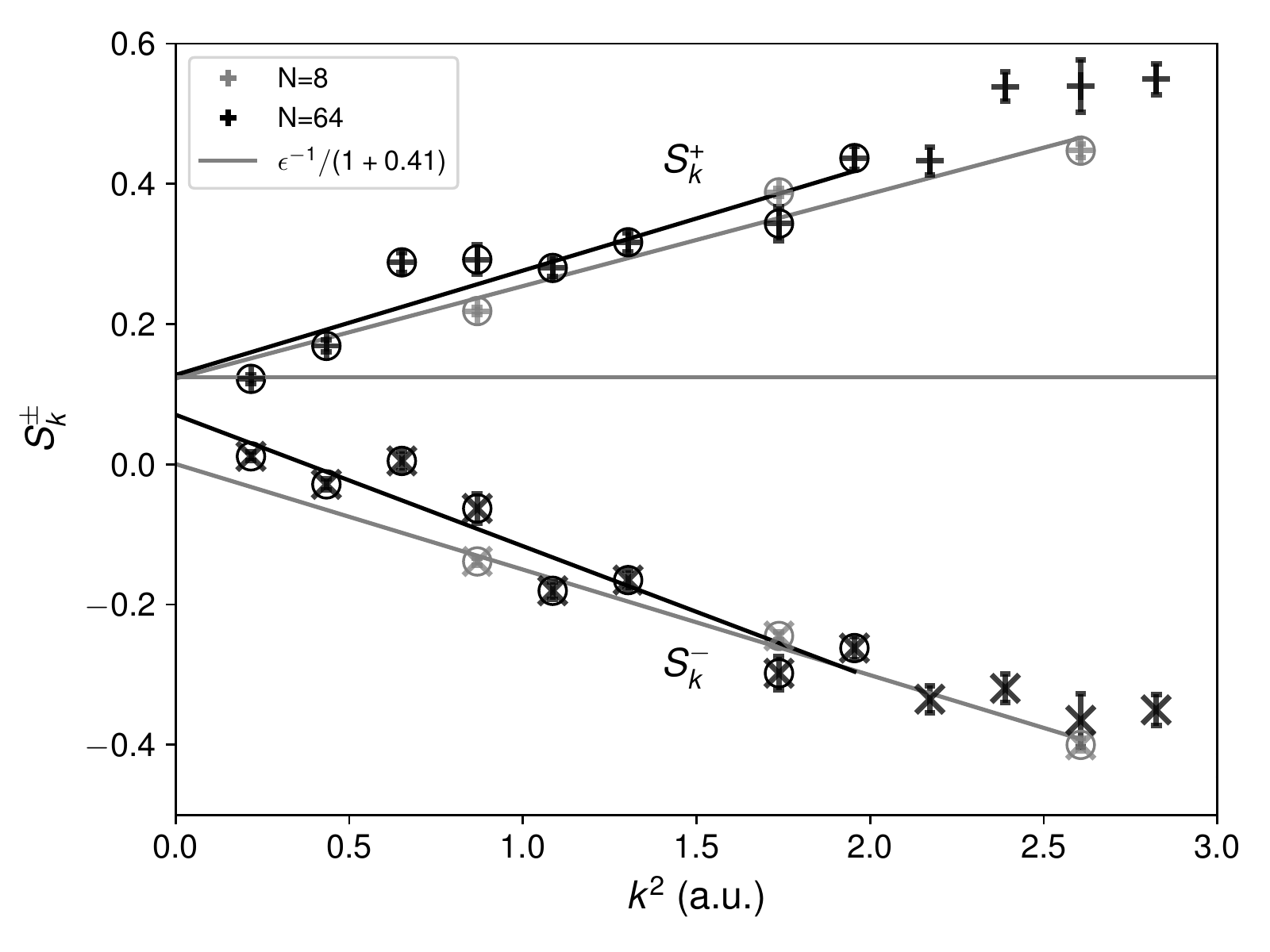}
(a) carbon
\end{minipage}
\begin{minipage}[b]{\columnwidth}
\includegraphics[width=\columnwidth]{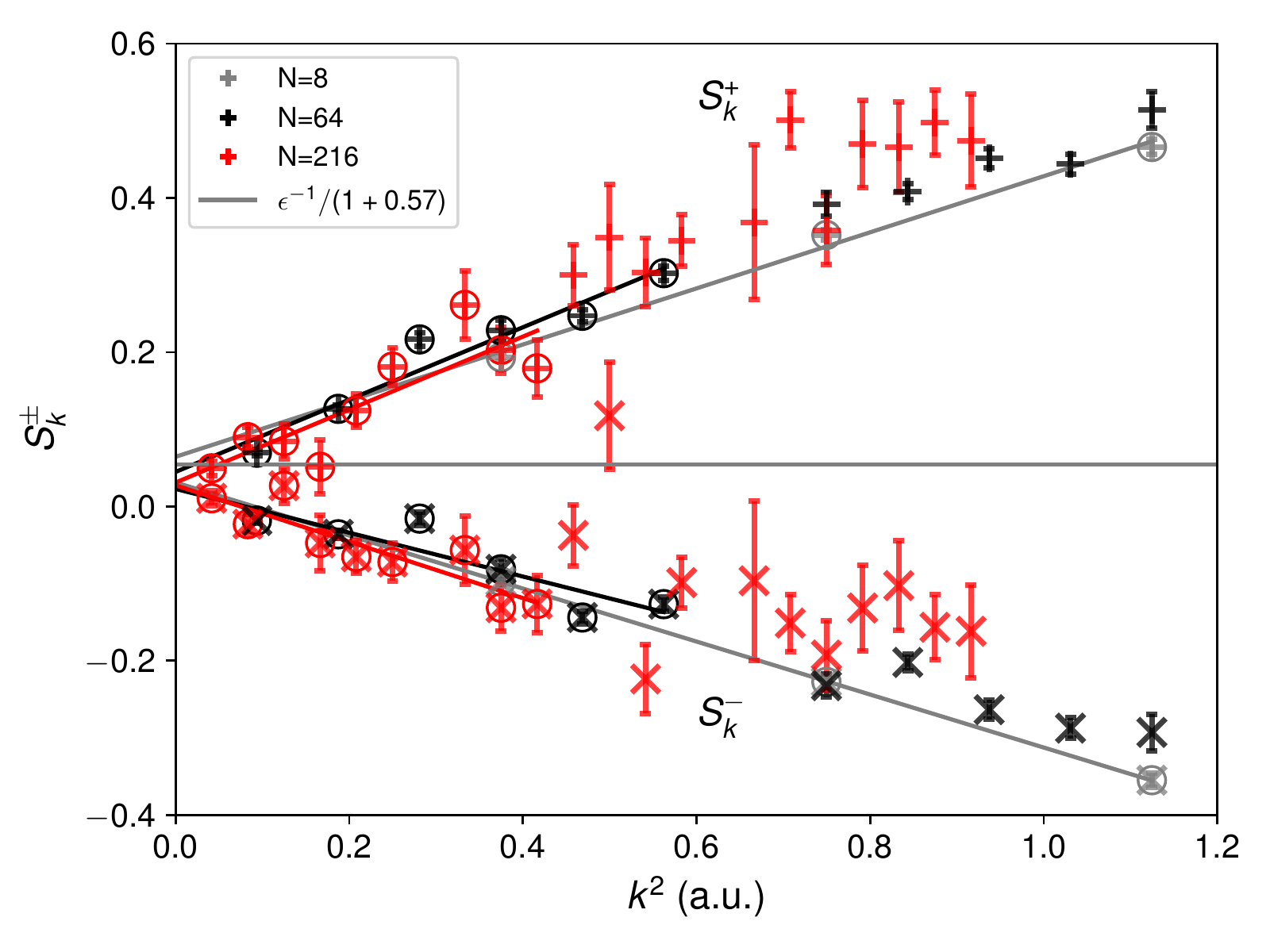}
(b) silicon
\end{minipage}
\caption{Change in the static structure factor as an electron (upper curves) or a hole (lower curves) is added to the insulating system with N atoms. 
The lines are fits to the data points.
The horizontal lines show the expected $k\rightarrow0$ limit based on the experimental dielectric constants. We have used $c=0.40$ for C and $c=0.57$ for Si.
\label{fig:c-si-skpm}}
\end{figure*}

\section{Computational methods\label{sec:methods}}

We have performed electronic QMC calculations on three insulating solids:
molecular hydrogen at high pressure,
and carbon and silicon in the diamond structure at zero pressure. Since we are interested in the
charge gap, we used an equal number of spin up and spin down electrons. We used a Slater-Jastrow  trial wave function with backflow corrections \cite{Holzmann2003,Trial}. The Jastrow and backflow functions were fully optimized within Variational Monte Carlo including the long-range (reciprocal lattice) contributions.
The orbitals in the Slater determinant were taken from DFT calculations using Quantum Espresso~\cite{QE2009, QE2017}. The carbon and silicon orbitals were generated using the LDA functional, whereas the hydrogen orbitals were generated using the PBE functional, which has been shown to provide a good trial QMC wave function \cite{Morales2012,Pierleoni2016}.

Molecular hydrogen was placed in the C2/c-24 structure \cite{Pickard2007} at two different densities ($r_s=1.38$ and $r_s=1.34$), roughly corresponding to pressures of 234GPa and 285 GPa, respectively.
Energies and structure factors were obtained from Reptation Quantum Monte Carlo calculations using the BOPIMC code \cite{BOPIMC}. For carbon and silicon, Diffusion Monte Carlo calculations have been performed with the QMCPACK code \cite{QMCPACK}
at the experimentally measured zero pressure valence densities, $r_s=1.318$ and $r_s=2.005$, respectively.
For hydrogen, the QMC calculations have been done with the bare Coulomb interaction. The PAW pseudopotential has been used for the DFT results shown in Fig.~\ref{fig:H_C2CP250}. For carbon and silicon, psuedopotentials were used to remove the core electrons: carbon ions modeled by the Burkatzki-Filippi-Dolg (BFD) pseudopotential~\cite{Burkatzki2007}, and silicon ions by the Trail-Needs (TN) pseudopotential~\cite{Trail2005}. These are considered good pseudopotentials for correlated calculations, but their use within DFT calculations produces slightly different results from the literature even with the same functional.
For hydrogen, we used a supercell with  $2 \times 2 \times 1$ primitive cells so that the supercell is nearly cubic and contained 96 protons. For carbon, we used two system sizes: the cubic cell containing 8 atoms and a $2\times 2\times 2$ supercell containing 64 atoms. For silicon, in addition to these systems, we used a $3\times 3\times 3$ supercell containing 216 atoms.
For hydrogen, the twist convergence has been achieved using a $8 \times 8 \times 8$ twist grid. For C and Si, the twist grid density decreases with increasing system size. The supplementary material contains the QMC calculated energies and variances of the insulating ground states
of the various systems obtained after twist averaging and two-body finite size corrections.

\section{Results\label{sec:results}}

For any single-particle theory, such as Kohn-Sham DFT,
the densities and energies, $n_e(\mu)$ and $e_0(\mu)$,
are obtained by occupying all single particle states below the chemical potential $\mu$.
By construction, the gap, as determined from the incompressible region of $n_e(\mu)$ or from the discontinuity in the derivative of $de_0/dn_e$ (see Fig.\ref{fig:H_C2CP250}), then coincides with the one obtained from the band structure.

\begin{figure*}
\begin{minipage}{\columnwidth}
\includegraphics[width=\linewidth]{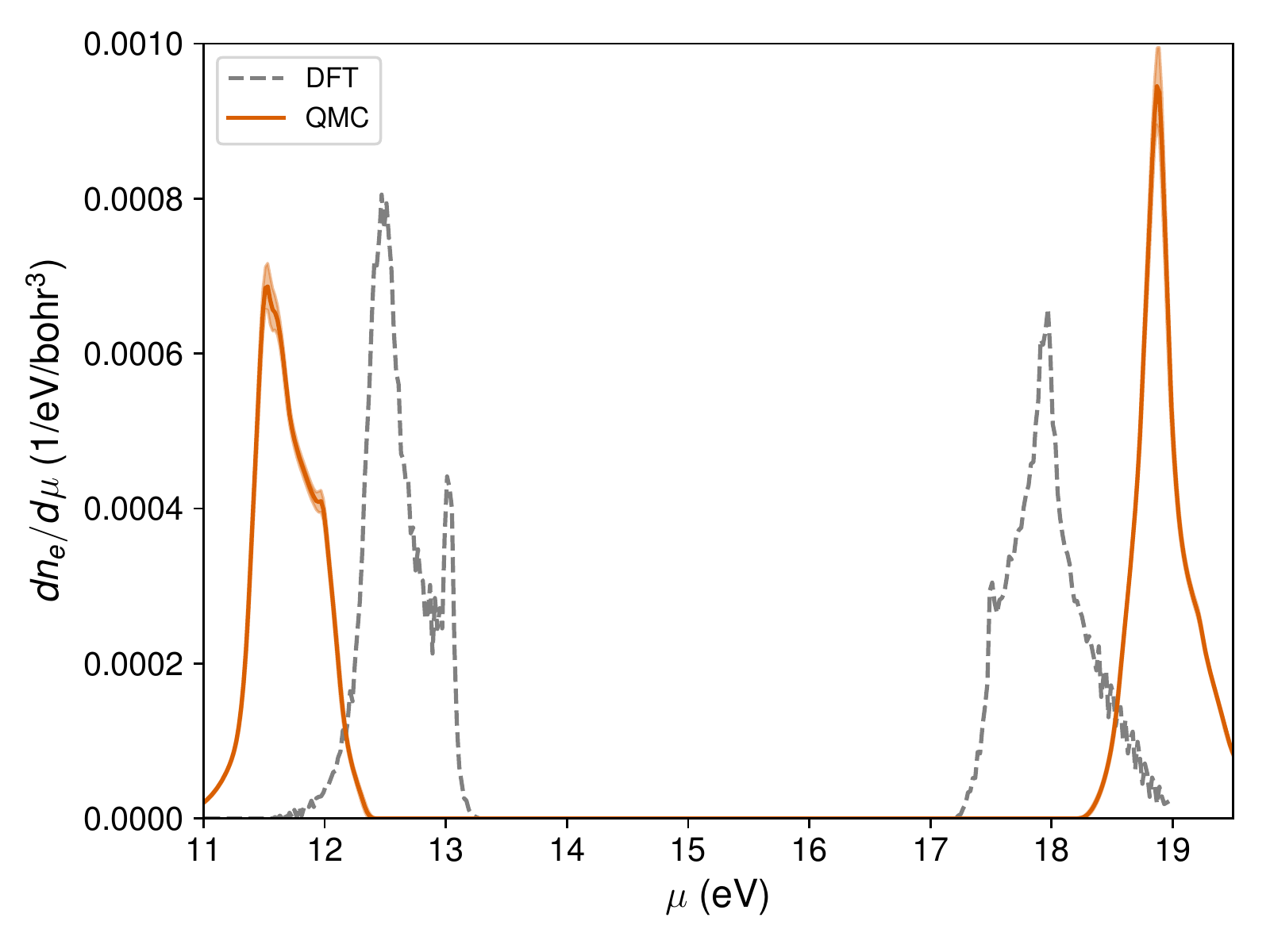}
(a) carbon
\end{minipage}
\begin{minipage}{\columnwidth}
\includegraphics[width=\linewidth]{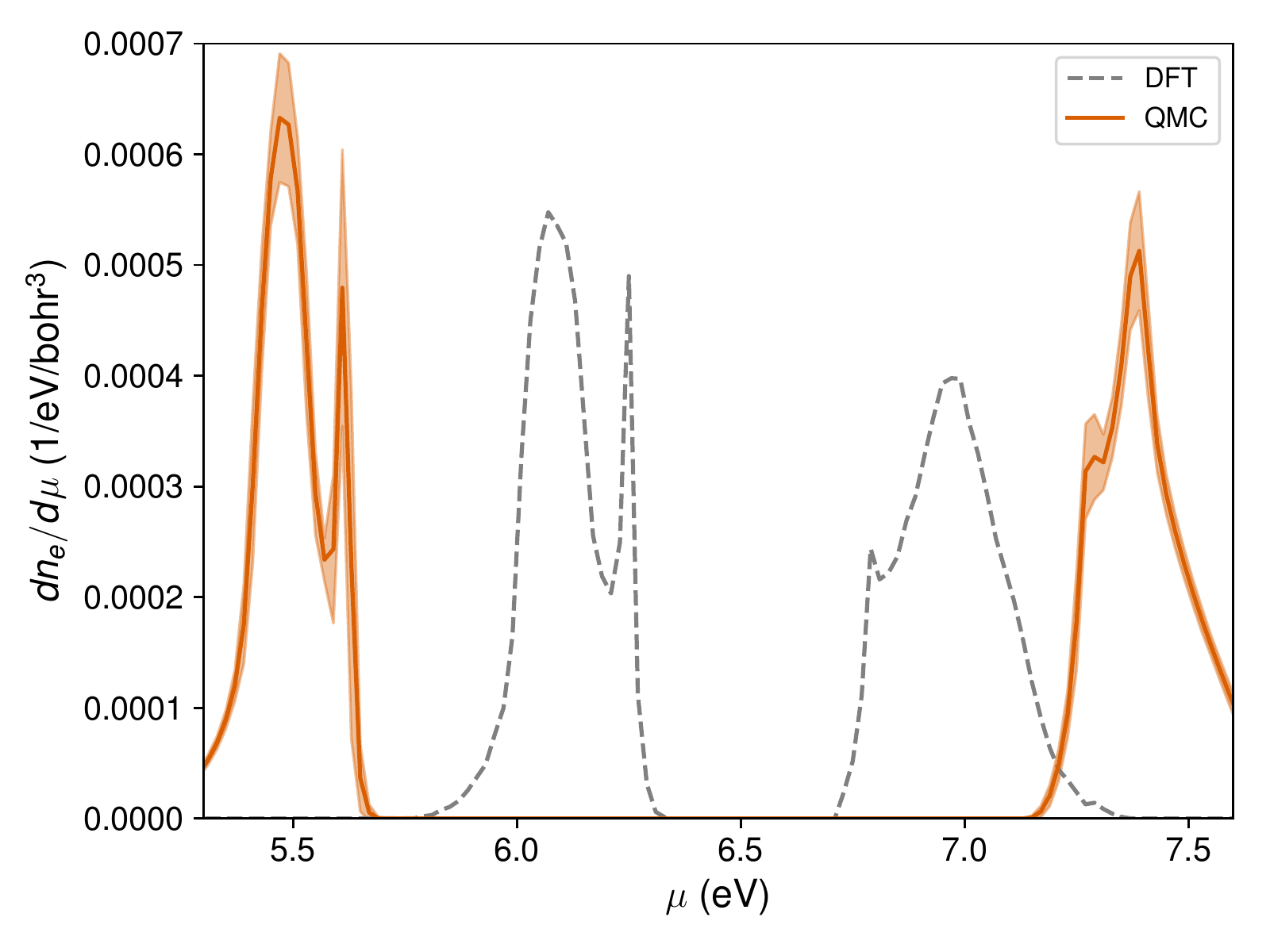}
(a) silicon
\end{minipage}
\caption{Density of states for carbon and silicon near the band edge. Each plot shows the derivative of the  mean electron density with respect to the chemical potential. This is the electronic density of states (DOS) in DFT, so the gap appears as a depleted region. 
The calculated DOS is only valid near the band edge because only the two bands closest to the gap are considered within DFT and QMC. The DFT bands (done in a primitive cell) have been folded into the Brillouin zone (BZ) of the 64-atom supercell to allow comparison with QMC. \label{fig:band_edges}}
\end{figure*}

To illustrate the equivalence of the two ways of computing the gap with LDA calculations. The LDA band gaps of carbon and silicon in the diamond structure, are indirect, and lie 
along the $\Gamma X$ direction where $\Gamma$ is the origin of the Brillouin zone and
$X$ the Brillouin boundary in the $(100)$ direction. By looking directly at the HOMO and LUMO states with LDA it is found that
the carbon gap  is 3.89 eV 
and the silicon gap is  0.34 eV.
The bands immediately above and below the gap can be fit
to a quadratic form which implies $e_0(\mu) = \mu^{\pm} n_e(\mu) + b^\pm n_e(\mu)^{5/3}$. Therefore, the derivative $de_0/dn_e = \mu^{\pm} + \frac{5b^{\pm}}{3} n_e^{2/3}$ has a discontinuity at $n_e=0$ and behaves as $ n_e^{2/3}$ above and below the gap. 
Applying our GCTABC procedure to a single-particle theory,
all states with energies below the chemical potential are occupied. Varying the chemical potential
thus scans the underlying density of states. The band gap is then determined by locating the band
edges, $\mu^\pm$, disregarding the location in the Brillouin zone \cite{BrillouinFootnote}.
Figure~\ref{fig:band_edges} illustrates the density of states
obtained from GCTABC giving an LDA gap of 3.95 eV for the carbon gap
and 0.38 eV for the silicon gap. 
The small differences ($\sim 0.05$ eV) from the values obtained before are due to the finite
resolution of the twist grid, and can be controlled by using denser grids.

\begin{figure}
\begin{minipage}{\columnwidth}
\includegraphics[width=\columnwidth]{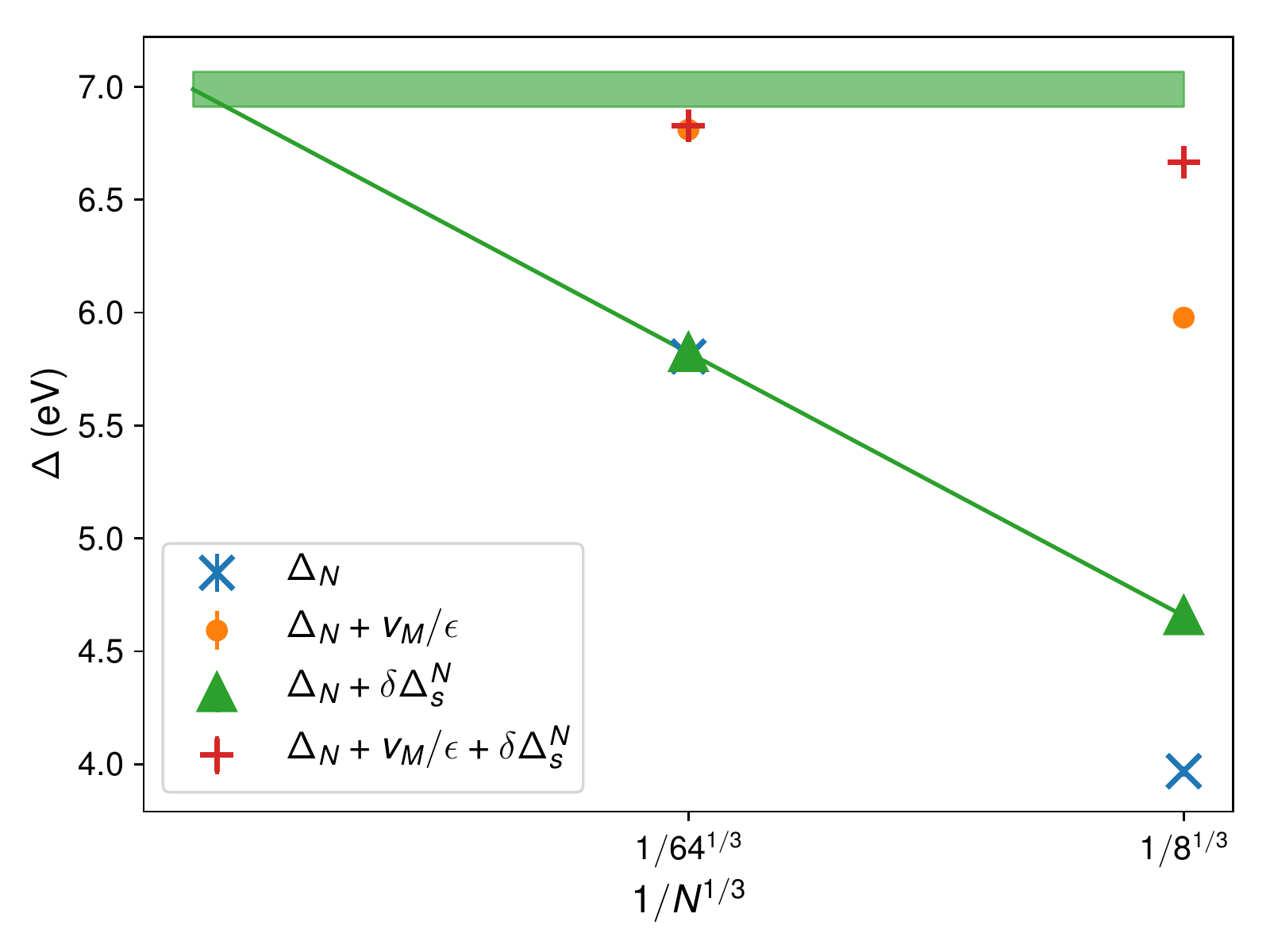}
(a) carbon
\end{minipage}
\begin{minipage}{\columnwidth}
\includegraphics[width=\columnwidth]{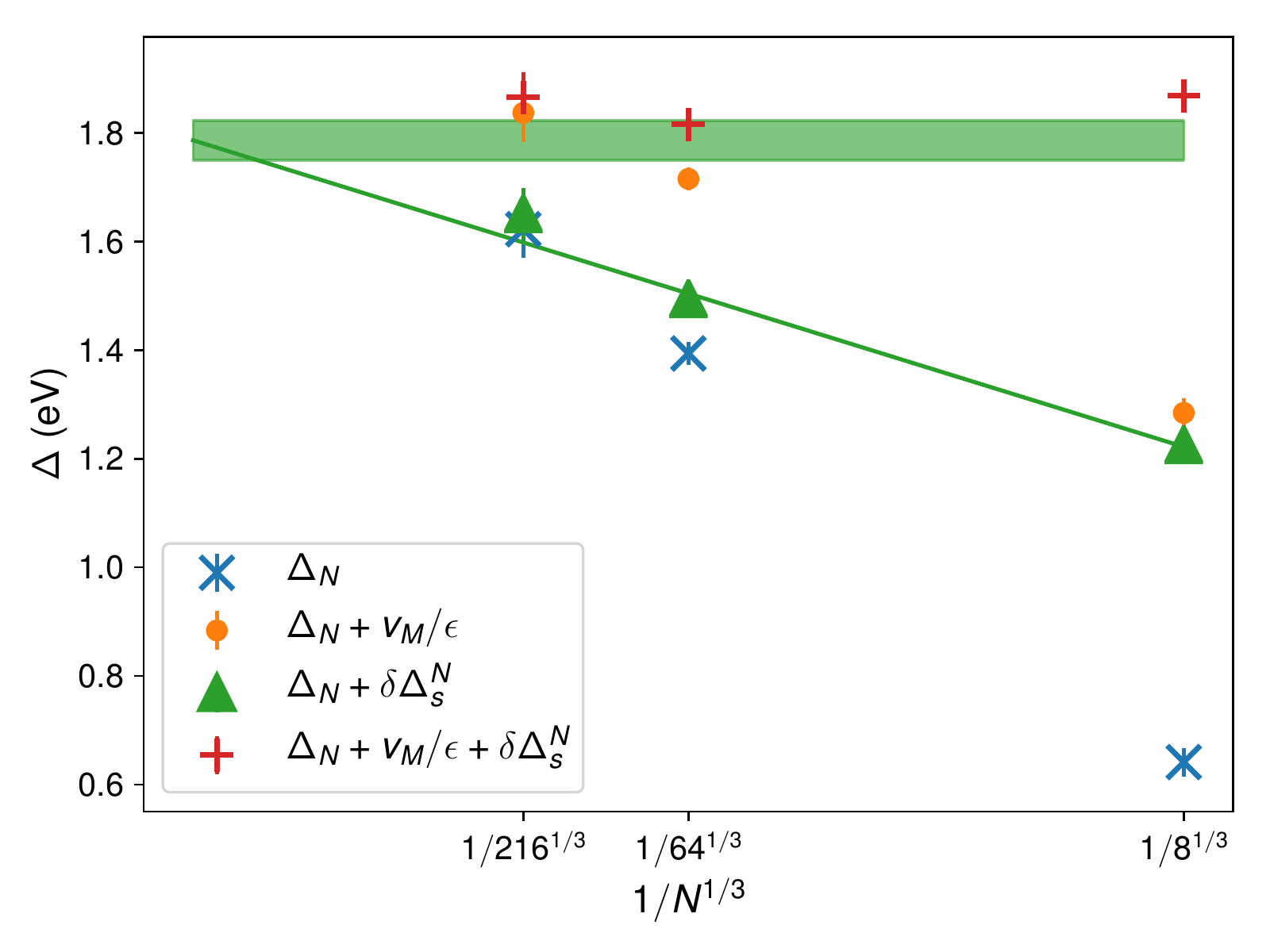}
(a) silicon
\end{minipage}
\caption{Fundamental gap before and after finite-size corrections. $\Delta_N$ is the DMC gap from a simulation with $N$ atoms in the supercell without any finite-size correction, $\delta\Delta^N_f$ is the leading-order Madelung correction using the experimental value of $\epsilon^{-1}$,
$\Delta^N_s$ is the next-to-leading-order density correction, which is related to the static part of the structure factor. The line is a fit to $\Delta^N_s$.\label{fig:gap-extrap}}
\end{figure}

As can be seen in the same figure, the effective band edge densities of states from GCTABC-DMC
have a similar functional form, but with a larger gap than the DFT ones.
The QMC computed gaps
for the different sizes of the supercell are summarized in table \ref{tab:dmc-gcta-gap}.
The results from different supercells clearly show the important bias 
on gap introduced by the finite size of the supercell. In Figure~\ref{fig:gap-extrap}, we show the bare gap, $\Delta_N$, the Madelung-corrected one, $\Delta_N+|v_M|/\epsilon$, and our best correction, $\Delta_\infty=\Delta_N+|v_M|/\epsilon+\delta \Delta_s$, for both systems against the linear size of the supercell where $N$ is the number of atoms in the supercell.
We see that the next-to-leading-order corrections are 
comparable to the leading-order one, in particular for the 8-atom supercell of Si,
whereas they rapidly decay for the larger sizes. 

The finite size corrected values, $\Delta_\infty$, of all different sizes C and Si supercells agree with
each other within the statistical uncertainty, yielding the DMC-SJ values $\Delta_\infty=6.8(1)$
and  $\Delta_\infty=1.8(1)$ for the C and Si gap, respectively. 
We further note, that these values also agree with a numerical
$N^{-1/3}$ extrapolation of the gap values corrected by $\delta \Delta_s$. For any numerical
$N^{-1/3}$ extrapolation, it is very important to reduce any bias due to higher order corrections
as much as possible, since the outcome of a fit is sensitive to the smallest system sizes
since they have the smallest statistical uncertainty. For Si,
a $N^{-1/3}$ extrapolation
of the bare $\Delta_N$ values yields an overestimation of $0.3$ eV compared to $\Delta_\infty$.

Since our finite-size corrected gaps show size-convergence for the smallest system size,
it is now feasible to address the systematic error due to the fixed node approximation.
In order to reduce this bias
we have added backflow (BF) correlations in the Slater orbitals. Our backflow correlations lower
the SJ gap by $0.1$ eV for both, C and Si.
Previous BF calculations \cite{Hunt} on Si have reported a $0.2$ eV lowering compared to SJ.
The difference might be due to a different functional form or optimization procedure.
A systematic study on the bias of the fixed-node approximation such as done with more general backflow correlations \cite{BFN,orbitalbf} or
multi-determinant trial wave functions \cite{Zhao19}, possible for small supercells,
could be done in the future.

So far, in our analysis of C and Si, we have imposed the experimentally known dielectric constant
in the leading order Madelung correction. As described in Sec.~\ref{sec:fse},
there is no need for any external knowledge to perform the size extrapolation as the value
of the Madelung correction can be obtained from the behavior of the static structure factor,
calculable within the same QMC run, see  Figs~\ref{fig:c-si-skpm} and \ref{fig:c-si-gk}. However, since
 the extrapolation involved introduces an additional uncertainty, we have preferred to use
the experimental values to benchmark our theory and better distinguish leading
from next-to-leading order size effects. 

Using the dielectric bound eq.~(\ref{davidbound}) on the ground-state structure factor to determine $\epsilon$, we get $\epsilon_0=6.2\pm0.4$ for C and $\epsilon_0=10.3\pm1.3$ for Si, which are compatible with the experimental values of $5.7$ and $11.7$. The corresponding leading-order finite-size corrections on the gap of the 64-atom system are then $0.92\pm0.06$ eV for C and $0.36\pm0.14$ eV for Si using the \emph{ab initio} $\epsilon^{-1}$, as opposed to $1.00$ eV for C and $0.32$ eV for Si based on the experimental values of $\epsilon^{-1}$.

As shown in Fig.~\ref{fig:c-si-skpm}, the asymptotic values of the finite sized structure factors,
$S_k^\pm$, are affected by a much larger uncertainty, introducing larger systematic bias when
used for \emph{ab-initio} size corrections. Still, already the extrapolation to a non-zero
value fixes the leading order size corrections to decay as $1/L$. This information alone can be
crucial as calculations for only two different supercell sizes will be sufficient to determine size effects, whereas more supercell sizes would be needed if the asymptotic form was not known.

\begin{table}
\caption{Energy gaps obtained from GCTAB QMC calculations in eV. The bare gap,
$\Delta_{N}$, was calculated from Eq.~(\ref{gap})
for a finite supercell containing $N$ atoms.
The leading-order finite-size corrections are given by the
screened Madelung constants $|v_M|/\epsilon$, the next-to-leading order by the twist correction of two particle density correlations, $\delta \Delta_s$.
We used the experimental value of $\epsilon$ for C and Si ($5.7$ and $11.7$, respectively) and 18.8 for H$_2$ extraced from S(k).
Finite size corrections were applied also to the band edges,  $\mu^{\pm}$.
The estimate of the gap in the thermodynamic limit is
 $\Delta_\infty=\Delta_{N_e}+|v_M|/\epsilon+\delta \Delta_s$.
From our LDA analysis, we estimate a systematic bias of $\sim 0.1$ eV from the finite twist grid. This bias is larger than the statistical error. SJ indicates Slater-Jastrow trial wave function,
BF backflow. \label{tab:dmc-gcta-gap}}
\begin{tabular}{lrrrrrrrr}
\hline\hline
 &$r_s$& $N$ & $\Delta_{N}$ & $|v_M|/\epsilon$ & $\delta \Delta_s$ & $\mu^-_{\infty}$ & $\mu^+_{\infty}$ & $\Delta_\infty$ \\
\hline
H$_2$ (BF) & 1.38  &  96 & 3.3(1)   & 0.40 & 0.020 & 6.9(1) & 10.7(1) & 3.8(1) \\
       & 1.34  &  96 & 2.4(1)   & 0.20 & 0.018 & 8.6(1) & 11.2(1) & 2.6(1) \\
\hline
C (BF) & 1.318 &   8 & 3.9(1) & 2.01 & 0.69 & 11.5(1) & 18.1(1) & 6.6(1) \\
\hline
C (SJ) & 1.318 &   8 & 4.0(1) & 2.01 & 0.69 & 11.5(1) & 18.2(1) & 6.7(1) \\
  &  &  64 & 5.8(1)   & 1.00 & 0.02 & 11.9(1) & 18.7(1) & 6.8(1) \\
\hline
Si (BF) & 2.005 &   8 & 0.6(1) & 0.64 & 0.55 & 5.2(1) & 6.9(1) & 1.7(1) \\
\hline
Si (SJ) & 2.005 &   8 & 0.6(1) & 0.64 & 0.58 &  5.2(1) &  7.0(1) & 1.9(1) \\
 &  &  64 & 1.4(1)   & 0.32 & 0.08 &  5.5(1) &  7.3(1) & 1.8(1) \\
 &  & 216 & 1.6(1)   & 0.21 & 0.01 &  5.6(1) &  7.4(1) & 1.8(1) \\
\hline
\hline
\end{tabular}
\end{table}

We have also computed the band gap of solid hydrogen using GCTABC in BF-RQMC calculations
for one of the possible
molecular structures predicted for phase III: C2/c-24 at $r_s = 1.38$ and $r_s=1.34$ (roughly corresponding  to pressures of 234 and 285 GPa respectively). The results, in table \ref{tab:dmc-gcta-gap} show that the gap and size effects decrease with increasing pressure. 
For these calculations, we use calculations for one supercell and use its structure factor to estimate the dielectric constant.  From Fig. \ref{fig:H_C2CP250}, we see that
 HSE DFT slightly underestimates the gap, however the deviations from the plateau on both sides are quite similar.

\section{Comparison with experiment}
\label{sec:compare}

Our best values for the fundamental electronic gap (BF-DMC)  significantly overestimate the experimentally measured values for C and Si  by 1.1 and 0.5 eV, respectively as shown in Table \ref{tab:c-si-gap-corr}. There are two main sources of systematic errors which need to be taken into account: the use of pseudo-potentials and the neglect of electron-phonon coupling.

The QMC values for C and Si presented above are based on pseudo-potentials to replace the 
core electrons of the atoms. Pseudo-potentials are usually designed for accurate prediction of
static structural quantities. Excitation spectra, in particular the single particle excitation gap,
may be less well described.
This has been found in many-body perturbation theory calculations within the $GW$ framework
where all-electron calculations have been shown to lower the gap of C and Si by $\sim -0.3$ eV
\cite{Wu02,GomezAbal08}
with respect to  pseudo-potentials calculations.
Although the actual pseudo-potentials of our QMC simulations differ from those used in the
$GW$ calculations, 
we expect that our QMC values will be shifted by a similar amount; we can roughly transfer
the all-electron correction of $GW$ to our QMC results. 

For lighter atoms, electron-phonon coupling leads to a further reduction of the
gap values, even at zero temperature, due to the presence of zero point motion of the ions in
the crystal. For C, $GW$ predicts a significant lowering of the gap by $-0.6$ eV \cite{Giustino10}, whereas a smaller shift between $-60$ meV \cite{Monserrat14} and $-0.1$ eV \cite{Lautenschlager85} is expected from DFT for Si.

Considering both, the bias due to the pseudo-potential approximation and the neglect of 
electron-phonon coupling, our BF-DMC for C and Si overestimate the gap by $\sim 0.1-0.2$ eV
(see table~\ref{tab:c-si-gap-corr}),
larger than our statistical uncertainty. This remaining offset to experiment may
either be due to residual bias of the
fixed-node approximation, or due to effects in pseudo-potential and
e-ph coupling beyond our simple estimations based on $GW$ and DFT.

For hydrogen, we do not compare to experiment since electron-phonon coupling is
expected to be very large, and the experimental results are not precise.
If we do not make  size corrections, our results are comparable to the Slater-Jastrow DMC calculations of Ref.~\cite{Azadi2017}
where the DFT band structure was corrected by a ``scissor operator'' based on QMC runs at the $\Gamma$ point of the supercell. However, no size effects been observed within the statistical
error in Ref.~\cite{Azadi2017}, so that their extrapolated results differ from ours
by $0.3-0.8$ eV (3.0 and 2.3 eV for 250 and 300 GPa).
Comparison to $GW$ values are also
not conclusive: 
whereas Ref.~\cite{Lebegue2012} provides smaller values of the gaps (1.8 and 1.0 eV for 250 and 300 GPa), the results  of Ref. \cite{McMinis2015} (3.7 and 2.8 eV for 250 and 313 GPa)  are close to our predictions.
However, we note that  the $GW$ calculations were done with
slightly different crystal structures.
In Ref.~\cite{Lebegue2012} PBE functional was used to optimize the lattice structure
in contrast to the vdW-DF1 functional of Ref.~\cite{McMinis2015} shown to be the most accurate functional at this density \cite{Clay2014}. The smaller gap can then
be seen as a consequence of  a larger bond length as it was shown that structures optimized with PBE functional have a larger bond length than the ones with vdW-DF1 \cite{McMinis2015}.

\begin{table}
\caption{Extrapolated band-gap
of Si and C from backflow DMC calculations, $\Delta_{BF}$
compared to the experimental values (exp).
We tabulated two main corrections:
the difference  between the gap of an all-electron (AE) and the pseudo-potential (PP)
calculation within GW calculations, and
the neglect of electron-phonon coupling (e-ph).
\label{tab:c-si-gap-corr}}
\begin{tabular}{lllll}
\hline\hline
& $\Delta_{BF}$ & AE - PP & e-ph  & exp \\
\hline
C & 6.6(2) &   $-0.26$ ($G_0W_0$) \cite{GomezAbal08} & $-0.6$ ($GW$)~~\cite{Giustino10} & $5.48$ \cite{exp}\\
\hline \\
Si &  1.7(1)  & $-0.25$ ($G_0W_0$)\cite{GomezAbal08}  & $-0.06$ ($DFT$) \cite{Monserrat14} & $1.17$ \cite{exp} \\
\hline\hline
\end{tabular}
\end{table}

\section{Conclusions}
\label{sec:conclude}

We have introduced a method to calculate the fundamental gap of insulators and semi-conductors
using QMC. Using grand-canonical twist averaging, 
the value of the gap can be determined at any point in the Brillouin zone whether the system has a direct or indirect gap. Although it is 
possible to map out the whole band structure, we have focused on the minimal, fundamental gap
in this paper. We have shown that for charged systems, finite size supercell calculations
are necessarily biased by a finite size error decaying as $1/L$, where the prefactor
is determined by the absolute value of the Madelung constant and the inverse dielectric constant.
We have pointed out that the $1/L$ functional form is encoded in the long wave length behavior of
the finite size structure factor extrapolating to a non-vanishing value at the origin.
Next-to-leading order effects can be corrected by proper use of twist-averaging in the
two-particle part of the static Coulomb potential. 

We have applied this procedure to determine the fundamental gap of molecular Hydrogen at high pressure
and carbon and silicon in the diamond structure at zero pressure. Our bare values for carbon and silicon  are larger than the experimental
ones. We have shown the bias is most likely due to the pseudo-potential approximation and the neglect of electron-phonon coupling.

We note that this procedure is not restricted to QMC calculations, but can be applied
within any method which calculates the many-body wave functions and ground state energies,
e.g. for coupled cluster methods~\cite{Gruber18}. Our results for C and Si demonstrate that
the bias due to the finite size supercell can be corrected for, so that
precise values in thermodynamic limit can be obtained for small supercells
without need for numerical extrapolation.

The procedure here has been developed for perfect crystals but can be generalized 
to systems with disorder, either due to thermal or quantum effects. Furthermore, the procedure provides a starting point 
to address optical --i. e.  charge neutral -- excitations. Although neutral excitations are expected to be less sensitive to finite size effects,
recent calculations \cite{Hunt,Frank19} have observed the same slow $1/L$ decay for the 
optical gap. Since it is often not practical to perform calculations for more than two significantly different supercell sizes, our method suggests that 
the asymptotic behavior of the structure factor provides
the needed insight to whether $1/L$ or $1/L^3$ should be used as functional form for the
size extrapolation. 

\begin{acknowledgments}

DMC and MH thank CCQ at the Flatiron Institute for support during the writing of this paper
and the Fondation NanoSciences (Grenoble) for support.
The Flatiron Institute is a division of the Simons Foundation.  DMC and YY were supported by DOE Grant NA DE-NA0001789. V.G. and C.P. were supported by the Agence Nationale de la Recherche (ANR) France, under the program ``Accueil de Chercheurs de Haut Niveau 2015'' project: HyLightExtreme. Computer time was provided by the PRACE Project 2016143296, ISCRAB the high-performance computer resources from Grand Equipement National de Calcul Intensif (GENCI) Allocations 2018-A0030910282 and by an allocation of the Blue Waters sustained petascale computing project, supported by the National Science Foundation (Award OCI 07- 25070) and the State of Illinois, and by GRICAD infrastructure (Grenoble) within the Equip@Meso project,
ANR-10-EQPX-29-01. 

\end{acknowledgments}


\begin{thebibliography}{99}
\bibitem{footnote} We can adapt the proof of Kohn \cite{Kohn58}
 by noting that 
the extra particle (hole) propagator does not interact with the other particles before $t=0$.
The equivalent graphs of class I contribute with $\pm 1$ for particle and hole excitations,
the equivalent graphs of class IIA remain, contributing $\pm (1/\epsilon -1)/2$, whereas graphs of
class IIB are absent.
\bibitem{BrillouinFootnote} The reciprocal space information is not lost. By keeping track of the
Bloch momentum of the states we can determine the locations of the HOMO and LUMO, but that has not been done here.
\end{thebibliography}
\end{document}


\title{Electronic band gaps from Quantum Monte Carlo methods (Supplemental Materials)}
\author{Yubo Yang}
\affiliation{Department of Physics, University of Illinois Urbana-Champaign, Urbana, Illinois 61801, USA}
\author{Vitaly Gorelov}
\affiliation{Maison de la Simulation, CEA, CNRS, Univ. Paris-Sud, UVSQ, Universit{\'e} Paris-Saclay, 91191 Gif-sur-Yvette, France}
\author{Carlo Pierleoni}
\affiliation{Maison de la Simulation, CEA, CNRS, Univ. Paris-Sud, UVSQ, Universit{\'e} Paris-Saclay, 91191 Gif-sur-Yvette, France}
\affiliation{Department of Physical and Chemical Sciences, University of L'Aquila, Via Vetoio 10, I-67010 L'Aquila, Italy}
\author{David M. Ceperley}
\affiliation{Department of Physics, University of Illinois Urbana-Champaign, Urbana, Illinois 61801, USA}
\author{Markus Holzmann} 
\affiliation{Univ. Grenoble Alpes, CNRS, LPMMC, 3800 Grenoble, France}
\affiliation{Institut Laue Langevin, BP 156, F-38042 Grenoble Cedex 9, France}
\maketitle

\subsection{Diagrammatic proof of the asymptotic behavior of the matrix elements}

In Ref.~\cite{Kohn58}, W. Kohn proved
that the static dielectric constant is given by the matrix elements, Eq.(10) in the main
paper,
\begin{equation}
\lim_{\qvec',\qvec \to 0,\qvec' \ne \qvec} \left\langle
\Psi_0(N_e\pm 1; \qvec) | \rho_{\qvec -\qvec'} | \Psi_0(N_e \pm 1;\qvec') \right\rangle
= \pm \frac{1}{\epsilon_0}
\end{equation}
based on diagrammatic perturbation theory where
$\Psi_0(N_e\pm 1; \qvec)$ denotes the fully insulating $N_e$ electron state with one
electron added/ removed in the particle/ hole orbital $\qvec$.
Here, we show how to adapt the diagrammatic proof of Ref.~\cite{Kohn58} to obtain the following matrix elements needed 
for the dominant size corrections of the band gap:
\bea
\lim_{\Kvec, \qvec \to 0}
\frac{\langle \Psi_{\Kvec}^{N+1} | \rho_\qvec a_{\Kvec-\qvec}^\dagger | \Psi_0^N \rangle}
{\langle \Psi_{\Kvec}^{N+1} |
 a_{\Kvec}^\dagger | \Psi_0^N \rangle}
=
\frac{1}{2}\left[\frac{1}{ \epsilon_0} +1 \right],
\label{particle}
\eea
and
\bea
\lim_{\Kvec, \qvec \to 0}
\frac{\langle \Psi_{\Kvec}^{N-1} | \rho_\qvec a_{\Kvec+\qvec} | \Psi_0^N \rangle}
{\langle \Psi_{\Kvec}^{N-1} |
 a_{\Kvec} | \Psi_0^{N} \rangle}
=
-\frac12 \left[1+ \frac{1}{\epsilon_0}\right].
\label{hole}
\eea
Both relations can be obtained directly from Kohn's approach by noting that 
the extra particle (hole) propagator does not interact with the other particles before $t=0$.
In the following, we sketch the argument for a particle excitation.

In figure \ref{normalization}, we show a typical graph of the
perturbation series contributing to the particle propagator in the denominator of eq.~(\ref{particle}), $\langle \Psi_{\Kvec}^{N+1} | a_{\Kvec}^\dagger | \Psi_0^N \rangle$. Interband transitions are suppressed in the limit of vanishing momenta, so we need to consider only diagrams where the particle excitation is in the conduction band.
Diagrams contributing to the numerator $\langle \Psi_{\Kvec}^{N+1} | \rho_\qvec a_{\Kvec-\qvec}^\dagger | \Psi_0^N \rangle$ fall into two classes.
In Class I, we simply attach the external density operator $\rho_{\Kvec'-\Kvec}$ to split a propagator line at $t=0$.
This is represented by the external dashed line $\Kvec'-\Kvec$, which ends at $X$ in Figure \ref{classI}. Insertion at $V_1$
contains intraband transitions for the matrix elements of  $\rho_{\Kvec'-\Kvec}$, giving $1$ in
the limit $\Kvec' \to \Kvec$, and the remaining term contributes exactly the same as the
equivalent diagram in the denominator. Insertion at $V_2$ cancels the contribution from an insertion at $V_3$ in the
limit of vanishing momenta, because they differ by only a minus sign due to the difference in the
number of hole propagators involved.
Therefore, all diagrams of class I contribute exactly $1$ to
the l.h.s. of Eq.~(\ref{particle}).

Diagrams of class II create an additional particle-hole pair at $t=0$, playing a similar role as one
side of the
Coulomb interaction vertex. A typical graph is shown in Figure \ref{classII}. In the limit of 
vanishing external momenta, these diagrams separate into a normalization diagram and a graph
involved in the static density response function, e.g. a simple electron-hole bubble attached at
point $V$ in Figure  \ref{classII}.
These diagrams exactly coincide with 
those of class IIA of Ref.~\cite{Kohn58}, whereas class IIB diagrams of Ref.~\cite{Kohn58} 
do not occur in our calculation because they
are built out of
dielectric graphs attached at points $V$ with $t<0$.
Following the line of argument presented in Ref.~\cite{Kohn58}, we see that
class II diagrams contribute with $\frac{1}{2}\left(\frac{1}{\epsilon_0} -1  \right)$
to  the  l.h.s. of Eq.~(\ref{particle}).

Since class I and II diagrams are the only non-vanishing graphs, we obtain Eq.~(\ref{particle}).
The same argument can be applied to the hole excitation diagrams, differing by only a global sign
due to the additional hole propagator involved, giving Eq.~(\ref{hole}).

\begin{figure}[h]
\setlength{\unitlength}{1.5cm}
\begin{picture}(4,2) \thicklines
\put(.75,.6){\line(-1,2){0.2}}
\put(.55,1.){\vector(1,-2){0.15}}
\put(1.,1.35){\vector(0,1){0.35}}
\put(1.,1.35){\line(0,1){0.5}}
\put(1.,1.35){\line(-1,-3){.25}}
\multiput(0.1,1)(0.06,0.){40}{\line(1,0){.02}}
\put(0.3,0.8){${}_{\Kvec}$}
\put(.7,1.7){${}_{\Kvec}$}
\put(2.5,.95){$t=0$}
\multiput(.75,0.6)(0.1,0.){10}{\line(1,0){.05}}
\multiput(1.,1.35)(0.1,0.){8}{\line(1,0){.05}}
\qbezier(1.72,.6)(2.12,.95)(1.72,1.35)
\qbezier(1.72,.6)(1.32,.95)(1.72,1.35)
\put(1.57,1.16){\vector(1,3){0.03}}
\put(1.86,1.16){\vector(1,-3){0.03}}

\qbezier(2.4,0.)(2.12,.25)(2.4,.5)
\qbezier(2.4,0.)(2.68,.25)(2.4,.5)
\multiput(2.4,.5)(0.1,0.){5}{\line(1,0){.05}}
\multiput(2.4,.)(0.1,0.){5}{\line(1,0){.05}}
\qbezier(2.9,0.)(2.62,.25)(2.9,.5)
\qbezier(2.9,0.)(3.18,.25)(2.9,.5)
\end{picture}
\caption{A graph occurring in the perturbation expansion of the normalization $\langle \Psi_\Kvec^{N+1} | a_\Kvec^\dagger |\Psi_0^N \rangle $. The disconnected part contains only ground state diagrams.}
\label{normalization}
\end{figure}

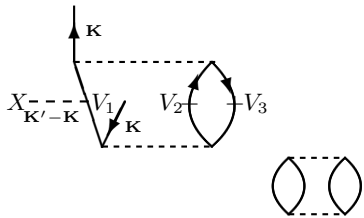
\begin{figure}
\setlength{\unitlength}{1.5cm}
\begin{picture}(4,2) \thicklines
\put(.75,.6){\line(1,2){0.2}}
\put(.95,1.){\vector(-1,-2){0.15}}
\put(.5,1.35){\vector(0,1){0.35}}
\put(.5,1.35){\line(0,1){0.5}}
\put(.75,.6){\line(-1,3){.25}}
\put(0.95,0.75){${}_{\Kvec}$}
\put(.6,1.6){${}_{\Kvec}$}
\multiput(.75,0.6)(0.1,0.){10}{\line(1,0){.05}}
\multiput(.5,1.35)(0.1,0.){13}{\line(1,0){.05}}
\qbezier(1.72,.6)(2.12,.95)(1.72,1.35)
\qbezier(1.72,.6)(1.32,.95)(1.72,1.35)
\put(1.57,1.16){\vector(1,3){0.03}}
\put(1.86,1.16){\vector(1,-3){0.03}}
\put(-0.1,.925){$X$}
\multiput(.1,1)(0.15,0.){4}{\line(1,0){.08}}
\put(.65,.925){$V_1$}
\put(1.44,.925){$+$}
\put(1.25,.925){$V_2$}
\put(1.84,.925){$+$}
\put(2.,.925){$V_3$}
\put(0.05,.85){${}_{\Kvec'- \Kvec}$}

\qbezier(2.4,0.)(2.12,.25)(2.4,.5)
\qbezier(2.4,0.)(2.68,.25)(2.4,.5)
\multiput(2.4,.5)(0.1,0.){5}{\line(1,0){.05}}
\multiput(2.4,.)(0.1,0.){5}{\line(1,0){.05}}
\qbezier(2.9,0.)(2.62,.25)(2.9,.5)
\qbezier(2.9,0.)(3.18,.25)(2.9,.5)
\end{picture}
\caption{A graph occurring in class I of the perturbation expansion of the numerator
$\langle \Psi_{\Kvec'}^{N+1} | \rho_{\Kvec' -\Kvec}  a_\Kvec^\dagger |\Psi_0^N \rangle $,
where $\rho_{\Kvec' -\Kvec}$ is inserted at the point $V_1$ at $t=0$.
}\label{classI}
\end{figure}

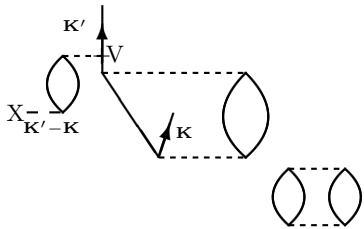
\begin{figure}[h]
\setlength{\unitlength}{1.5cm}
\begin{picture}(4,2) \thicklines
\put(1.25,0.6){\vector(1,3){0.1}}
\put(1.25,0.6){\line(1,3){0.13}}
\put(.75,1.44){\vector(0,1){0.35}}
\put(.75,1.35){\line(0,1){0.6}}
\put(0.75,1.35){\line(2,-3){.5}}
\multiput(0.08,1)(0.2,0.){2}{\line(1,0){.1}}
\multiput(0.4,1.5)(0.1,0.){4}{\line(1,0){.05}}
\put(.78,1.45){V}
\put(.66,1.445){+}
\qbezier(0.4,1.)(0.12,1.25)(0.4,1.5)
\qbezier(0.4,1.)(0.68,1.25)(0.4,1.5)
\put(1.4,0.8){${}_{\Kvec}$}
\put(.4,1.75){${}_{\Kvec'}$}
\put(-0.1,.925){X}
\put(0.05,.85){${}_{\Kvec'- \Kvec}$}
\multiput(1.25,0.6)(0.1,0.){8}{\line(1,0){.05}}
\multiput(.75,1.35)(0.1,0.){13}{\line(1,0){.05}}
\qbezier(2.02,.6)(2.42,.95)(2.02,1.35)
\qbezier(2.02,.6)(1.62,.95)(2.02,1.35)

\qbezier(2.4,0.)(2.12,.25)(2.4,.5)
\qbezier(2.4,0.)(2.68,.25)(2.4,.5)
\multiput(2.4,.5)(0.1,0.){5}{\line(1,0){.05}}
\multiput(2.4,.)(0.1,0.){5}{\line(1,0){.05}}
\qbezier(2.9,0.)(2.62,.25)(2.9,.5)
\qbezier(2.9,0.)(3.18,.25)(2.9,.5)
\end{picture}
\caption{A graph occurring in class II of the perturbation expansion of the numerator
$\langle \Psi_{\Kvec'}^{N+1} | \rho_{\Kvec' -\Kvec}  a_\Kvec^\dagger |\Psi_0^N \rangle $.}
\label{classII}
\end{figure}

\subsection{Simulation cell}

For hydrogen, we studied the C2/c-24 structure of molecular solid hydrogen at two different densities $r_s=1.38$ and $r_s=1.34$, corresponding to pressures of 234GPa and 285GPa, respectively as estimated with QMC. The crystalline structures have been optimized by variable cell structural relaxation with DFT vdW-DF1 at pressures of 250GPa and 300GPa, respectively. (QMC and DFT-vdW-DF1 pressures differ by $\sim$10-20GPa \cite{Clay2014,Gorelov2019}).

For carbon and silicon, we studied the diamond lattice at experimental densities at ambient conditions $r_s=1.318$ and $r_s=2.005$, respectively. Then the lattice constant of the conventional unit cell was  $L=6.74065$ bohr and $L=10.2622$ bohr for carbon and silicon, respectively. The ideal simple cubic Madelung constant for the conventional cell is $v_M\equiv [\sum-\int] v_k=-2.837297479/L\text{ ha/e}$. 

\subsection{Insulating state total energy, structure factor, and finite-size corrections}

In Table~\ref{tab:E0}, we show total energies and variances of the insulating state of all systems studied. The hydrogen total energies are upper bounds to the exact ground-state energy at the corresponding density and system size because no pseudopotential was used. The carbon and silicon energies and variances are comparable only to calculations that use the same pseudopotentials. More accurate trial wavefunctions have a lower energy and variance as can be seen in the change from SJ to BF wavefunction for carbon and silicon.

\begin{table}[h]
\caption{RQMC/DMC energies per electron (in Ha), $E_0/N_e$,
for H$_2$, C and S crystals with
Slater-Jastrow (SJ) trial wave functions and backflow transformations (BF). $\sigma^2_E/N_e$ is the DMC energy variance. The RQMC energy variance depends on projection time and are not shown. $(E_0/N_e)_{\infty}$ denotes the finite-size
corrected results including two-body potential and kinetic energy corrections. 
\label{tab:E0}
}
\begin{tabular}{llrrrrr}
\hline\hline
& $r_s$ &  & $N_e$ & $E_0/N_e$ & $\sigma_E^2/N_e$ & $(E_0/N_e)_{\infty}$ \\
\hline
H$_2$ & 1.38  & BF  & 96   & -0.539206(4) &  & -0.536056(4) \\
      & 1.34  & BF  & 96   & -0.531888(5) &  & -0.528548(5) \\
\hline
C   & 1.318 & BF  & 32   & -1.435550(5)  & 0.016676(8)  &        \\
    &       & SJ  & 32   & -1.434641(6)  & 0.022406(8)  & -1.424061(6) \\
    &       &     & 256  & -1.425975(1)  & 0.026937(3)  & -1.424635(1) \\
\hline
Si  & 2.005 & BF  & 32   & -0.989922(4)  & 0.005179(4)  &       \\
    &       & SJ  & 32   & -0.989240(7)  & 0.008349(4)  & -0.982740(7) \\
 &  &             & 256  & -0.983847(1)  & 0.008709(2)  & -0.983017(1) \\
 &  &             & 864  & -0.983296(1)  & 0.009115(2)  & -0.983046(1)\\
\hline
\end{tabular}
\end{table}

\begin{figure}[ht!]
\begin{subfigure}{0.48\textwidth}
\includegraphics[width=0.9\textwidth]{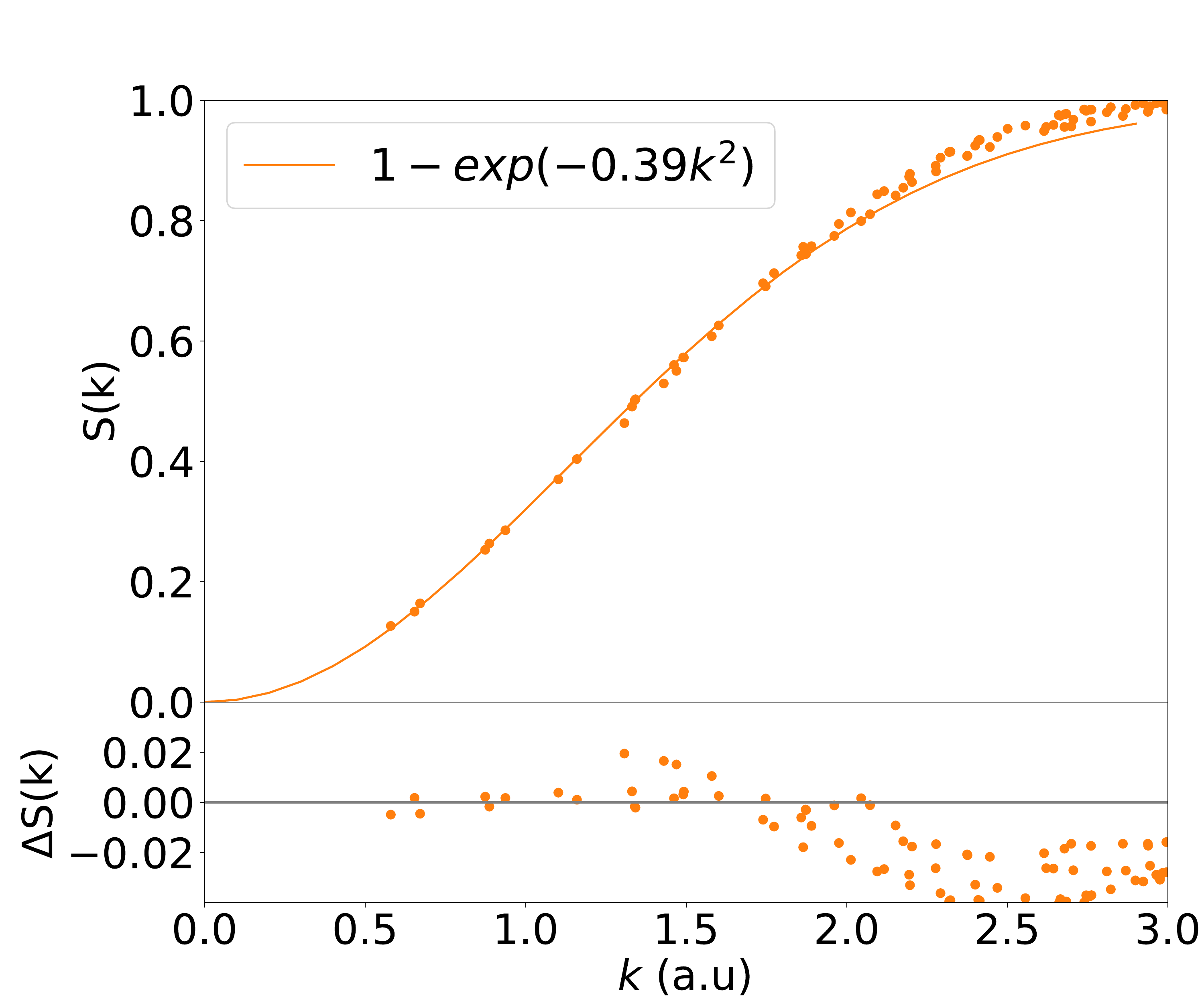}
\caption{H$_2$ $r_s=1.34$ 285 GPa}
\end{subfigure}
\begin{subfigure}{0.48\textwidth}
\includegraphics[width=0.9\textwidth]{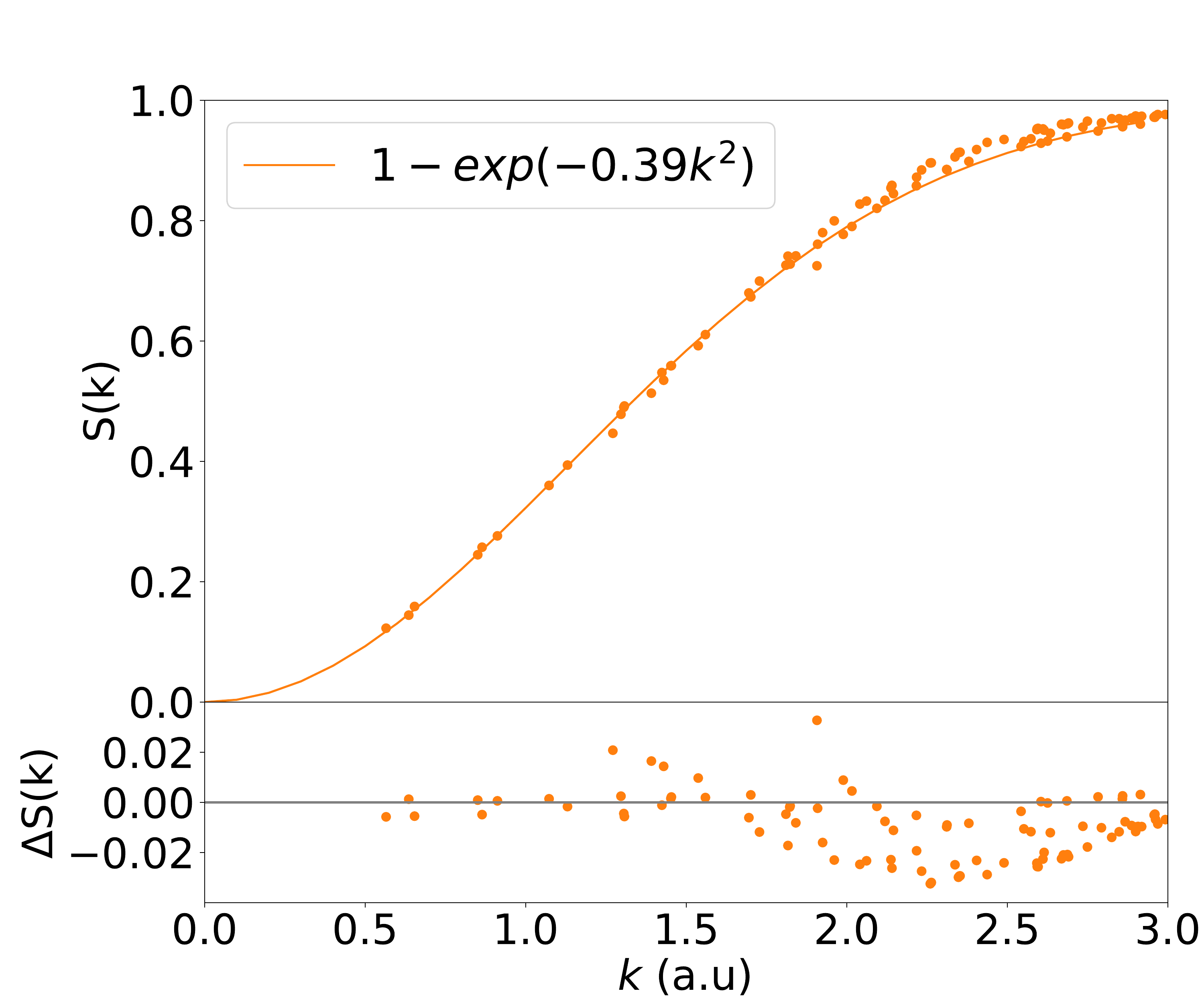}
\caption{H$_2$ $r_s=1.38$ 234 GPa}
\end{subfigure}
\caption{Structure factor S(k) of molecular hydrogen. The dots are RQMC data. The lines are fits to $1-e^{-\alpha k^2}$. The bottom panels show the fit residues. \label{fig:hsk}}
\end{figure}

\begin{figure}[ht!]
\begin{subfigure}{0.48\textwidth}
\includegraphics[width=\columnwidth]{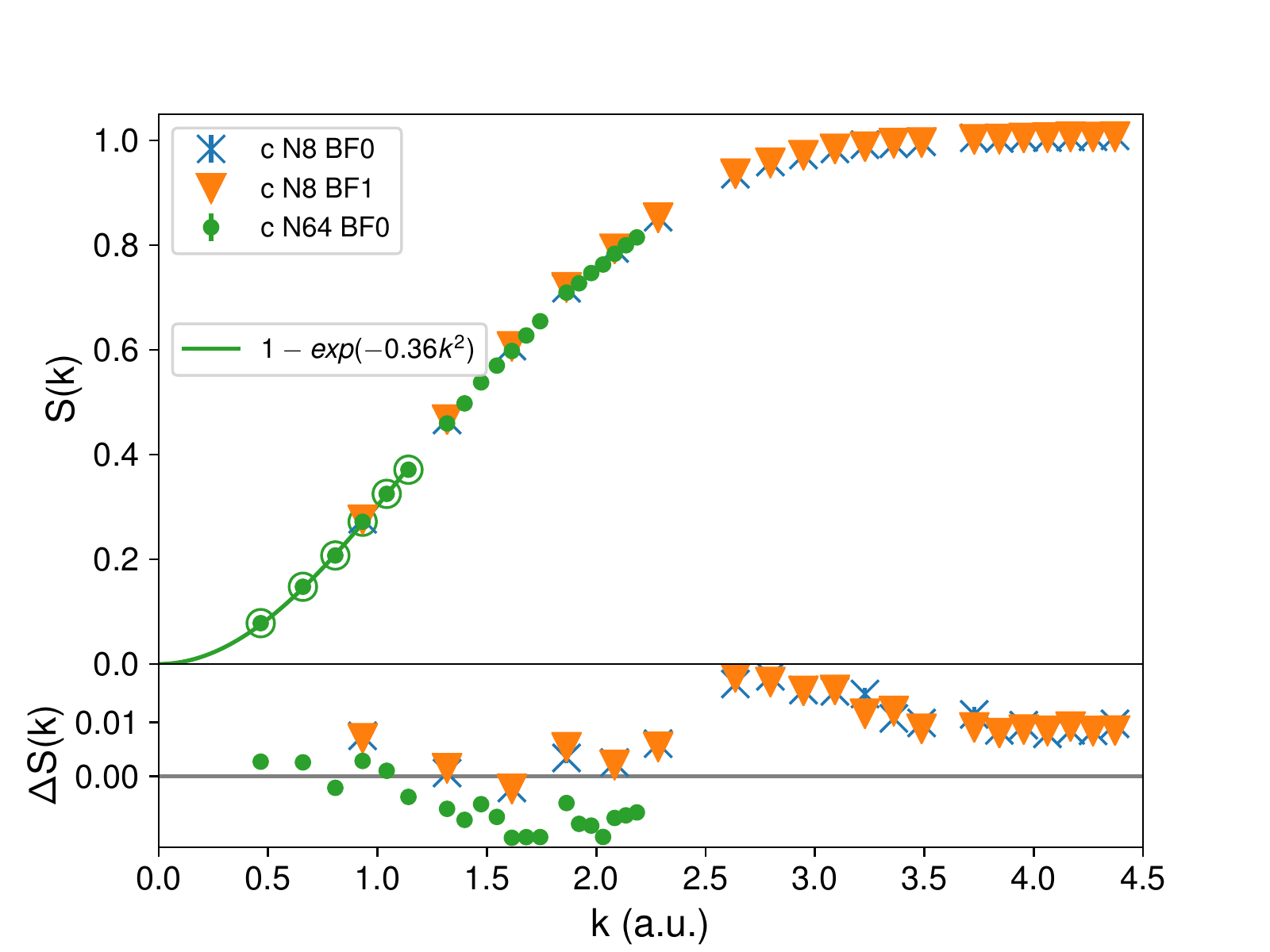}
\caption{C structure factor S(k)}
\end{subfigure}
\begin{subfigure}{0.48\textwidth}
\includegraphics[width=\columnwidth]{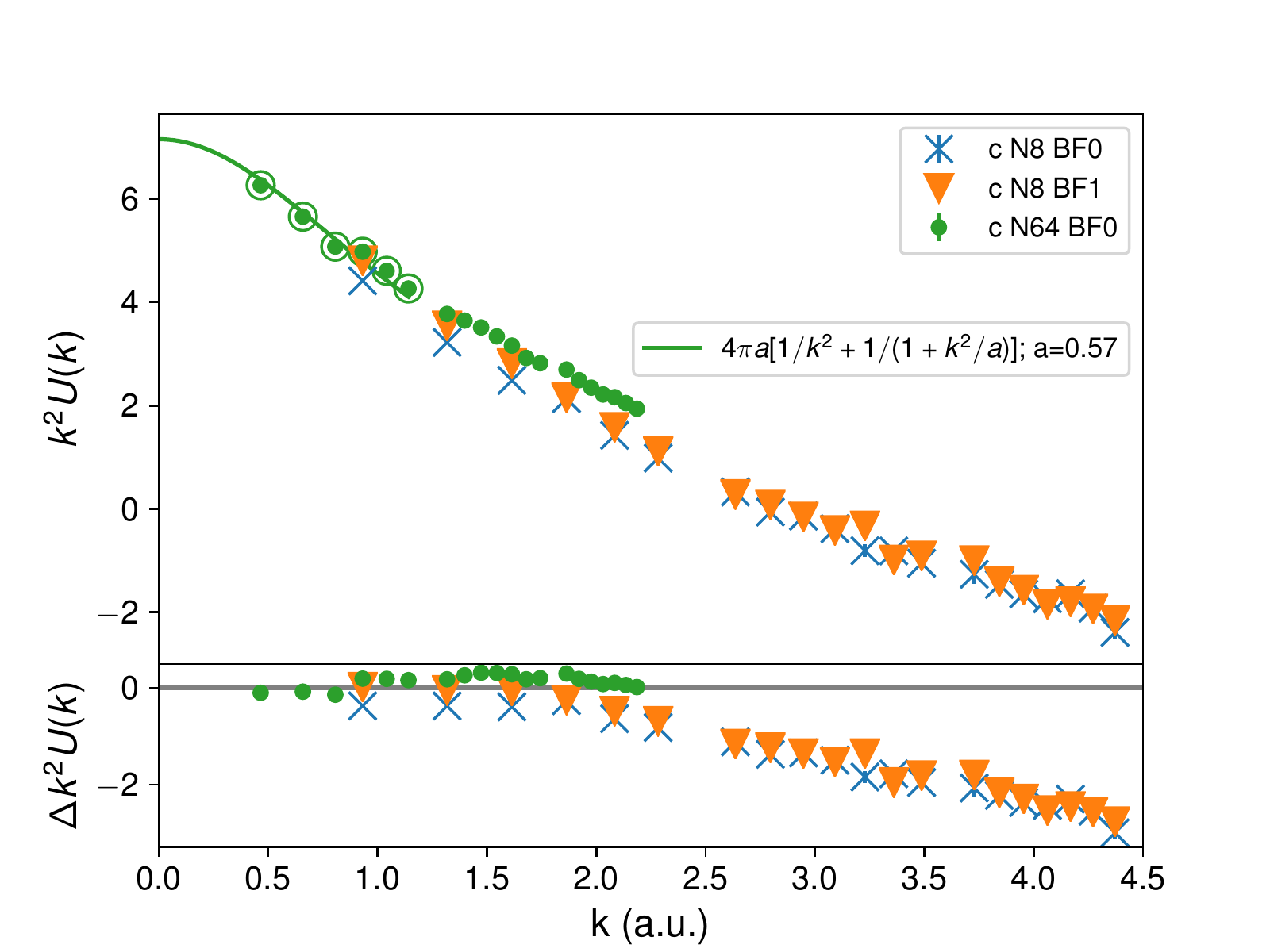}
\caption{C Jastrow potential U(k)}
\end{subfigure}
\begin{subfigure}{0.48\textwidth}
\includegraphics[width=\columnwidth]{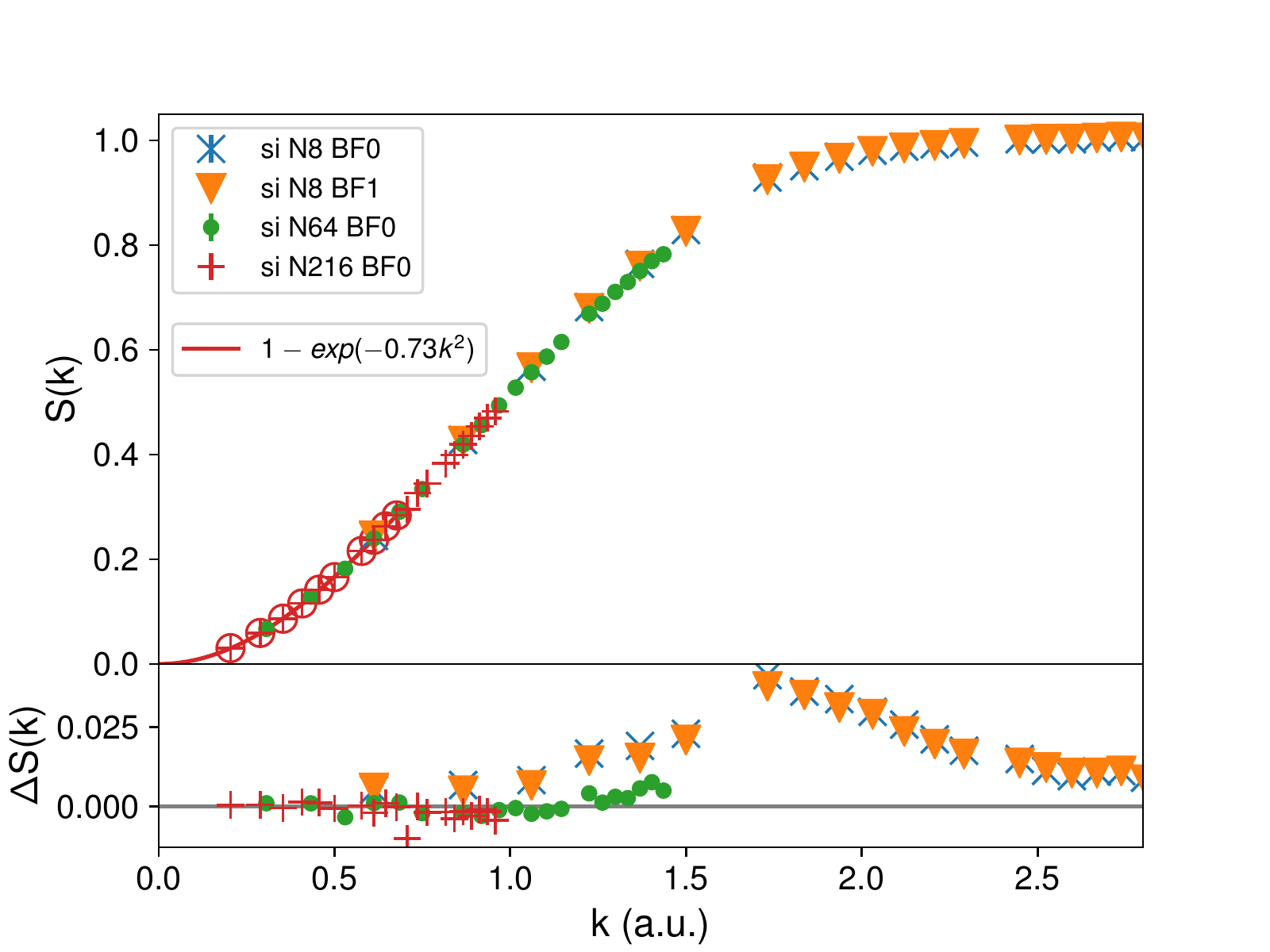}
\caption{Si structure factor S(k)}
\end{subfigure}
\begin{subfigure}{0.48\textwidth}
\includegraphics[width=\columnwidth]{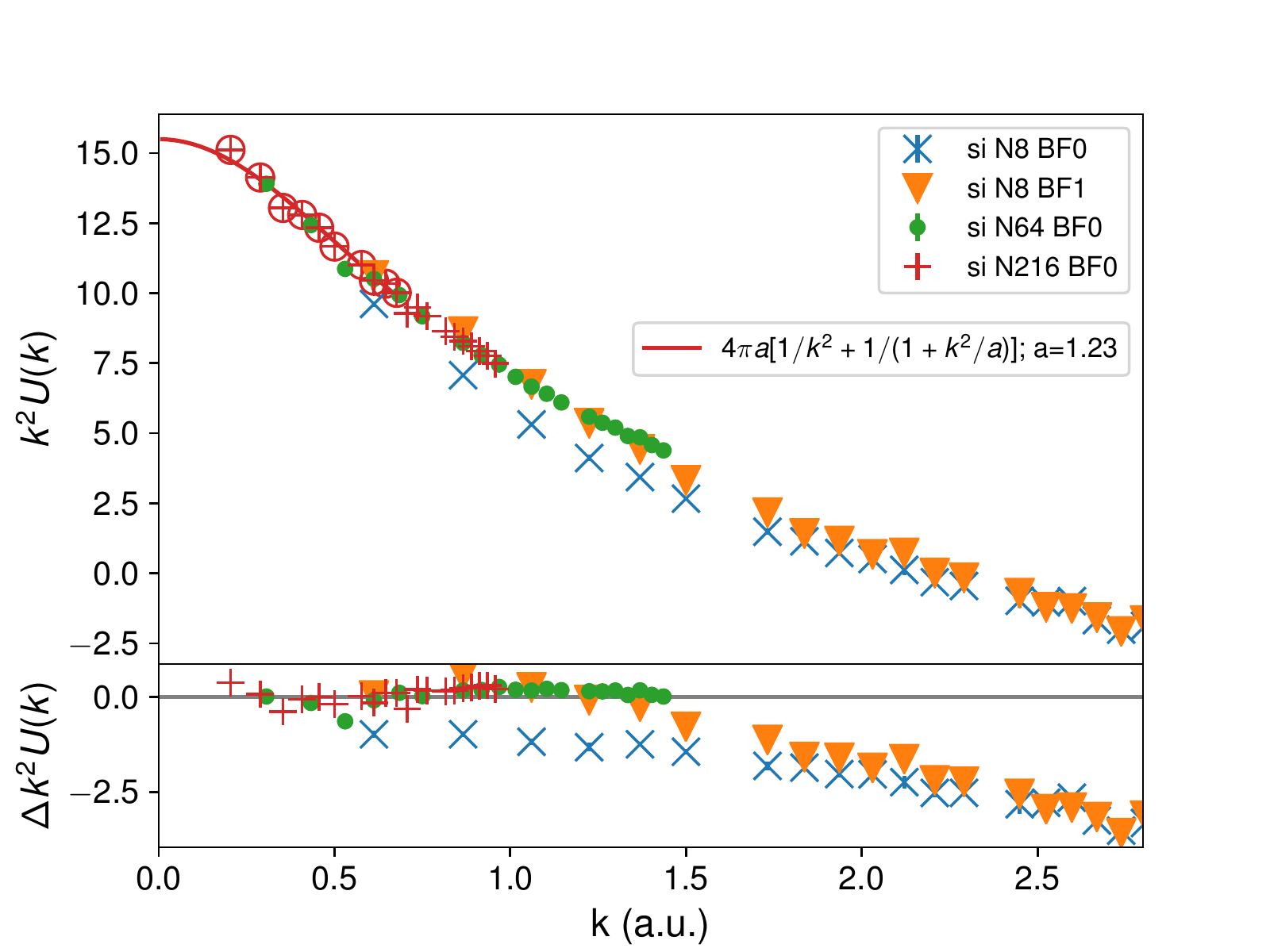}
\caption{Si Jastrow potential U(k)}
\end{subfigure}
\caption{\small Structure factor S(k) and Jastrow pair potential U(k) for carbon and silicon. The markers in the $S(k)$ plot are QMC data linearly extrapolated to remove mixed-estimator error. The markers in the $U(k)$ plot are approximations to the fixed-node Jastrow pair potential $U(k)= U_T(k)+\frac{S_{DMC}^{-1}(k)-S_{VMC}^{-1}(k)}{n}$, where $n$ is the electron density. The lines in the upper panels show $S(k)$ and $U(k)$ models fit to data from the largest system (circled points of the same color).\label{fig:skuk}}
\end{figure}

The fluctuating part of the static structure factors $S(k)$ from our simulations are shown in Fig.~\ref{fig:hsk} and ~\ref{fig:skuk}. The small-k region of $S(k)$ is fit to the one-parameter model $(1-\exp(-\alpha k^2))$ at the largest system size available for each structure. As shown in the bottom panels in Fig.~\ref{fig:skuk}, the fit residue is small and flat as $k\rightarrow0$. Further, the same fit matches the data from all sizes rather well. We therefore use one $S(k)$ for each element to correct the potential energy finite-size errors at all system sizes. Also shown in Fig.~\ref{fig:skuk} are the Jastrow pair potential for carbon and silicon $U(k)$. The $1/k^2$ divergence of $U(k)$ is used to estimate the kinetic energy correction. Similar to $S(k)$, we fit the small-k region of $U(k)$ to the one-parameter model $4\pi a\left(\frac{1}{k^2}+\frac{1}{k^2+1/a}\right)$. We correct both $S(k)$ and $U(k)$ for mixed-estimator bias to calculate unbiased finite-size corrections for fixed-node potential and kinetic energies. In Fig.~\ref{fig:skuk}, the fixed-node static structure factor is approximated by linear extrapolation $S(k)=2S_{DMC}-S_{VMC}$. The fixed-node Jastrow pair potential $U(k)= U_T(k)+\frac{S_{DMC}^{-1}(k)-S_{VMC}^{-1}(k)}{n}$, where $n$ is the electron density. The Jastrow potentials of the hydrogen structures are not shown, because the asymptotic behavior of the analytical trial wavefunction for hydrogen is exactly known~\cite{Holzmann2003}. Further, RQMC has no mixed-estimator error, so these corrections do not apply. If the RPA is exact, then $u_k = \frac{4\pi}{\omega_p k^2}$, where the plasmon frequency in 3D $\omega_p=\sqrt{\frac{3}{r_s^3}}$.
This implies that for hydrogen  $a=\frac{1}{\omega_p}$. Assuming $\lim_{k\rightarrow0} u_k=\frac{4\pi a}{k^2}$, the $c$ variable defined after eq.~(14) in the main text can be shown to be $c=(a\omega_p)^2=1$.

\begin{table}[h]
\caption{Fit parameters of $S(k)=1-e^{-\alpha k^2}$ and $U(k)=4\pi a(\frac{1}{k^2}+\frac{1}{k^2+1/a})$, $c=3a^2/r_s^3$ and total energy finite-size correction calculated from $\rho(k)$ ($\delta V_N^s$), $S(k)$ ($\delta V_N^f$), and $u(k)$ ($\delta T_N$) as compared to that from 1/N extrapolation. All energies are in units of mha/e.}
\label{tab:E0_FS}
\begin{tabular}{llrlllllcrll}
\toprule
& wf & $N_e$ & $\alpha$ & $a$ & $\delta V_N^f$ & $\delta V_N^s$ & $\delta T^f_N$ & $c$ & $\delta T_N^{BF}$ & $\delta E_N$ & $\delta E_N^{extrap.}$ \\
\midrule
H$_2$ 1.38& BF &  96 &  0.386 & 0.936 & 2.59 & -0.047 & 2.59 &     1 & -1.98 & 3.15 &          \\
\hfill1.34& BF &  96 &  0.390 & 0.896 & 2.74 & -0.056 & 2.74 &     1 & -2.08 & 3.34 &          \\
C         & BF &  32 &  0.365 & 0.572 & 7.47 & -0.08 & 3.10 & 0.428 & &     &          \\
          & SJ &  32 &  0.370 & 0.544 & 7.46 & -0.08 & 2.81 & 0.388 & 0 & 10.27& 9.904(6) \\
          & SJ & 256 &  0.357 & 0.555 & 0.92 & -0.00 & 0.37 & 0.403 & 0 & 1.29 & 1.234(2) \\
\midrule
Si & BF &  32 & 0.74 & 1.25 & 4.32 & -0.10 & 2.46 & 0.581 & & & \\
   & SJ &  32 & 0.75 & 1.07 & 4.34 & -0.10 & 1.80 & 0.427 & 0 & 6.04 & 6.176(7) \\
   & SJ & 256 & 0.73 & 1.00 & 0.53 & -0.00 & 0.19 & 0.372 & 0 & 0.72 & 0.783(2) \\
   & SJ & 864 & 0.72 & 1.25 & 0.16 & -0.00 & 0.09 & 0.577 & 0 & 0.25 & 0.232(2) \\
\bottomrule
\end{tabular}
\end{table}

\begin{figure}[h]
\begin{subfigure}{0.48\textwidth}
\includegraphics[width=\columnwidth]{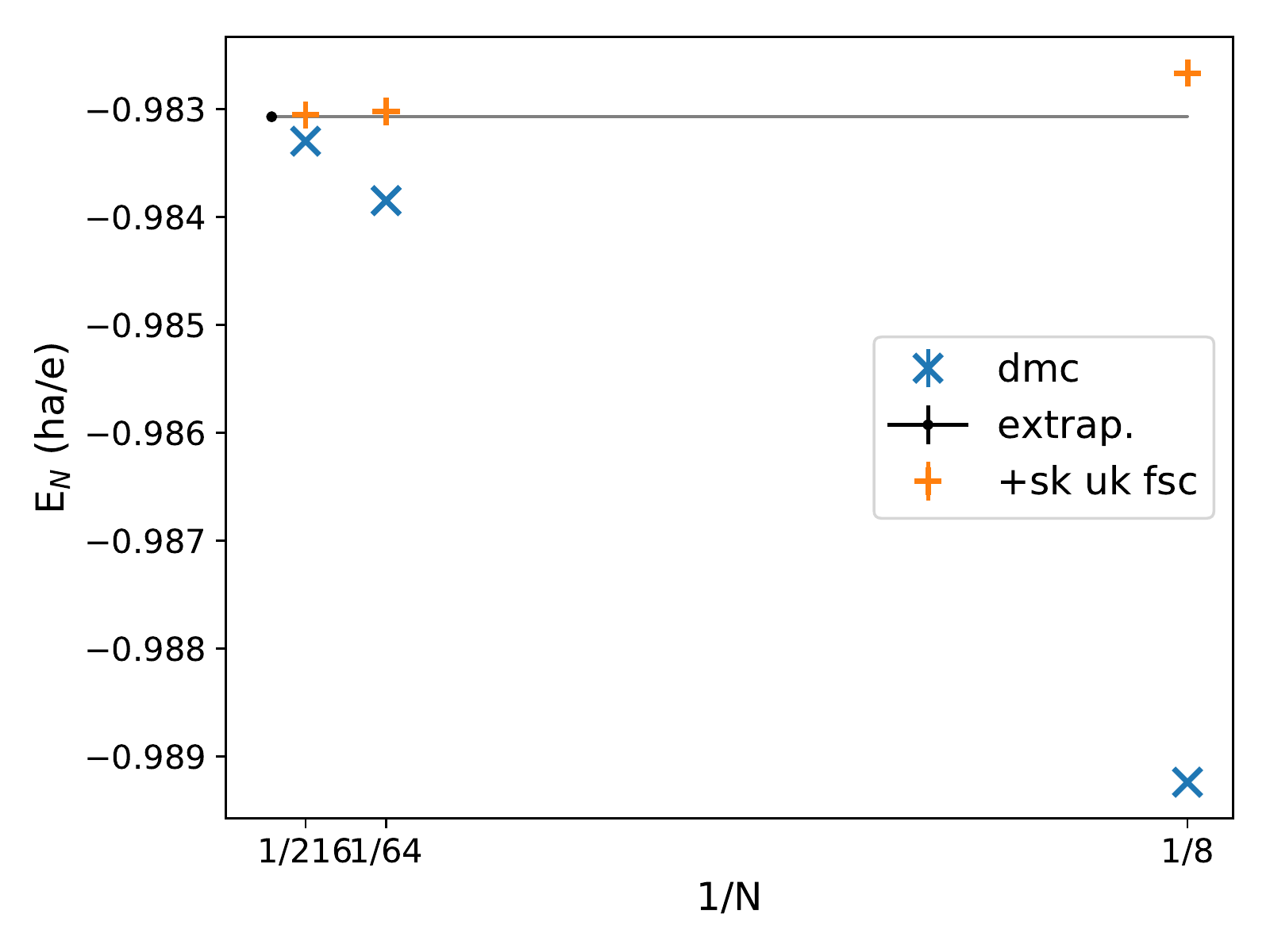}
\caption{Si total energy\label{fig:si-etot-extrap}}
\end{subfigure}
\begin{subfigure}{0.48\textwidth}
\includegraphics[width=\columnwidth]{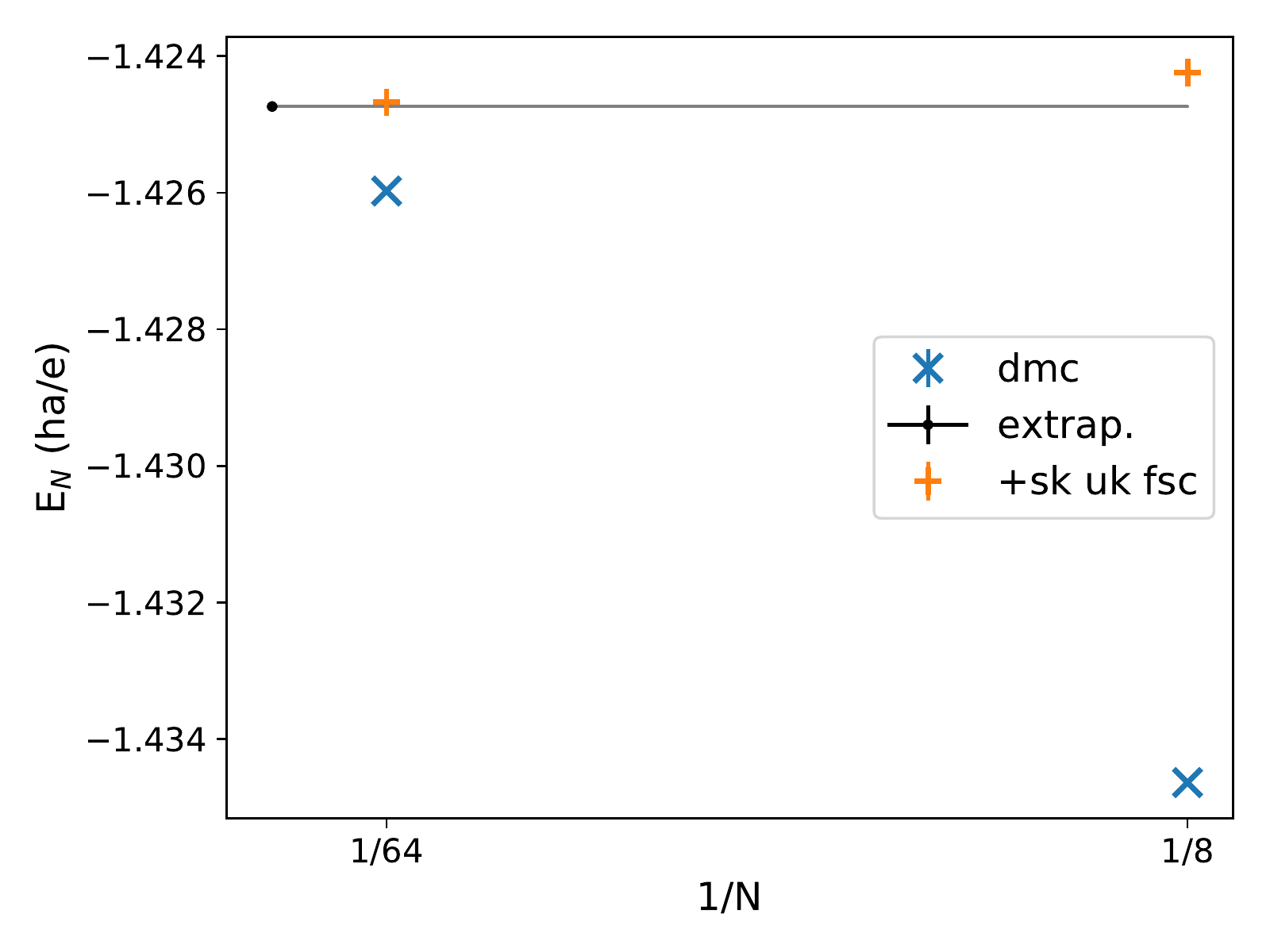}
\caption{C total energy\label{fig:c-etot-extrap}}
\end{subfigure}
\caption{Total energy finite-size correction}
\end{figure}

\subsection{Extract dielectric constant from structure factor}

\begin{figure*}[h]
\begin{minipage}[b]{.4\columnwidth}
\includegraphics[width=\columnwidth]{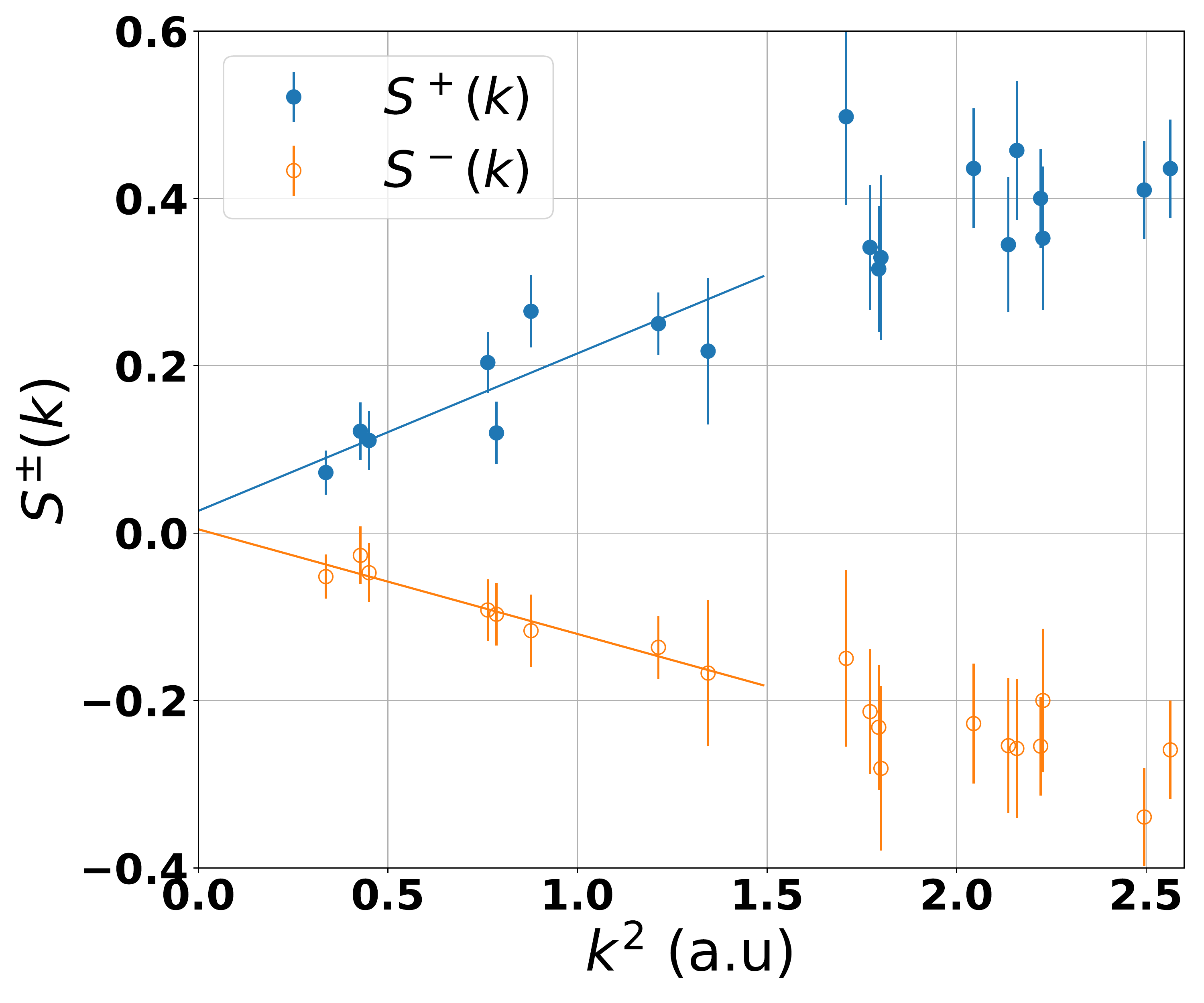}
(a) H$_2$ $r_s=1.34$ 285GPa
\end{minipage}
\begin{minipage}[b]{.4\columnwidth}
\includegraphics[width=\columnwidth]{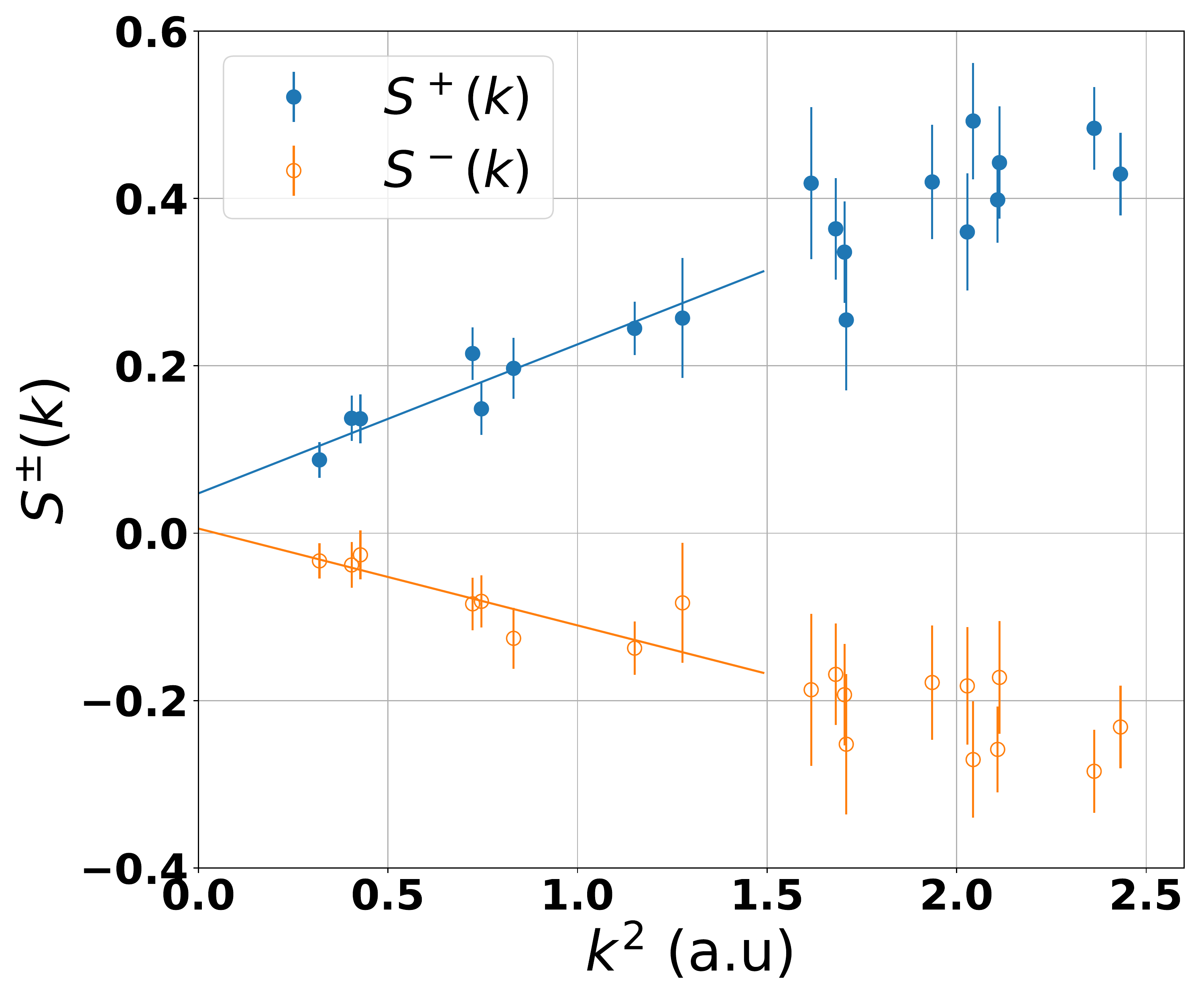}
(b) H$_2$ $r_s=1.38$ 234GPa
\end{minipage}
\caption{Change to static structure factor as an electron (blue filled circle) or a hole (orange open circle) is added to the neutral system. The lines are the fits to obtain asymptotic values at $k\to0$.
 \label{fig:Sofk}}
\end{figure*}

\begin{figure*}[h]
\begin{minipage}[b]{.45\columnwidth}
\includegraphics[width=\columnwidth]{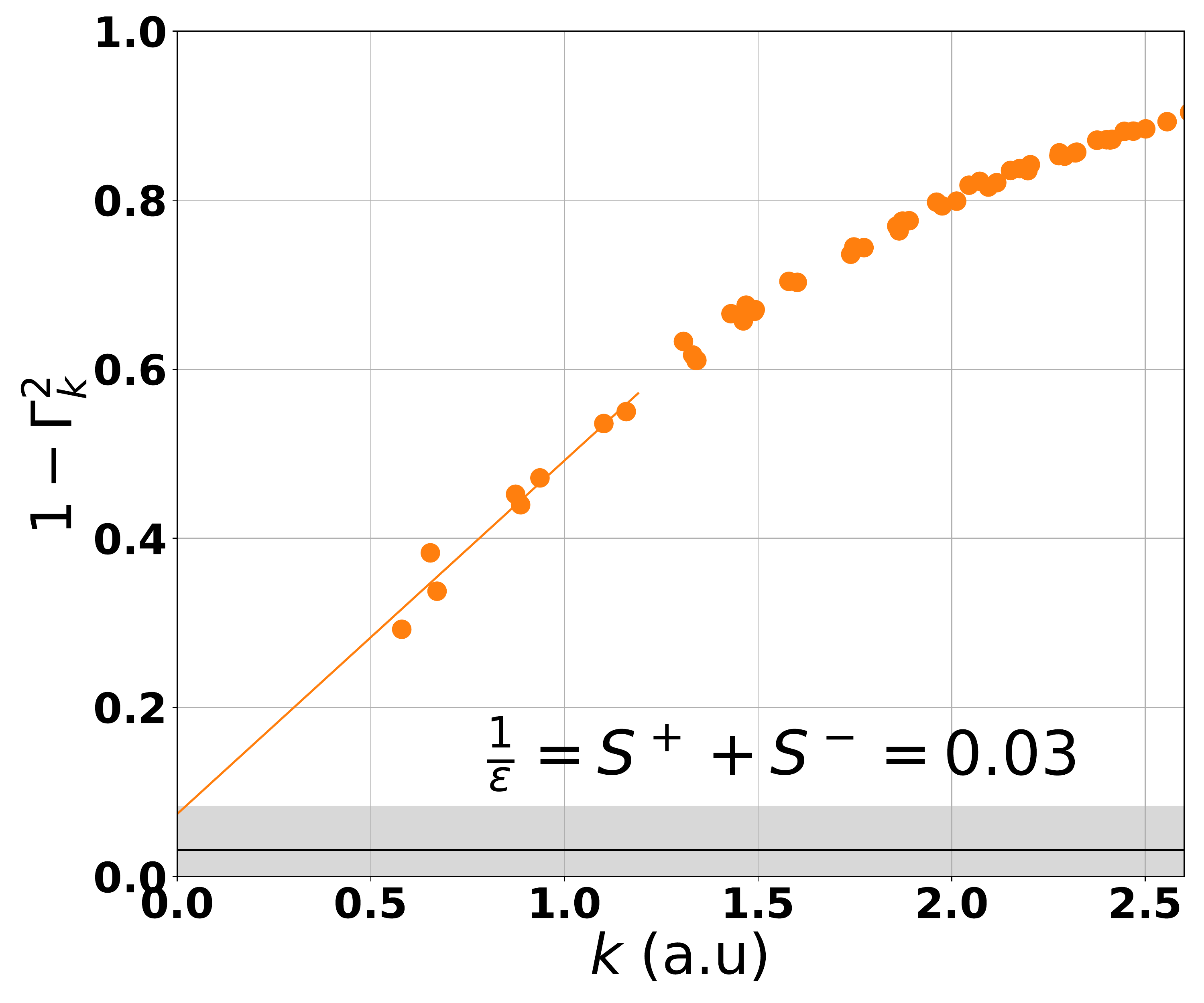}
(a) $r_s=1.34$, 285GPa
\end{minipage}
\begin{minipage}[b]{.45\columnwidth}
\includegraphics[width=\columnwidth]{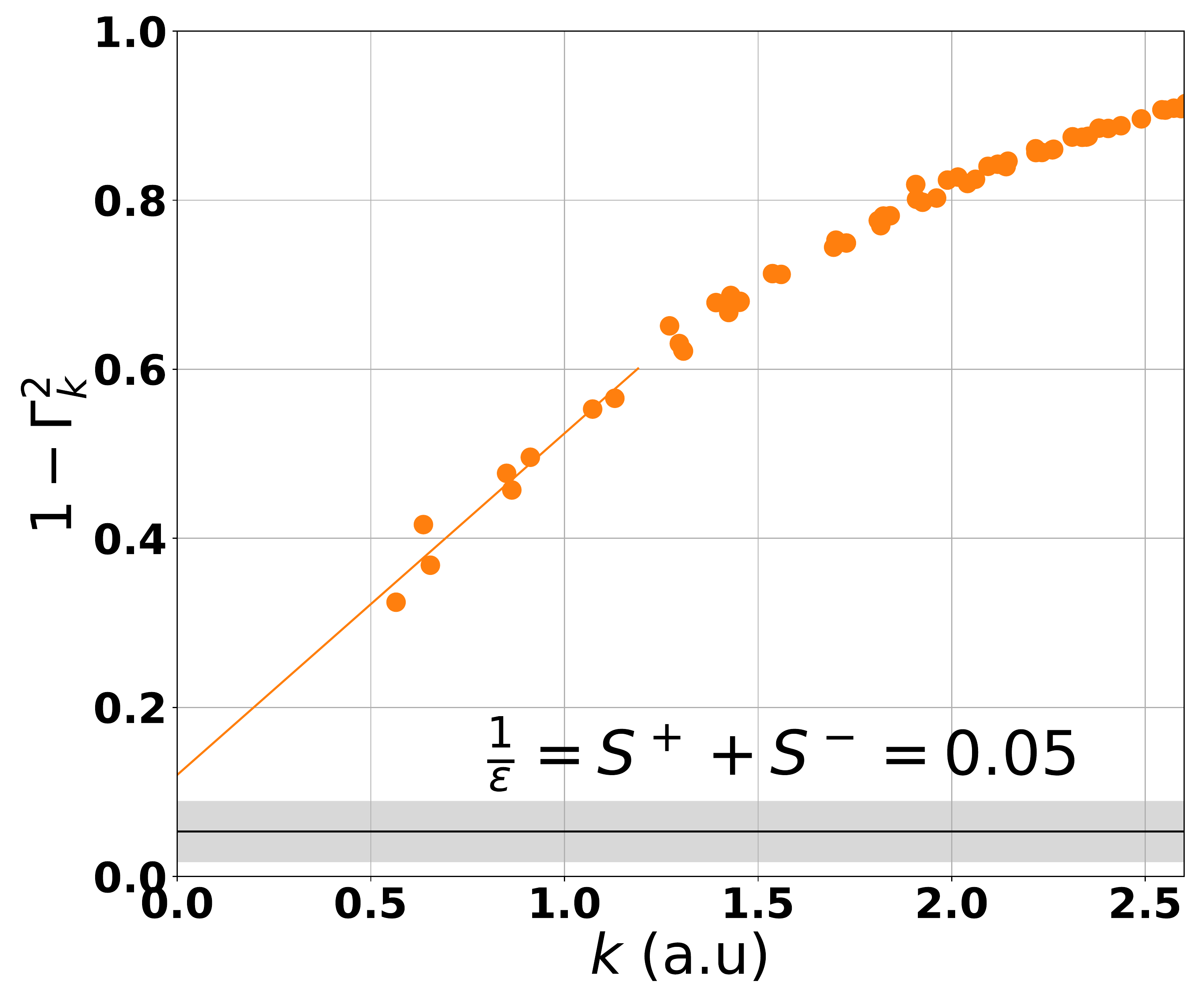}
(b) $r_s=1.38$, 234GPa
\end{minipage}
\caption{Inverse dielectric constant upper bound for solid H$_2$, where $\Gamma_k\equiv\frac{2m\omega_pS_{N_e}(k)}{\hbar k^2}$. The line is the fit to obtain asymptotic values at $k\to0$. Black line indicates the inverse dielectric constant extracted from asymptotic behavior of $S^{\pm}_k$ (Fig. \ref{fig:Sofk}) and considering constant $c=1$ (see table \ref{tab:E0_FS}). Gray shaded area is the error of the fit of $S^{\pm}_k$.
 \label{fig:Gamma}}
\end{figure*}

For hydrogen, it is difficult to access experimental information on the static dielectric constant. Therefore, we use the asymptotic values of the structure factors $S^{\pm}_k$ (Fig.~\ref{fig:Sofk}) as \textit{ab-initio} size corrections of the excitation energies. From Fig.~\ref{fig:Sofk} it can be noticed that the size effects are decreasing with increasing pressure. Fig. \ref{fig:Gamma} shows the upper bound on the inverse dielectric constant based on Eq. (6) of the main text. We see that for hydrogen the above inequality holds for both densities, however the values of inverse dielectric constants determined from the asymptotic behavior of $S^{\pm}_k$ are considerably lower that the bound. As the $S^{\pm}_k$ explicitly contain information on the excitation energies, we used it to correct the gap values. The equivalent of Fig.~\ref{fig:Sofk} and \ref{fig:Gamma} for carbon and silicon are shown in Fig.~2 and 3 in the main article.

\subsection{GCTA vs. CBM-VBM approaches to the gap}

\begin{figure}[h]
\begin{subfigure}{0.48\textwidth}
\includegraphics[width=\textwidth]{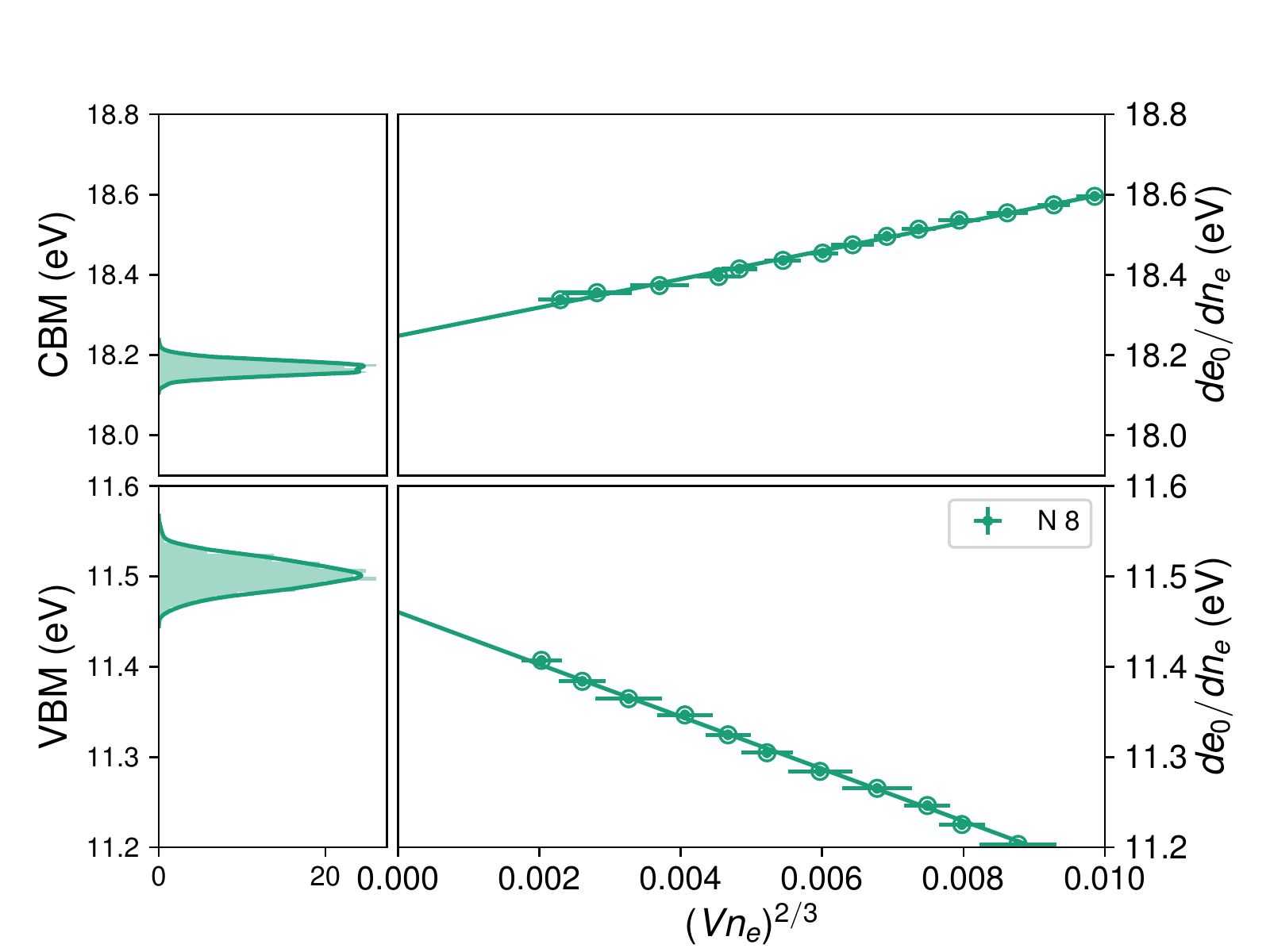}
\caption{C N8}
\end{subfigure}
\begin{subfigure}{0.48\textwidth}
\includegraphics[width=\textwidth]{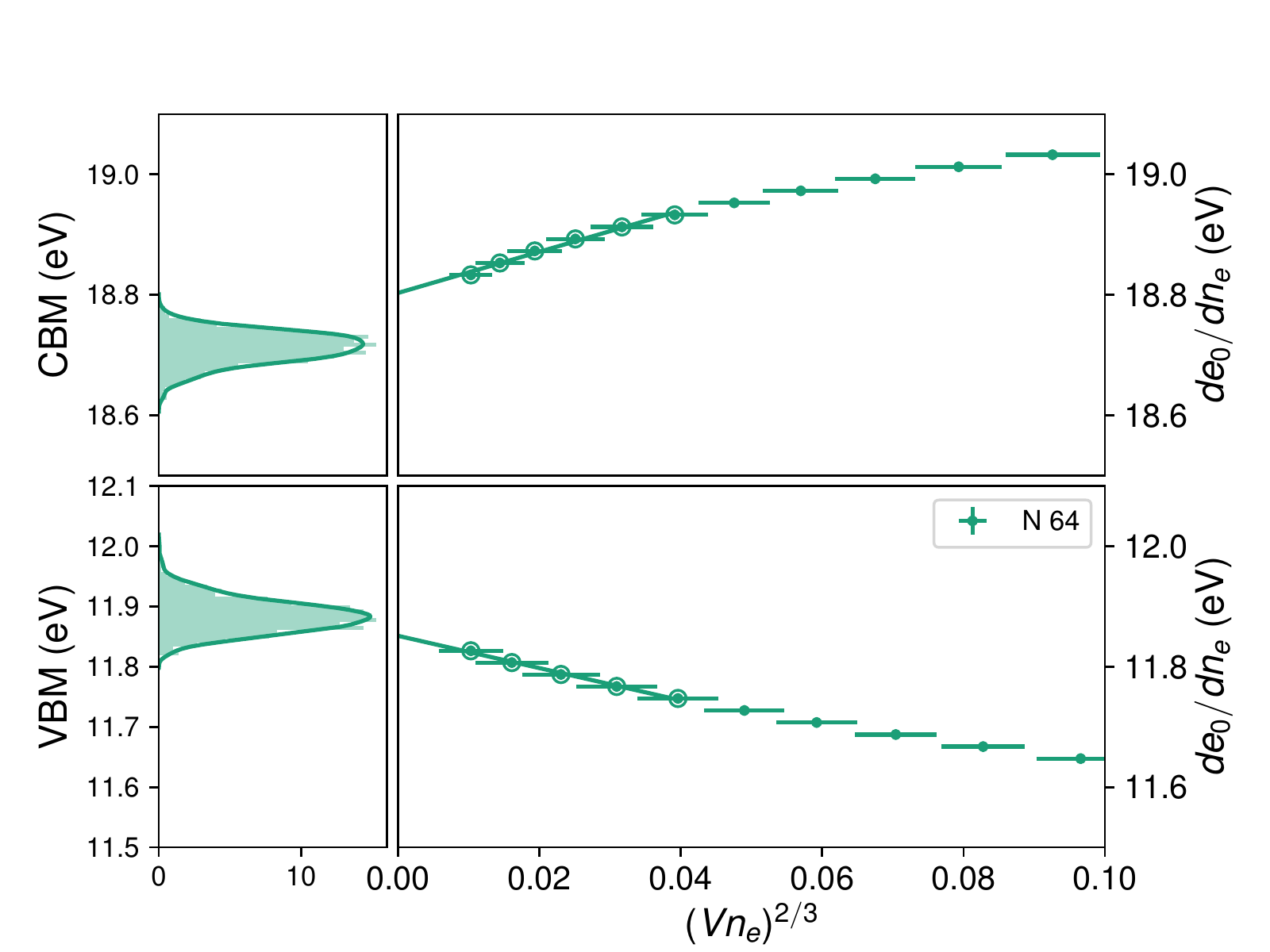}
\caption{C N64}
\end{subfigure}
\caption{\small Finite-size corrected gap of carbon from conduction band minimum (CBM), valence band maximum (CBM) and grand-canonical twist average (GCTA) discontinuity of $de_0/dn_e$. We linearly interpolate $\mu^{+/-}$ on a $64\times64\times64$ uniform grid in the Brillouin zone of the simulation cell. The left panels show the CBM and VBM of the interpolated grid. The right panels show $de_0/dn_e$ calculated on the interpolated grid. The statistical errors of DMC data are propagated using 1024 bootstrap samples.\label{fig:gcta-c}}
\end{figure}

\begin{figure}[h]
\begin{subfigure}{0.48\textwidth}
\includegraphics[width=\textwidth]{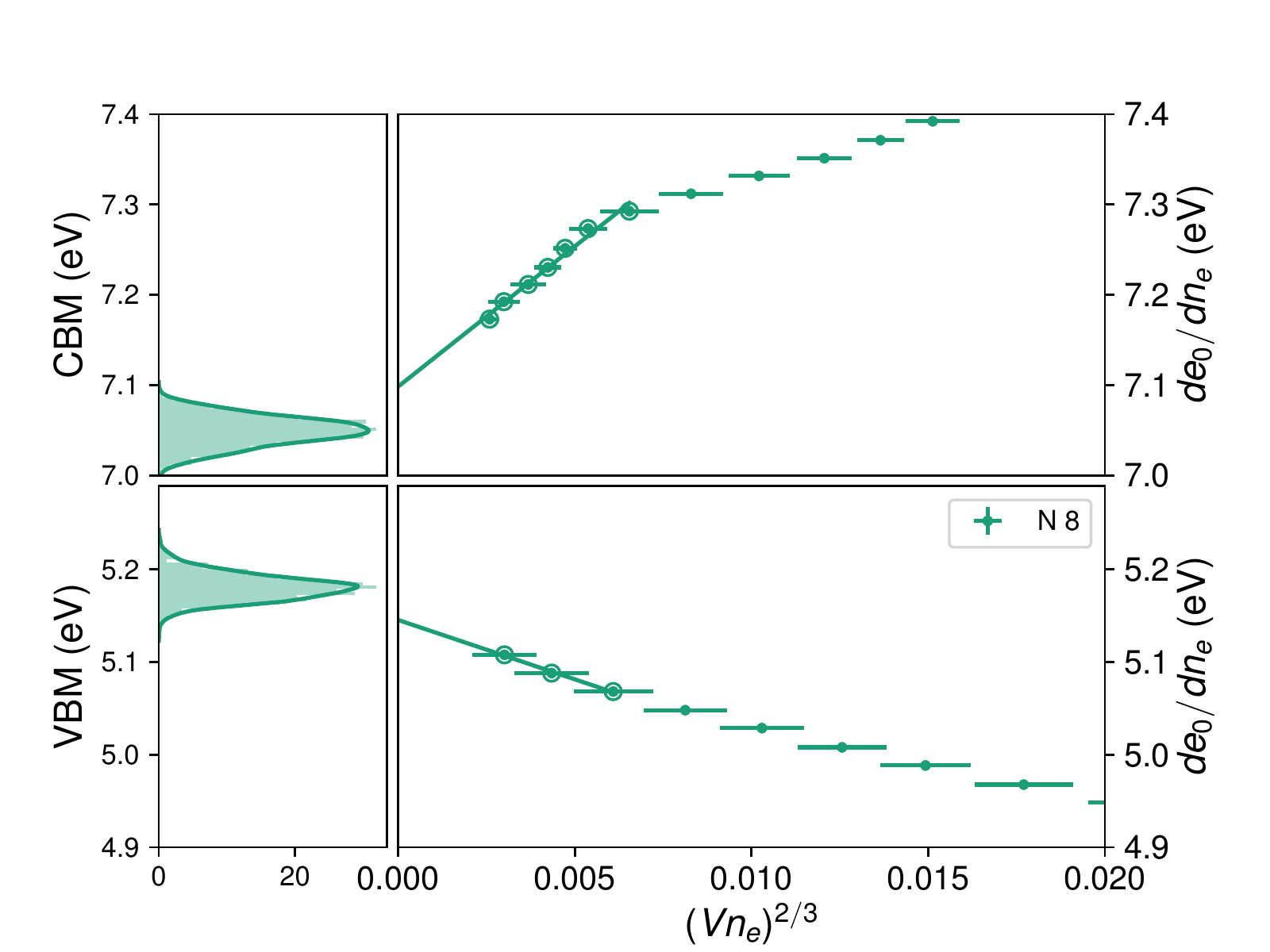}
\caption{Si N8}
\end{subfigure}
\begin{subfigure}{0.48\textwidth}
\includegraphics[width=\textwidth]{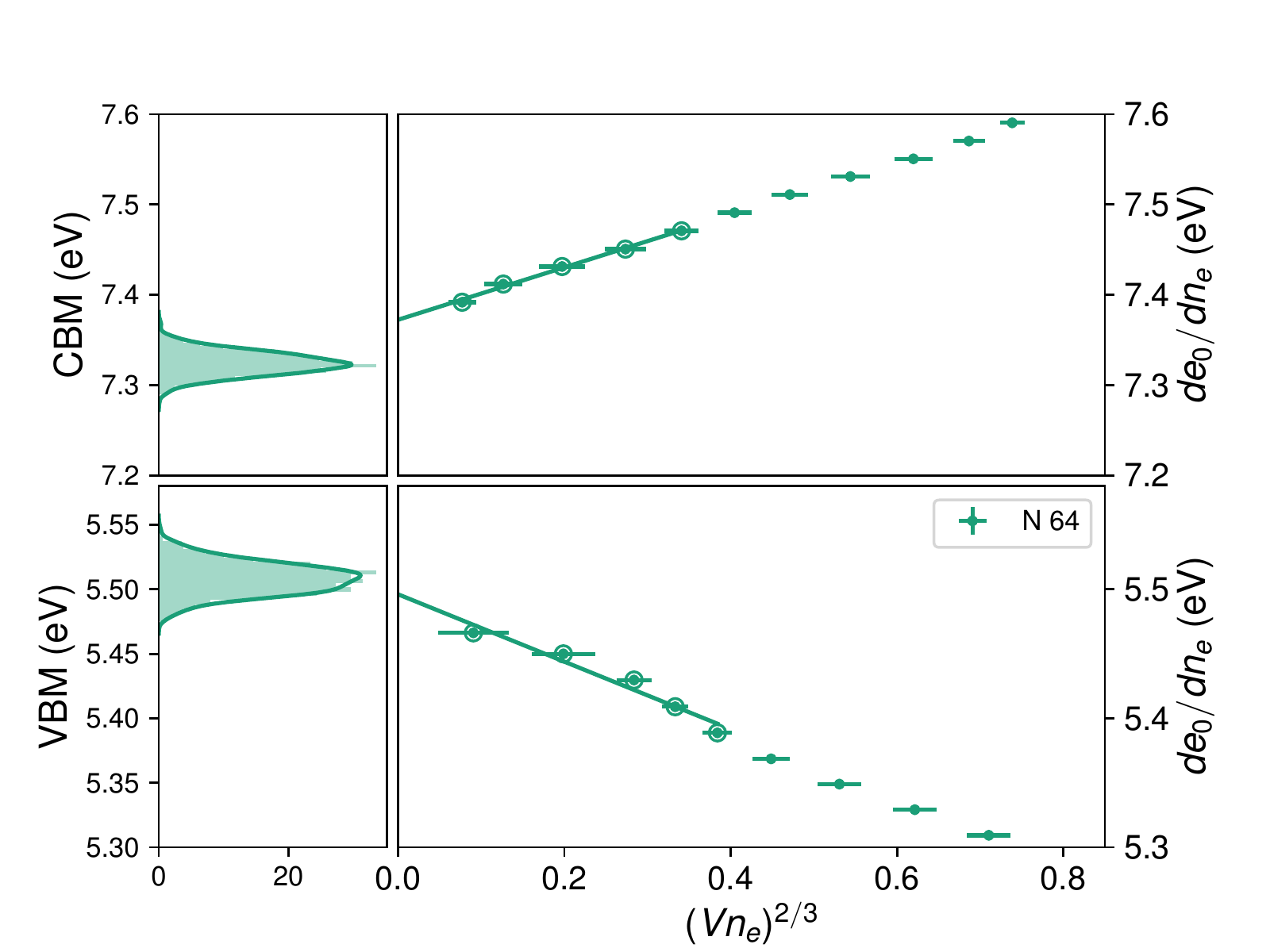}
\caption{Si N64}
\end{subfigure}
\begin{subfigure}{0.48\textwidth}
\includegraphics[width=\textwidth]{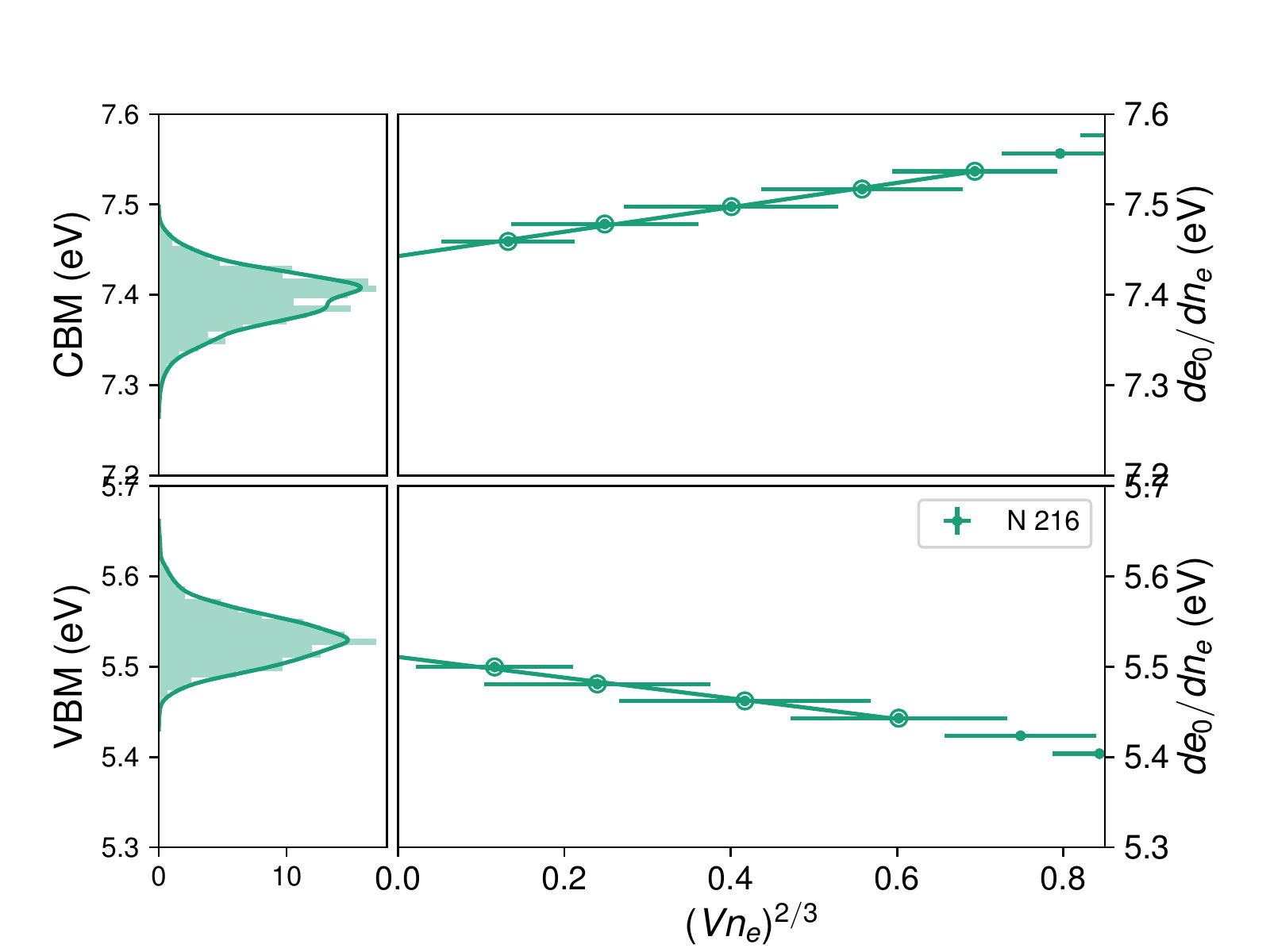}
\caption{Si N216}
\end{subfigure}
\caption{\small Finite-size corrected gap of silicon. Similar to Fig.~\ref{fig:gcta-c}. \label{fig:gcta-si}}
\end{figure}

In Fig.~\ref{fig:gcta-c} and \ref{fig:gcta-si}, we compare the GCTA approach to the traditional CBM-VBM approach to the gap. The right panels show $de_0/dn_e$ from the GCTA approach, similar to the inset of Fig.~1b in the main text. The discontinuity at zero is the GCTA estimate of the gap. The left panels show the CBM and VBM values on the grid of electron addition and removal energies ($\mu^{+/-}$). The two methods give the same results within 0.1 eV at all sizes for carbon and silicon. Therefore, we have assigned 0.1 eV as the systematic error in our estimates of the gap in Table I of the main article.

\subsection{Electron addition/removal energy tables}

Table~\ref{tab:dvs-gap} and \ref{tab:si-c-cbm-vbm} summarizes information on conduction-band minimum (CBM) and valence-band maximum (VBM) only for those k-vectors that give the fundamental gap. It is clear that the gap for both densities is indirect. More detailed information for all k-vectors for both densities is presented in tables \ref{tab:dvs-38-first} - \ref{tab:dvs-34-last}. From these tables one can notice that the energies corrections from static charge density are largest around the $\Gamma$ point (first entries of tables \ref{tab:dvs-38-first} and \ref{tab:dvs-34-first}).

\begin{table}[ht!]
\centering
\caption{Removal and addition energies ($\mu^{\pm}$) and corrections static charge density ($\mu^{\pm}_s$) for hydrogen. Shown only for those twists that give CBM-VBM gap (bold). Quantities with an overline are twist-average corrected. $\Delta_k$ is the vertical gap at twist value $k$. K-vectors of CBM and VBM are in units of $2\pi/a$.}
\label{tab:dvs-gap}
\small
\begin{tabular}{lcccccccccccc}
\toprule
 &$r_s$ & $k_x$ & $k_y$ & $k_z$ & $\mu^+$ & $\mu^-$ & $\delta\mu^+_s$ & $\delta\mu^-_s$ & $\overline{\mu}^+$ & $\overline{\mu}^-$ & $\Delta_k$ & $\overline{\Delta}_k$   \\
\midrule
H$_2$ & 1.38 & 0.062 & -0.062 & 0.062 & 10.64(12) & 6.35(16) & 0.015 & 0.066 & \textbf{10.65(12)} & 6.41(16) & 4.29(20) & 4.24(20) \\
   &1.38 & 0.437 & -0.188 & 0.312 & 11.65(11) & 6.97(16) & 0.009 & 0.006 & 11.66(11) & \textbf{6.97(16)} & 4.68(19) & 4.69(19) \\
   &1.34 & 0.063 & -0.062 & 0.063 & 11.23(14) & 7.79(19) & 0.005 & 0.045 & \textbf{11.23(14)} & 7.83(19)& 3.44(24) & 3.40(24) \\
   &1.34 & 0.063 & -0.062 & 0.437 & 11.52(14) & 8.51(20) & 0.006 & 0.013 & 11.53(14) & \textbf{8.53(20)} & 3.01(24) & 3.00(24) \\
\bottomrule
\end{tabular}
\end{table}

\begin{table}[ht!]
\centering
\caption{Removal and addition energies ($\mu^{\pm}$) and corrections static charge density ($\mu^{\pm}_s$) for carbon and silicon. Shown only for those twists that give CBM-VBM gap (bold). Quantities with an overline are twist-average corrected. $\Delta_k$ is the vertical gap at twist value $k$. K-vectors of CBM and VBM are fractional units.}
\label{tab:si-c-cbm-vbm}
\begin{tabular}{lcccccccccccc}
\toprule
 & N & $k_x$ & $k_y$ & $k_z$ & $\mu^+$ & $\mu^-$ & $\delta\mu^+_s$ & $\delta\mu^-_s$ & $\overline{\mu}^+$ & $\overline{\mu}^-$ & $\Delta_k$ & $\overline{\Delta}_k$   \\
\midrule
Si &8&  0.000 &  0.000 &  0.000 &        7.09(2) &        6.03(2) &             0.03 &            -0.54 &            7.12(2) &            \textbf{5.49(2) }&        1.06(3) &               1.63(3) \\
&8&  0.000 &  0.000 &  0.250 &        6.68(2) &        5.01(2) &             0.06 &            -0.12 &            \textbf{6.75(2)} &            4.89(2) &        1.67(3) &               1.86(3) \\
& 64 &  0.0 &  0.0 &  0.0 &        7.41(2) &        5.74(2) &             0.01 &            -0.10 &            7.42(2) &            \textbf{5.64(2)} &        1.67(3) &               1.77(3) \\
& 64 &  0.0 &  0.0 &  0.2 &        7.14(2) &        5.63(2) &             0.01 &            -0.05 &            \textbf{7.15(2)} &            5.57(2) &        1.51(2) &               1.57(2) \\
&216 &  0.0 &  0.0 &  0.0 &        7.71(4) &        5.39(4) &              0.0 &            -0.02 &            7.71(4) &            5.37(4) &        2.32(6) &               2.34(6) \\
&216 &  0.0 &  0.0 &  0.2 &        7.27(5) &        5.64(6) &              0.0 &            -0.01 &            \textbf{7.27(5)} &            \textbf{5.63(6)} &        1.63(7) &               1.65(7) \\
\midrule
C & 8  &  0.000 &  0.000 &  0.000 &       17.94(2) &       13.17(2) &            -0.05 &            -0.67 &           17.89(2) &           \textbf{12.51(2)} &        4.77(2) &               5.38(2) \\
\bottomrule
\end{tabular}
\end{table}

\begin{table}[h]
\caption{C SJ N8 gap. Conduction-band minimum (CBM) $\min(\mu^+)$=17.14(3) eV at twist 2. Valence-band maximum (VBM) $\max(\mu^-)$=13.17(2) eV at twist 0. CBM-VBM gap $\Delta=$3.97(4) eV.}
\tiny
\clearpage{}\begin{tabular}{lrrrllrrllll}
\toprule
{} &     kx &     ky &     kz &        $\mu^+$ &        $\mu^-$ &  $\delta\mu^+_s$ &  $\delta\mu^-_s$ & $\overline{\mu}^+$ & $\overline{\mu}^-$ &     $\Delta_v$ & $\overline{\Delta}_v$ \\
\midrule
0  &  0.000 &  0.000 &  0.000 &       17.94(2) &       13.17(2) &            -0.05 &            -0.67 &           17.89(2) &           12.51(2) &        4.77(2) &               5.38(2) \\
1  &  0.000 &  0.000 &  0.125 &       17.35(2) &       12.69(2) &            -0.06 &            -0.46 &           17.30(2) &           12.23(2) &        4.67(2) &               5.07(2) \\
2  &  0.000 &  0.000 &  0.250 &       17.14(3) &       11.39(3) &             0.02 &            -0.09 &           17.16(3) &           11.30(3) &        5.75(4) &               5.86(4) \\
3  &  0.000 &  0.000 &  0.375 &       17.25(1) &        9.69(2) &             0.10 &             0.13 &           17.35(1) &            9.82(2) &        7.56(2) &               7.54(2) \\
4  &  0.000 &  0.000 & -0.500 &       17.63(2) &        8.04(2) &             0.15 &             0.04 &           17.78(2) &            8.07(2) &        9.59(3) &               9.71(3) \\
5  &  0.000 &  0.125 &  0.125 &       18.13(2) &       12.74(2) &            -0.06 &            -0.37 &           18.07(2) &           12.36(2) &        5.40(2) &               5.71(2) \\
6  &  0.000 &  0.125 &  0.250 &       17.82(2) &       11.92(2) &             0.00 &            -0.11 &           17.82(2) &           11.81(2) &        5.90(2) &               6.00(2) \\
\bottomrule
\end{tabular}
\clearpage{}
\end{table}

\begin{table}[h]
\caption{C SJ N64 gap. Conduction-band minimum (CBM) $\min(\mu^+)$=18.25(2) eV at twist 2. Valence-band maximum (VBM) $\max(\mu^-)$=12.27(4) eV at twist 0. CBM-VBM gap $\Delta=$5.98(4) eV.}
\clearpage{}\begin{tabular}{lrrrllrrllll}
\toprule
{} &    kx &    ky &    kz &        $\mu^+$ &        $\mu^-$ &  $\delta\mu^+_s$ &  $\delta\mu^-_s$ & $\overline{\mu}^+$ & $\overline{\mu}^-$ &     $\Delta_v$ & $\overline{\Delta}_v$ \\
\midrule
0 &  0.00 &  0.00 &  0.00 &       18.81(5) &       12.27(4) &            -0.03 &            -0.07 &           18.78(5) &           12.19(4) &        6.54(6) &               6.59(6) \\
1 &  0.00 &  0.00 &  0.25 &       18.27(2) &       12.06(2) &            -0.00 &            -0.02 &           18.27(2) &           12.04(2) &        6.22(2) &               6.24(2) \\
2 &  0.00 &  0.00 & -0.50 &       18.25(2) &       10.69(2) &             0.01 &             0.02 &           18.26(2) &           10.71(2) &        7.57(3) &               7.55(3) \\
3 &  0.00 &  0.25 &  0.25 &       18.91(1) &       12.30(1) &            -0.01 &            -0.01 &           18.91(1) &           12.29(1) &        6.61(2) &               6.62(2) \\
4 &  0.00 &  0.25 & -0.50 &       18.78(1) &       11.58(1) &             0.01 &             0.02 &           18.79(1) &           11.59(1) &        7.21(2) &               7.20(2) \\
5 &  0.00 & -0.50 & -0.50 &       19.88(3) &       11.44(2) &             0.01 &             0.01 &           19.89(3) &           11.45(2) &        8.44(4) &               8.44(4) \\
6 &  0.25 &  0.25 &  0.25 &       18.95(2) &       11.75(2) &             0.02 &            -0.00 &           18.97(2) &           11.74(2) &        7.20(2) &               7.23(2) \\
\bottomrule
\end{tabular}
\clearpage{}
\end{table}

\begin{table}[h]
\caption{C BF N8 gap. Conduction-band minimum (CBM) $\min(\mu^+)$=17.06(2) eV at twist 2. Valence-band maximum (VBM) $\max(\mu^-)$=13.13(1) eV at twist 0. CBM-VBM gap $\Delta=$3.92(2) eV.}
\tiny
\clearpage{}\begin{tabular}{lrrrllrrllll}
\toprule
{} &     kx &     ky &     kz &        $\mu^+$ &        $\mu^-$ &  $\delta\mu^+_s$ &  $\delta\mu^-_s$ & $\overline{\mu}^+$ & $\overline{\mu}^-$ &     $\Delta_v$ & $\overline{\Delta}_v$ \\
\midrule
0  &  0.000 &  0.000 &  0.000 &       17.76(1) &       13.13(1) &            -0.05 &            -0.67 &           17.71(1) &           12.47(1) &        4.63(2) &               5.25(2) \\
1  &  0.000 &  0.000 &  0.125 &       17.25(1) &       12.66(2) &            -0.06 &            -0.46 &           17.19(1) &           12.20(2) &        4.59(3) &               4.99(3) \\
2  &  0.000 &  0.000 &  0.250 &       17.06(2) &       11.38(2) &             0.02 &            -0.09 &           17.08(2) &           11.28(2) &        5.68(2) &               5.79(2) \\
3  &  0.000 &  0.000 &  0.375 &       17.19(1) &        9.67(2) &             0.10 &             0.13 &           17.30(1) &            9.80(2) &        7.52(2) &               7.50(2) \\
4  &  0.000 &  0.000 & -0.500 &       17.54(1) &        8.20(2) &             0.15 &             0.04 &           17.69(1) &            8.24(2) &        9.34(2) &               9.46(2) \\
5  &  0.000 &  0.125 &  0.125 &       17.92(2) &       12.69(2) &            -0.06 &            -0.37 &           17.87(2) &           12.32(2) &        5.23(2) &               5.54(2) \\
6  &  0.000 &  0.125 &  0.250 &       17.73(1) &       11.90(1) &             0.00 &            -0.11 &           17.73(1) &           11.79(1) &        5.82(2) &               5.93(2) \\
\bottomrule
\end{tabular}
\clearpage{}
\end{table}

\begin{table}[h]
\caption{Si SJ N8 gap. $\delta\mu_s^{\pm}$ is finite-size correction of electron addition/removal energy from static charge density. Quantities with an overline are twist-average corrected. Conduction-band minimum (CBM) $\min(\mu^+)=6.68(2)$ eV at twist 2, valence-band maximum (VBM) $\max(\mu^-)=6.03(2)$ eV at twist 0, CBM-VBM gap $\Delta=0.65(3)$ eV.}
\tiny
\clearpage{}\begin{tabular}{lrrrllrrllll}
\toprule
{} &     kx &     ky &     kz &        $\mu^+$ &        $\mu^-$ &  $\delta\mu^+_s$ &  $\delta\mu^-_s$ & $\overline{\mu}^+$ & $\overline{\mu}^-$ &     $\Delta_v$ & $\overline{\Delta}_v$ \\
\midrule
0  &  0.000 &  0.000 &  0.000 &        7.09(2) &        6.03(2) &             0.03 &            -0.54 &            7.12(2) &            5.49(2) &        1.06(3) &               1.63(3) \\
1  &  0.000 &  0.000 &  0.125 &        6.69(2) &        5.68(2) &             0.04 &            -0.35 &            6.73(2) &            5.33(2) &        1.01(3) &               1.40(3) \\
2  &  0.000 &  0.000 &  0.250 &        6.68(2) &        5.01(2) &             0.06 &            -0.12 &            6.75(2) &            4.89(2) &        1.67(3) &               1.86(3) \\
3  &  0.000 &  0.000 &  0.375 &        6.89(2) &        4.33(2) &             0.09 &            -0.02 &            6.97(2) &            4.31(2) &        2.55(3) &               2.66(3) \\
4  &  0.000 &  0.000 & -0.500 &        7.18(2) &        3.74(2) &             0.10 &             0.00 &            7.28(2) &            3.74(2) &        3.44(3) &               3.55(3) \\
5  &  0.000 &  0.125 &  0.125 &        7.07(2) &        5.81(2) &             0.01 &            -0.31 &            7.09(2) &            5.50(2) &        1.26(3) &               1.59(3) \\
6  &  0.000 &  0.125 &  0.250 &        7.12(2) &        5.36(2) &             0.01 &            -0.08 &            7.13(2) &            5.27(2) &        1.76(3) &               1.86(3) \\
\bottomrule
\end{tabular}
\clearpage{}
\end{table}

\begin{table}[h]
\caption{Si SJ N64 gap. Conduction-band minimum (CBM) $\min(\mu^+)=7.14(2)$ eV at twist 1, valence-band maximum (VBM) $\max(\mu^-)=5.74(2)$ eV at twist 0, CBM-VBM gap $\Delta=1.40(2)$ eV.}
\clearpage{}\begin{tabular}{lrrrllrrllll}
\toprule
{} &   kx &   ky &   kz &        $\mu^+$ &        $\mu^-$ &  $\delta\mu^+_s$ &  $\delta\mu^-_s$ & $\overline{\mu}^+$ & $\overline{\mu}^-$ &     $\Delta_v$ & $\overline{\Delta}_v$ \\
\midrule
0 &  0.0 &  0.0 &  0.0 &        7.41(2) &        5.74(2) &             0.01 &            -0.10 &            7.42(2) &            5.64(2) &        1.67(3) &               1.77(3) \\
1 &  0.0 &  0.0 &  0.2 &        7.14(2) &        5.63(2) &             0.01 &            -0.05 &            7.15(2) &            5.57(2) &        1.51(2) &               1.57(2) \\
2 &  0.0 &  0.0 &  0.4 &        7.20(2) &        5.07(2) &             0.02 &            -0.00 &            7.22(2) &            5.07(2) &        2.13(3) &               2.15(3) \\
3 &  0.0 &  0.2 &  0.2 &        7.34(2) &        5.70(2) &             0.01 &            -0.03 &            7.35(2) &            5.68(2) &        1.64(3) &               1.67(3) \\
4 &  0.0 &  0.2 &  0.4 &        7.39(2) &        5.46(2) &             0.01 &             0.02 &            7.39(2) &            5.47(2) &        1.93(3) &               1.92(3) \\
5 &  0.0 &  0.4 &  0.4 &        7.64(1) &        5.54(1) &             0.01 &             0.02 &            7.65(1) &            5.55(1) &        2.10(2) &               2.09(2) \\
6 &  0.2 &  0.2 &  0.2 &        7.37(1) &        5.53(1) &             0.01 &            -0.02 &            7.39(1) &            5.51(1) &        1.84(2) &               1.88(2) \\
\bottomrule
\end{tabular}
\clearpage{}
\end{table}

\begin{table}[h]
\caption{Si SJ N216 gap. Conduction-band minimum (CBM) $\min(\mu^+)=7.27(5)$ eV at twist 1, valence-band maximum (VBM) $\max(\mu^-)=5.64(2)$ eV at twist 1, CBM-VBM gap $\Delta=1.63(7)$ eV.}
\clearpage{}\begin{tabular}{lrrrllrrllll}
\toprule
{} &   kx &   ky &   kz &        $\mu^+$ &        $\mu^-$ &  $\delta\mu^+_s$ &  $\delta\mu^-_s$ & $\overline{\mu}^+$ & $\overline{\mu}^-$ &     $\Delta_v$ & $\overline{\Delta}_v$ \\
\midrule
0 &  0.0 &  0.0 &  0.0 &        7.71(4) &        5.39(4) &              0.0 &            -0.02 &            7.71(4) &            5.37(4) &        2.32(6) &               2.34(6) \\
1 &  0.0 &  0.0 &  0.2 &        7.27(5) &        5.64(6) &              0.0 &            -0.01 &            7.27(5) &            5.63(6) &        1.63(7) &               1.65(7) \\
2 &  0.0 &  0.0 &  0.4 &        7.41(4) &        5.20(4) &              0.0 &            -0.00 &            7.41(4) &            5.20(4) &        2.21(5) &               2.21(5) \\
3 &  0.0 &  0.2 &  0.2 &        7.50(5) &        5.60(5) &              0.0 &            -0.01 &            7.50(5) &            5.59(5) &        1.90(7) &               1.90(7) \\
4 &  0.0 &  0.2 &  0.4 &        7.37(3) &        5.53(4) &              0.0 &             0.00 &            7.37(3) &            5.53(4) &        1.84(5) &               1.84(5) \\
5 &  0.0 &  0.4 &  0.4 &        7.62(4) &        5.47(4) &             -0.0 &             0.00 &            7.62(4) &            5.48(4) &        2.15(6) &               2.14(6) \\
6 &  0.2 &  0.2 &  0.2 &        7.40(5) &        5.56(4) &              0.0 &            -0.00 &            7.40(5) &            5.55(4) &        1.84(6) &               1.85(6) \\
\bottomrule
\end{tabular}
\clearpage{}
\end{table}

\begin{table}[h]
\caption{Si BF N8 gap. $\delta\mu_s^{\pm}$ is finite-size correction of electron addition/removal energy from static charge density. Quantities with an overline are twist-average corrected. Conduction-band minimum (CBM) $\min(\mu^+)=6.60(2)$ eV at twist 2, valence-band maximum (VBM) $\max(\mu^-)=6.03(1)$ eV at twist 0, CBM-VBM gap $\Delta=0.56(2)$ eV.}
\tiny
\clearpage{}\begin{tabular}{lrrrllrrllll}
\toprule
{} &     kx &     ky &     kz &         $\mu^+$ &         $\mu^-$ &  $\delta\mu^+_s$ &  $\delta\mu^-_s$ & $\overline{\mu}^+$ & $\overline{\mu}^-$ &     $\Delta_v$ & $\overline{\Delta}_v$ \\
\midrule
0  &  0.000 &  0.000 &  0.000 &         6.87(1) &         6.03(1) &             0.03 &            -0.54 &            6.90(1) &            5.49(1) &        0.83(2) &               1.41(2) \\
1  &  0.000 &  0.000 &  0.125 &         6.60(2) &         5.65(1) &             0.04 &            -0.35 &            6.64(2) &            5.30(1) &        0.94(2) &               1.34(2) \\
2  &  0.000 &  0.000 &  0.250 &         6.60(1) &         4.99(1) &             0.06 &            -0.12 &            6.67(1) &            4.87(1) &        1.62(1) &               1.80(1) \\
3  &  0.000 &  0.000 &  0.375 &         6.80(1) &         4.36(2) &             0.09 &            -0.02 &            6.88(1) &            4.34(2) &        2.43(2) &               2.54(2) \\
4  &  0.000 &  0.000 & -0.500 &         7.15(1) &         3.82(2) &             0.10 &             0.00 &            7.26(1) &            3.82(2) &        3.34(2) &               3.44(2) \\
5  &  0.000 &  0.125 &  0.125 &         6.94(1) &         5.84(1) &             0.01 &            -0.31 &            6.95(1) &            5.53(1) &        1.10(2) &               1.43(2) \\
6  &  0.000 &  0.125 &  0.250 &         7.01(1) &         5.35(1) &             0.01 &            -0.08 &            7.03(1) &            5.27(1) &        1.66(2) &               1.76(2) \\
\bottomrule
\end{tabular}
\clearpage{}
\end{table}

\begin{table}[ht!]
\centering
\caption{H$_2$ $\mathbf{r_s=1.38}$. Removal and addition energies corrections ($\mu^{\pm}_s$) from static charge density for $8\times8\times8$ k-grid shifted to exclude the $\Gamma$ point. Part I.}
\label{tab:dvs-38-first}
\tiny
\clearpage{}\begin{tabular}{cccccccccccc}
\hline\hline
  $k_x$ & $k_y$ & $k_z$ & $\mu^+$ & $\mu^-$ & $\delta\mu^+_s$ & $\delta\mu^-_s$ & $\overline{\mu}^+$ & $\overline{\mu}^-$ & $\Delta_k$ & $\overline{\Delta}_k$   \\
\hline
0.062 & 0.062 & 0.062 & 11.17(12) & 6.07(16) & 0.024 & 0.103 & 11.19(12) & 6.17(16) & 5.10(20) & 5.02(20) \\ 
0.062 & 0.062 & 0.187 & 11.19(11) & 5.86(16) & 0.008 & 0.041 & 11.20(11) & 5.91(16) & 5.32(20) & 5.29(20) \\ 
0.062 & 0.062 & 0.312 & 11.54(11) & 6.42(16) & 0.005 & 0.032 & 11.54(11) & 6.46(16) & 5.12(20) & 5.09(20) \\ 
0.062 & 0.062 & 0.437 & 10.81(12) & 6.76(16) & 0.002 & 0.015 & 10.81(12) & 6.78(16) & 4.05(20) & 4.04(20) \\ 
0.062 & 0.062 & -0.437 & 11.09(11) & 6.52(16) & 0.008 & 0.025 & 11.10(11) & 6.54(16) & 4.58(19) & 4.56(19) \\ 
0.062 & 0.062 & -0.312 & 11.66(11) & 6.23(16) & 0.009 & 0.021 & 11.67(11) & 6.25(16) & 5.42(19) & 5.41(19) \\ 
0.062 & 0.062 & -0.187 & 11.24(11) & 5.86(16) & 0.008 & 0.050 & 11.24(11) & 5.91(16) & 5.38(20) & 5.33(20) \\
\hline\hline
\end{tabular}\clearpage{}
\end{table}

\begin{table}[ht!]
\centering
\caption{H$_2$ $\mathbf{r_s=1.38}$. Part II.}
\tiny
\clearpage{}\begin{tabular}{cccccccccccc}
\hline\hline
  $k_x$ & $k_y$ & $k_z$ & $\mu^+$ & $\mu^-$ & $\delta\mu^+_s$ & $\delta\mu^-_s$ & $\overline{\mu}^+$ & $\overline{\mu}^-$ & $\Delta_k$ & $\overline{\Delta}_k$   \\
\hline
0.188 & 0.062 & 0.062 & 12.06(11) & 6.16(16) & -0.011 & 0.040 & 12.05(11) & 6.20(16) & 5.90(20) & 5.85(20) \\ 
0.188 & 0.062 & 0.187 & 12.28(11) & 6.50(16) & -0.001 & 0.027 & 12.28(11) & 6.53(16) & 5.78(20) & 5.75(20) \\ 
0.188 & 0.062 & 0.312 & 11.48(11) & 6.40(16) & -0.002 & 0.013 & 11.47(11) & 6.41(16) & 5.08(19) & 5.06(19) \\ 
0.188 & 0.062 & 0.437 & 10.95(11) & 6.76(16) & -0.003 & -0.000 & 10.95(11) & 6.76(16) & 4.18(19) & 4.18(19) \\ 
0.188 & 0.062 & -0.437 & 11.29(11) & 6.68(15) & 0.003 & 0.001 & 11.30(11) & 6.68(15) & 4.62(19) & 4.62(19) \\ 
0.188 & 0.062 & -0.312 & 11.30(11) & 6.46(16) & 0.005 & 0.011 & 11.31(11) & 6.47(16) & 4.85(19) & 4.84(19) \\ 
0.188 & 0.062 & -0.187 & 12.51(11) & 6.53(16) & 0.004 & 0.031 & 12.52(11) & 6.56(16) & 5.98(19) & 5.95(19) \\
\hline\hline
\end{tabular}\clearpage{}
\end{table}

\begin{table}[ht!]
\centering
\caption{H$_2$ $\mathbf{r_s=1.38}$. Part III.}
\tiny
\clearpage{}\begin{tabular}{cccccccccccc}
\hline\hline
  $k_x$ & $k_y$ & $k_z$ & $\mu^+$ & $\mu^-$ & $\delta\mu^+_s$ & $\delta\mu^-_s$ & $\overline{\mu}^+$ & $\overline{\mu}^-$ & $\Delta_k$ & $\overline{\Delta}_k$   \\
\hline
0.312 & 0.062 & 0.062 & 13.15(12) & 6.38(16) & -0.009 & 0.045 & 13.14(12) & 6.43(16) & 6.76(20) & 6.71(20) \\ 
0.312 & 0.062 & 0.187 & 12.44(11) & 6.47(16) & -0.006 & 0.026 & 12.44(11) & 6.50(16) & 5.97(19) & 5.94(19) \\ 
0.312 & 0.062 & 0.312 & 11.70(11) & 6.48(16) & -0.000 & 0.015 & 11.70(11) & 6.49(16) & 5.22(19) & 5.21(19) \\ 
0.312 & 0.062 & 0.437 & 11.78(11) & 6.19(16) & -0.001 & 0.021 & 11.78(11) & 6.21(16) & 5.59(19) & 5.57(19) \\ 
0.312 & 0.062 & -0.437 & 11.89(11) & 6.27(16) & 0.004 & 0.009 & 11.89(11) & 6.28(16) & 5.61(20) & 5.61(20) \\ 
0.312 & 0.062 & -0.312 & 11.64(11) & 6.36(16) & 0.003 & 0.023 & 11.64(11) & 6.38(16) & 5.28(19) & 5.26(19) \\ 
0.312 & 0.062 & -0.187 & 12.56(11) & 6.62(16) & -0.001 & 0.018 & 12.56(11) & 6.63(16) & 5.95(19) & 5.93(19) \\
\hline\hline
\end{tabular}\clearpage{}
\end{table}

\begin{table}[ht!]
\centering
\caption{H$_2$ $\mathbf{r_s=1.38}$. Part IV.}
\tiny
\clearpage{}\begin{tabular}{cccccccccccc}
\hline\hline
  $k_x$ & $k_y$ & $k_z$ & $\mu^+$ & $\mu^-$ & $\delta\mu^+_s$ & $\delta\mu^-_s$ & $\overline{\mu}^+$ & $\overline{\mu}^-$ & $\Delta_k$ & $\overline{\Delta}_k$   \\
\hline
0.437 & 0.062 & 0.062 & 13.60(11) & 6.22(16) & -0.003 & 0.014 & 13.59(11) & 6.23(16) & 7.38(20) & 7.36(20) \\ 
0.437 & 0.062 & 0.187 & 12.57(11) & 6.80(16) & -0.004 & 0.023 & 12.57(11) & 6.82(16) & 5.77(19) & 5.75(19) \\ 
0.437 & 0.062 & 0.312 & 12.03(11) & 6.95(16) & 0.001 & 0.013 & 12.03(11) & 6.96(16) & 5.08(19) & 5.07(19) \\ 
0.437 & 0.062 & 0.437 & 12.81(12) & 6.57(16) & 0.002 & 0.006 & 12.82(12) & 6.57(16) & 6.25(20) & 6.24(20) \\ 
0.437 & 0.062 & -0.437 & 12.70(12) & 6.65(16) & 0.004 & 0.009 & 12.70(12) & 6.66(16) & 6.05(20) & 6.05(20) \\ 
0.437 & 0.062 & -0.312 & 12.11(12) & 6.89(16) & -0.006 & 0.011 & 12.10(12) & 6.90(16) & 5.22(20) & 5.20(20) \\ 
0.437 & 0.062 & -0.187 & 12.47(11) & 6.84(15) & -0.004 & 0.031 & 12.46(11) & 6.87(15) & 5.63(19) & 5.60(19) \\
\hline\hline
\end{tabular}\clearpage{}
\label{tab:dvs-38-last}
\end{table}

\begin{table}[ht!]
\centering
\caption{H$_2$ $\mathbf{r_s=1.34}$. Removal and addition energies corrections ($\mu^{\pm}_s$) from static charge density for $8\times8\times8$ k-grid shifted to exclude the $\Gamma$ point. Part I.}
\label{tab:dvs-34-first}
\tiny
\clearpage{}\begin{tabular}{cccccccccccc}
\hline\hline
  $k_x$ & $k_y$ & $k_z$ & $\mu^+$ & $\mu^-$ & $\delta\mu^+_s$ & $\delta\mu^-_s$ & $\overline{\mu}^+$ & $\overline{\mu}^-$ & $\Delta_k$ & $\overline{\Delta}_k$   \\
\hline
0.063 & 0.062 & 0.063 & 11.80(14) & 7.77(20) & 0.023 & 0.064 & 11.82(14) & 7.83(20) & 4.02(24) & 3.98(24) \\ 
0.063 & 0.062 & 0.187 & 11.82(14) & 7.48(18) & 0.009 & 0.039 & 11.83(14) & 7.52(18) & 4.34(23) & 4.31(23) \\ 
0.063 & 0.062 & 0.313 & 12.33(14) & 7.72(19) & 0.019 & 0.024 & 12.35(14) & 7.75(19) & 4.61(23) & 4.60(23) \\ 
0.063 & 0.062 & 0.437 & 11.56(14) & 8.06(19) & -0.002 & 0.011 & 11.56(14) & 8.07(19) & 3.50(24) & 3.49(24) \\ 
0.063 & 0.062 & -0.437 & 11.51(14) & 8.24(20) & 0.009 & 0.016 & 11.52(14) & 8.26(20) & 3.26(24) & 3.26(24) \\ 
0.063 & 0.062 & -0.313 & 12.24(14) & 7.91(19) & 0.006 & 0.022 & 12.24(14) & 7.93(19) & 4.33(24) & 4.31(24) \\ 
0.063 & 0.062 & -0.187 & 11.77(14) & 7.27(19) & 0.002 & 0.039 & 11.78(14) & 7.31(19) & 4.50(24) & 4.46(24) \\
\hline\hline
\end{tabular}\clearpage{}
\end{table}

\begin{table}[ht!]
\centering
\caption{H$_2$ $\mathbf{r_s=1.34}$. Part II.}
\tiny
\clearpage{}\begin{tabular}{cccccccccccc}
\hline\hline
  $k_x$ & $k_y$ & $k_z$ & $\mu^+$ & $\mu^-$ & $\delta\mu^+_s$ & $\delta\mu^-_s$ & $\overline{\mu}^+$ & $\overline{\mu}^-$ & $\Delta_k$ & $\overline{\Delta}_k$   \\
\hline
0.187 & 0.062 & 0.063 & 12.53(14) & 7.84(19) & -0.003 & 0.040 & 12.52(14) & 7.88(19) & 4.69(24) & 4.64(24) \\ 
0.187 & 0.062 & 0.187 & 13.36(14) & 7.99(19) & 0.004 & 0.039 & 13.37(14) & 8.02(19) & 5.38(24) & 5.34(24) \\ 
0.187 & 0.062 & 0.313 & 12.16(14) & 7.64(19) & 0.001 & 0.017 & 12.16(14) & 7.65(19) & 4.52(24) & 4.51(24) \\ 
0.187 & 0.062 & 0.437 & 11.98(14) & 7.98(19) & -0.005 & 0.002 & 11.97(14) & 7.98(19) & 4.00(24) & 3.99(24) \\ 
0.187 & 0.062 & -0.437 & 11.94(14) & 8.11(19) & 0.003 & 0.013 & 11.95(14) & 8.12(19) & 3.84(24) & 3.83(24) \\ 
0.187 & 0.062 & -0.313 & 12.08(14) & 7.84(19) & 0.010 & 0.007 & 12.09(14) & 7.85(19) & 4.24(24) & 4.24(24) \\ 
0.187 & 0.062 & -0.187 & 13.32(14) & 7.81(19) & 0.011 & 0.044 & 13.34(14) & 7.85(19) & 5.52(24) & 5.49(24) \\
\hline\hline
\end{tabular}\clearpage{}
\end{table}

\begin{table}[ht!]
\centering
\caption{H$_2$ $\mathbf{r_s=1.34}$. Part III.}
\tiny
\clearpage{}\begin{tabular}{cccccccccccc}
\hline\hline
  $k_x$ & $k_y$ & $k_z$ & $\mu^+$ & $\mu^-$ & $\delta\mu^+_s$ & $\delta\mu^-_s$ & $\overline{\mu}^+$ & $\overline{\mu}^-$ & $\Delta_k$ & $\overline{\Delta}_k$   \\
\hline
0.312 & 0.062 & 0.063 & 13.88(14) & 7.59(19) & 0.002 & 0.035 & 13.89(14) & 7.63(19) & 6.29(24) & 6.26(24) \\ 
0.312 & 0.062 & 0.187 & 12.98(14) & 8.43(19) & -0.007 & 0.021 & 12.97(14) & 8.45(19) & 4.55(23) & 4.52(23) \\ 
0.312 & 0.062 & 0.313 & 12.25(14) & 7.87(19) & 0.001 & 0.010 & 12.25(14) & 7.88(19) & 4.38(24) & 4.37(24) \\ 
0.312 & 0.062 & 0.437 & 12.65(14) & 7.58(19) & -0.002 & 0.010 & 12.64(14) & 7.59(19) & 5.06(24) & 5.05(24) \\ 
0.312 & 0.062 & -0.437 & 12.38(14) & 7.75(19) & 0.003 & 0.013 & 12.38(14) & 7.76(19) & 4.63(24) & 4.62(24) \\ 
0.312 & 0.062 & -0.313 & 12.17(14) & 7.86(20) & 0.004 & 0.020 & 12.18(14) & 7.88(20) & 4.31(24) & 4.29(24) \\ 
0.312 & 0.062 & -0.187 & 13.30(14) & 8.11(19) & 0.005 & 0.031 & 13.30(14) & 8.14(19) & 5.19(24) & 5.16(24) \\
\hline\hline
\end{tabular}\clearpage{}
\end{table}

\begin{table}[ht!]
\centering
\caption{H$_2$ $\mathbf{r_s=1.34}$. Part IV.}
\tiny
\clearpage{}\begin{tabular}{cccccccccccc}
\hline\hline
  $k_x$ & $k_y$ & $k_z$ & $\mu^+$ & $\mu^-$ & $\delta\mu^+_s$ & $\delta\mu^-_s$ & $\overline{\mu}^+$ & $\overline{\mu}^-$ & $\Delta_k$ & $\overline{\Delta}_k$   \\
\hline
0.438 & 0.062 & 0.063 & 14.63(14) & 7.54(20) & -0.010 & 0.016 & 14.62(14) & 7.55(20) & 7.10(24) & 7.07(24) \\ 
0.438 & 0.062 & 0.187 & 13.32(14) & 8.24(19) & 0.005 & 0.021 & 13.32(14) & 8.26(19) & 5.08(23) & 5.07(23) \\ 
0.438 & 0.062 & 0.313 & 12.80(14) & 8.38(19) & 0.002 & 0.027 & 12.80(14) & 8.40(19) & 4.43(24) & 4.40(24) \\ 
0.438 & 0.062 & 0.437 & 13.41(14) & 8.30(20) & 0.005 & 0.015 & 13.42(14) & 8.32(20) & 5.11(24) & 5.10(24) \\ 
0.438 & 0.062 & -0.437 & 13.54(14) & 8.12(19) & 0.003 & 0.010 & 13.55(14) & 8.13(19) & 5.42(24) & 5.42(24) \\ 
0.438 & 0.062 & -0.313 & 12.71(14) & 8.36(19) & 0.001 & 0.007 & 12.72(14) & 8.36(19) & 4.36(24) & 4.35(24) \\ 
0.438 & 0.062 & -0.187 & 13.04(14) & 8.28(19) & -0.005 & 0.009 & 13.03(14) & 8.29(19) & 4.76(23) & 4.75(23) \\
\hline\hline
\end{tabular}\clearpage{}
\label{tab:dvs-34-last}
\end{table}

\FloatBarrier